\documentclass[12pt,preprint,a4paper]{article}

\usepackage{amsmath}
\usepackage{amssymb}
\usepackage{mathrsfs}
\usepackage{graphicx,subcaption}
\usepackage{dsfont}
\usepackage{cite} 
\usepackage[colorlinks=true,linkcolor=black,citecolor=black,urlcolor=black]{hyperref}
\usepackage{enumerate}
\usepackage{epstopdf}

\newcommand{\df}[1]{\boldsymbol{#1}}

\renewcommand\d{\partial}

\begin{document}

\title{\begin{flushright}\vspace{-1in}
       \end{flushright}
       \vskip 20pt
Galilean Geometry in Condensed Matter Systems}

\date{\today}

\author{
Michael Geracie
\thanks{\href{mailto:mgeracie@uchicago.edu}         {mgeracie@uchicago.edu}}  \\
   {\it Kadanoff Center for Theoretical Physics,}\\
   {\it   University of Chicago, Chicago, IL 60637 USA}
} 

\maketitle

\begin{abstract}
We present a systematic means to impose Galilean invariance within field theory. We begin by defining the most general background geometries consistent with Galilean invariance and then turn to applications within effective field theory, fluid dynamics, and the quantum Hall effect.
\end{abstract}

~ \\
~ \\

\begin{center}{\it \large
A dissertation submitted to \\
the faculty of the Division of the Physical Sciences \\
in cadidacy for the degree of \\
Doctor of Philosophy }
\end{center}

\newpage
\section*{Acknowledgments}
This thesis wouldn't exist without the care and mentorship of my advisor, Dam Son, whose help over the years has been so crucial to my development as a physicist. My sincerest thanks also go out to my collaborators, Matt Roberts and Kartik Prabhu, whose work is no small part of this thesis. David Schuster has also been of immense help through the years in helping me to find a place and a research direction I am happy with, and his approachability and encouragement will be greatly missed.

Most importantly I would like to acknowledge the support of my family and dear friends, who make the work and the effort worthwhile, and who I couldn't hope to all list here. However, I would most especially like to acknowledge my parents, Frank and Mary Ann, for their excitement and support, my sister Danielle, for all the fun we've had together, my trombone instructor of many years, Mark Hoelscher, for teaching me the important life lessons that lead to success inside and outside the practice room, and finally, my dearest friends Phil Bridge, Dale Cheng, and Jennifer Lin. I had so much fun talking about physics with you guys.

\newpage
\tableofcontents
\newpage

\section{Introduction}\label{chap:Intro}

Spacetime symmetries are among the most powerful principles in physics. The prime example, Lorentz invariance, is pervasive in high energy physics and is efficiently implemented within the framework of pseudo-Riemannian geometry. Recent years have shown a great deal of activity in trying to impose similar restrictions in condensed matter settings. In this case however, relativistic symmetry is generically not available.\footnote{Among the most notable exceptions is of course the low energy theory graphene, which supports a Dirac mode with a relativistic dispersion relation.}  Rather Galilean invariance is the relevant symmetry of the problem.

As one might expect from the relativistic example, applying Galilean invariance has proven quite fruitful.
Early work by Greiter, Witten, and Wilczek \cite{Greiter:1989qb} implemented Galilean invariance by demanding the equality of the momentum and charge currents
\begin{align}\label{wittenGalInv}
	\tau^{0 i} = \frac{m}{e} j^i 
\end{align}
to restrict the low energy effective theory of superconductors, including a proof of the London Hall effect.

The subject found renewed interest following work by Son and Wingate \cite{Son:2005rv}, who demonstrated that Galilean invariance dramatically reduced the number of phenomenological parameters allowed in the effective theory of the unitary Fermi gas. A flurry of activity resulted over the next decade in which their method was applied especially to quantum Hall physics. In \cite{Hoyos:2011ez} it was demonstrated that the $| \mathbf q |^2$ part of the Hall conductivity $\sigma_H ( \omega , \mathbf q )$ is determined entirely by the shift $\mathcal S$ and the equation of state.\footnote{In an alternate approach, this was shown later \cite{Bradlyn:2012ea} to be the result of non-relativistic Ward identities. As we shall see, these follow from coordinate invariance on a background with Galilean symmetry.} A similar analysis in \cite{Gromov:2014gta} found the chiral central charge could be determined by measuring the density response to applied curvature. 
Non-relativistic symmetry in conjunction with the lowest Landau level limit was used to construct a dynamical theory of the quantum Hall effect in \cite{Son:2013} and used to demonstrate universal behavior of the polarization tensor up to order $| \mathbf q |^2$. This program was furthered in \cite{Golkar:2013gqa}, which proved a number of spectral sum rules for the traceless part of the stress tensor. Newton-Cartan geometry also features prominently in the work of \cite{Carter:1993aq,Carter:2003im,Carter:2004fw,Carter:2003qn} on the superfluid dynamics of neutron stars.
In short, Galilean invariance has a number of powerful consequences and it's worth understanding how to efficiently implement it.

An approach to Galilean invariance exists in the general relativity literature, pioneered by Cartan in the 1920's with the goal of providing a geometric picture of Newtonian gravity a la general relativity, now called Newton-Cartan geometry \cite{Cartan:1923zea,Cartan:1924yea}. We give a brief overview of this subject in section \ref{sec:NewtonCartanGeo}. For applications to non-relativistic supergravity theories, see \cite{Andringa:2013mma,Bergshoeff:2015ija,Bergshoeff:2014gja,Knodel:2015byb}. Though used in varying degrees by the condensed matter references listed above, it is never adopted exclusively since this approach is poorly adapted to effective field theory in a few respects.

The first is that the formalism, as presented by Cartan, is not convenient for the construction of invariant actions, a matter that we will discuss later in this introduction. The second is that, as we shall see, torsion is an essential ingredient in a number of condensed matter applications, particularly fluid dynamics and linear response theory. Cartan's approach did not involve torsion because it did not need to for the task of describing Newtonian gravity. Furthermore, in contrast to the relativistic case, it also proved surprisingly difficult to adapt to a torsionful setting. We will discuss both of these matters in detail in section \ref{sec:NewtonCartanGeo}.

The point of this thesis is to present a reformulation of Cartan's work that easily incorporates torsion in a Galilean invariant way and is well adapted to applications in condensed matter physics, allowing for automatic implementation of Galilean invariance within field theory in the same manner that Lorentz invariance is typically realized in the relativistic case. In their common domain of application this formalism reduces entirely to that articulated by Cartan. The formalism, which we call the extended vielbein formalism, is similar to the first-order formulation of pseudo-Riemannian geometry, the key difference being that the vielbein does not lie in the defining representation of gauge group, but rather a higher-dimensional representation which we call the extended representation. The vielbein then contains additional information beyond a basis for the tangent space. This information precisely encodes Newtonian gravitational effects.

We present this formalism in chapter \ref{chap:Formalism}, which principally draws from the work of this author and collaborators in \cite{GPR_fluids,GPR_geometry}. The rest of the thesis is then dedicated to demonstrating how it is applied in practice. Chapter \ref{chap:Ward} defines the set of currents available in any non-relativistic theory and the Ward identities they satisfy by virtue of coordinate and Galilean invariance. This treatment collects the non-relativistic Ward identities found in \cite{Son:2008ye,Andreev:2013qsa} into a single, manifestly Galilean covariant presentation and generalizes them to the curved, torsionful backgrounds established in chapter \ref{chap:Formalism}, a step that will be essential for the fluid analyses later on. The material of chapter \ref{chap:Ward} was developed over the course of several papers, beginning with \cite{Geracie:2014nka}, which presented the Ward identities for the first time in a manifestly diffeomorphism covariant form but did not deal with Galilean invariance. A manifestly Galilean invariant formulation followed in \cite{GPR_fluids}, which was then extended to the spinful case, with careful attention paid to the difference between the Cauchy and unphysical currents, in \cite{GPR_improv}.

Chapter \ref{chap:Examples} gives simple examples, most of which also appear in \cite{GPR_improv}, of how to construct Galilean invariant field theories and ends with a general effective action for quantum Hall systems that encodes response to an unrestricted Bargmann geometry. The additional freedom this provides gives us access to new quantized coefficients characterizing the topological phase that have not previously been considered. Chapter \ref{chap:Apps} concludes with a number of applications within non-relativistic fluid dynamics, including linear response theory and the quantum Hall effect, drawing from material in \cite{Geracie:2014zha,GPR_fluids}.

Before we begin, a few words are due regarding the relevance of Galilean invariance in condensed matter. We have seen that it has a number of interesting an powerful consequences, but to be clear, Galilean invariance is often broken in the condensed matter setting by a variety of sources such as lattice effects and disorder. This however does not diminish the utility of a formalism that makes the Galilean invariance of the spacetime itself manifest. One may simply add the relevant Galilean breaking effects as additional data. In this thesis we will usually restrict ourselves to idealized systems with true Galilean invariance as a first step, in some cases breaking it with a finite temperature and chemical potential and/or applied electromagnetic fields as the need arises, keeping this in mind. Additional Galilean breaking effects may be considered on top of this if desired, but we will not consider them.

Our index conventions will be as follows. Greek letters $\mu , \nu , \dots$ represent spacetime indices and can take on either temporal or spatial values $t, i , j, \dots$. Indices that transform in the vector and covector representation of $SO(d)$ are denoted with lower case Latin letters $a, b, \dots$ while those in the vector or covector representation of $Gal(d)$ use upper case Latin letters $A,B, \dots$ which can take on temporal or spatial values $0, a, b, \dots$. Indices in the extended or dual extended representations of $Gal(d)$ are denoted $I,J, \dots$. We will often convert between coordinate indices and internal Galilean indices using the frame $e^\mu_A$ or coframe $e^A_\mu$ without note.

\subsection{Newton Cartan-Geometry}\label{sec:NewtonCartanGeo}
We begin with a brief overview of Newton-Cartan geometry and then motivate the need to reformulate it. Our treatment is a much condensed version of Malament \cite{Malament} which we refer the reader to for details and a complete list of references.

A Newton-Cartan geometry is a $(d+1)$-dimensional manifold $\mathcal M$ with a closed non-vanishing one-form $n_\mu$, a positive semidefinite symmetric tensor $h^{\mu \nu}$ whose kernel is spanned by $n_\mu$
\begin{align}\label{NC1}
	h^{\mu \nu} n_\nu = 0,
\end{align}
and a derivative operator $\nabla$ compatible with both structures
\begin{align}\label{compatibility}
	\nabla_\mu n_\nu = 0,
	&&\nabla_\lambda h^{\mu \nu} = 0 .
\end{align} 
This data has the following interpretation. The one-form $\df n$ determines the rate at which clocks tick and is called the clock form. An observer traversing a path $c$ between two events measures an elapsed time
\begin{align}\label{elapsedTime}
	\Delta t = \int_c \df n 
\end{align}
which all such observers agree upon since $\df n$ is closed. It also determines a temporal orientation of $\mathcal M$. Those vectors $\xi^\mu$ for which $\xi^\mu n_\mu >0$ are said to be future directed timelike and those for which $\xi^\mu n_\mu <0$ are past directed timelike.\footnote{See Malament for a discussion of non-time orientable Newton-Cartan geometries \cite{Malament}. We shall have no cause to consider them.}
Since $\df n$ is closed and non-vanishing, it is at least locally derived from a local time function $\df n = dt$ so that we have a preferred time coordinate that foliates the manifold into spatial slices.\footnote{We shall always assume our spacetime is simply connected so that this may be taken to be a global time function.} Of course, the data contained in $\df n$ and $t$ is equivalent, but we'll soon need to relax the condition $d \df n =0$, in which case the only notion of elapsed time is given by (\ref{elapsedTime}) and is path dependent.

The tensor $h^{\mu \nu}$ serves as a spatial metric in the following sense. A vector $\xi^\mu$ is said to spatial if $\xi^\mu n_\mu = 0$ and the set of spatial vectors form a subspace $T^s_p \mathcal M$ of the tangent space at a point. Any spatial vector can be written as a one-form ``raised'' with $h^{\mu \nu}$, though this expression is not unique.
\begin{align}\label{ambiguity}
	\xi^\mu = h^{\mu \nu} \mu_\nu = h^{\mu \nu} ( \mu_\nu + f n_\nu ) ,
	&&\text{for any function} ~ f .
\end{align}
One then defines the length of $ \xi \in T^s_p \mathcal M$ as
\begin{align}
	|| \xi || = ( h^{\mu \nu} \mu_\mu \mu_\nu)^{\frac 1 2 }
\end{align}
which is not affected by the ambiguity (\ref{ambiguity}).

Why does one not simply use a metric with lowered indices? A rank-$d$ symmetric matrix with lowered indices defines a unique non-vanishing vector field $v^\mu$ with zero eigenvalue. This is most certainly not what we want in a theory of non-relativistic gravity as it defines a preferred class of observers with velocity $v^\mu$ and would explicitly break the principle of Galilean relativity. In what follows we must be careful not to at any point invoke a preferred vector field $v^\mu$ unless it is provided for by additional physical data or we can demonstrate that our results do not depend on that choice.

The connection $\nabla$ defines a notion of parallel transport and, as one would expect, encodes gravity. However, the manner in which it does so is qualitatively different from the relativistic case. In general relativity, the connection is totally equivalent to the data found in the metric as there is a single torsion free, metric compatible connection. On the contrary, there are many connections satisfying (\ref{compatibility}) at fixed torsion. Unlike GR, gravity has nothing to do with the metric data $n_\mu$ and $h^{\mu \nu}$, but is instead contained in the unfixed part of the connection.

For the time being, we consider only the torsionless case. Any two derivative operators $\nabla$ and $\nabla'$ differ by a tensor field $C^\lambda{}_{\mu \nu} = C^\lambda{}_{( \mu \nu )}$. For example
\begin{align}
	(\nabla_\mu - \nabla'_\mu) \xi^\nu = C^\nu{}_{\mu \lambda} \xi^\lambda .
\end{align}
For both to be compatible with (\ref{compatibility}), this tensor must be of the form
\begin{align}
	C^\lambda{}_{\mu \nu} = n_{(\mu} \kappa_{\nu )}{}^{\lambda}
\end{align}
where $\kappa_{\mu \nu}$ is a two-form whose index we have raised with $h^{\mu \nu}$. We will often use this convention, though one should be careful since $h^{\mu \nu}$ is not invertible. An index, once raised, cannot be lowered and many standard manipulations in the pseudo-Riemannian case are forbidden.

To show how this is related to gravity, we for the moment break covariance to follow the textbook treatment by Misner, Thorne, and Wheeler which we suspect is most familiar to readers \cite{MTW}. Consider a Newton-Cartan geometry with a flat clock form and spatial metric in the sense that there exists a coordinate system where
\begin{align}\label{flatBackground}
	n_\mu =
	\begin{pmatrix}
		1 & 0 \\
	\end{pmatrix},
	&&
	h^{\mu \nu} = 
	\begin{pmatrix}
		0 & 0 \\
		0& \delta^{ij}
	\end{pmatrix} .
\end{align}
The $\Gamma^k{}_{t\mu}$ components of the Christoffel symbol are left undetermined by metric compatibility. If we declare that they are of the form
\begin{align}\label{MTWChristoffel}
	\Gamma^k{}_{tt} = \partial^k \phi,
	&&\Gamma^k{}_{ti} = 0,
\end{align}
then the geodesic equation reads
\begin{align}
	\xi^\nu \nabla_\nu \xi^\mu = 0,
	&&\iff
	&& \ddot x^i = - \partial^i \phi .
\end{align}
Here $\xi^\mu = \frac{d x^\mu}{dt}$ is the velocity vector of a causal curve (affinely parametrized in the sense $n_\mu \xi^\mu = 0$).
That is, free fall in this Newton-Cartan geometry is equivalent to motion in a Newtonian gravitational potential. In the presence of a matter distribution $\rho$, the potential is fixed by the Poisson equation $\partial^2 \phi = 4 \pi \rho$. One may check using (\ref{MTWChristoffel}) that this is equivalent to a non-relativistic version of the Einstein equations in which stress energy is dominated by the mass density $\rho$
\begin{align}
	R_{00} = 4 \pi \rho ,
\end{align}
with all other components of $R_{\mu \nu}$ equal to zero.

The covariant version of this is the Trautman recovery theorem \cite{Malament}, which says that, under certain conditions, a Newton-Cartan geometry may always be thought of as encoding a Newtonian gravitational potential.

{\bf Theorem:}
	Let $( \mathcal M , n_\mu , h^{\mu \nu} , \nabla )$ be a Newton-Cartan geometry that satisfies the curvature conditions
	\begin{gather}
		R_{\mu \nu} = 4 \pi \rho n_\mu n_\nu , \\
		R^\mu{}_\nu{}^\lambda{}_\rho = R^\lambda{}_\rho{}^\mu{}_\nu , \\
		R^{\mu \nu}{}_{\lambda \rho} = 0 .
	\end{gather}
	For each point $p \in \mathcal M$ there exists a neighborhood $O$ of $p$ and a derivative operator $\nabla'$ and scalar field $\phi$ on $O$ such that
	\begin{enumerate}
		\item $\nabla'$ is compatible with $n_\mu$ and $h^{\mu \nu}$.
		\item $\nabla'$ is flat $R'^\lambda{}_{\rho \mu \nu} = 0$.
		\item For causal curves with velocity $\xi^\mu$
		\begin{align}
			\xi^\nu \nabla_\nu \xi^\mu = 0
			&& \iff
			&& \xi^\nu \nabla'_\nu \xi^\mu = - \nabla'^\mu \phi .
		\end{align}
		\item $\phi$ satisfies the Poisson equation
		\begin{align}
			\nabla'^2 \phi = 4 \pi \rho .
		\end{align}
	\end{enumerate}

The derivative operator $\nabla$ is said to be a geometrization of $(\nabla', \phi )$. If $\rho$ has compact support and we require that $\phi$ go to zero at spatial infinity, then $\nabla$ is in fact the geometrization of a unique Newtonian potential.
Before moving on, we note that while the above discussion may make it seem as though flatness is essential for an interpretation of Newton-Cartan geometry as a geometrization of Newtonian gravity, this is not the case. Even the torsionful geometries we will construct using the extended vielbein formalism can be interpreted in this way, though additional forces will also be present.

\subsection{Action Principles}\label{sec:ActionPrinciples}

This formalism as presented is completely sufficient for the task it was designed for, that of describing non-relativistic gravity in a covariant fashion and that which we propose will reduce to it in the torsionless case. However, for condensed matter applications we will find it at best maladapted and at worst inadequate. Let's take a minute to summarize the issues, beginning with it's use in effective field theory.

One of the advantages of a geometric formulation of relativistic kinematics is it's utility in constructing action principles consistent with Lorentz invariance. Any term constructed by contracting fields and their derivatives with the metric is admissible. Spinor fields of course require reference to a vielbein and spin connection, but this is well understood and there is no essential complication. In contrast, there is no manifestly invariant form for even the simplest example of a non-relativistic field theory, the Schr\"{o}dinger equation
\begin{align}\label{SchrodL}
	\mathcal L =  \frac i 2 \psi^\dagger \overset{\leftrightarrow} \partial_t \psi - \frac{\delta^{ij}}{2m} \partial_i \psi^\dagger \partial_j \psi - m \phi \psi^\dagger \psi ,
\end{align}
that refers only to the data presented in (\ref{NC1}) and (\ref{compatibility}). The Schr\"odinger Lagrangian is not  invariant under boosts\footnote{In presenting the standard textbook treatment of Galilean invariance for the Schr\"odinger equation we have not been careful to distinguish internal boosts and diffeomorphisms, a mistake we will soon rectify.}
\begin{align}
	t' = t,
	&&x'^i = x^i - k^i t,
\end{align}
unless one also specifies that $\psi$ transforms projectively, sending plane waves of momentum $p^i$ to those of momentum $p^i - m k^i$
\begin{align}\label{projTransf}
	\psi' (t', x') = e^{i \frac 1 2 m k^2 t - i m k_i x^i} \psi (t, x) .
\end{align} 
In this treatment then, Galilean invariance relies on a careful cancellation between the time derivative and kinetic energy terms under a projectively realized boost transformation.

The generalization of this treatment to curved spacetimes is perhaps somewhat less familiar
\begin{align}\label{CovSchrodL}
	\mathcal L = \frac i 2 v^\mu \psi^\dagger \overset{\leftrightarrow} D_\mu \psi - \frac{h^{\mu \nu}}{2m} D_\mu \psi^\dagger D_\nu \psi,
	&&\text{where}
	&&D_\mu = \nabla_\mu - i m a_\mu .
\end{align}
Here we have subsumed the Newtonian potential into a Newtonian gravitational vector potential $a_\mu$ whose temporal component is $- \phi$. This Lagrangian also appears to break Galilean invariance since it refers to a preferred future-directed timelike vector field $v^\mu$ (again, normalized so that $n_\mu v^\mu = 1$), which is not available in the Newton-Cartan data. However, the action does not depend on this choice if we demand that $a_\mu$ transforms when we make a different selection of $v^\mu$
\begin{align}\label{aTransf}
	v^\mu \rightarrow v^\mu + k^\mu ,
	&&a_\mu \rightarrow a_\mu + k_\mu - \frac 1 2 k^2 n_\mu ,
\end{align}
where $k_\mu$ is an arbitrary one-form such that $v^\mu k_\mu = 0$. Note that in this picture, $\psi$ does not transform at all under boosts. To our knowledge, this treatment of boost invariance, as well as the diffeomorphism invariant form of the Lagrangian (\ref{CovSchrodL}), first appeared in \cite{DK}.

Since $a_\mu$ has $d$ components that are pure gauge, it in fact contains no more information than the Newtonian potential and we shall call the gauge in which $\df a = - \phi \df n$ Newtonian gauge.\footnote{Note that this gauge is not $U(1)_M$ gauge invariant since $\df a \rightarrow \df a + d f$.} Moreover, we see in this treatment that the coupling of a massive field to gravity follows directly from Galilean invariance, which further implies that the inertial mass (the mass the appears in the denominator of the kinetic energy) is equal to gravitational mass (the mass with which the density couples to $\phi$), which is rather satisfying. 

As one would hope, the two approaches just given are equivalent in the case of flat metric data. In this case, one can choose a coordinate system such that (\ref{CovSchrodL}) reduces to (\ref{SchrodL}) by selecting coordinates so that
\begin{align}
	v^\mu = \left( \frac{\partial}{\partial t}\right)^\mu,
	&&h^{\mu \nu} = \delta^{ij} \left( \frac{\partial}{\partial x^i}\right)^\mu \otimes  \left( \frac{\partial}{\partial x^j}\right)^\mu ,
	&&a_\mu = - \phi (dt)_\mu.
\end{align}
If one made a different selection $v'^\mu = v^\mu + k^\mu$, then one must select different coordinates
\begin{align}
	t'=t,
	&&x'^i = x^i - k^i t
\end{align}
to preserve the form of $v^\mu$ and $h^{\mu \nu}$. Of course there is now a new mass gauge field 
\begin{align}
	a'_\mu = \left( - \phi - \frac 1 2 k^2 \right) (dt)_\mu + k_i (d x^i)_\mu ,
\end{align}
but one can bring $a_\mu$ back into the desired form by a $U(1)_M$ gauge transformation
\begin{align}
	a_\mu \rightarrow a_\mu + \partial_\mu \alpha,
	&&\psi \rightarrow e^{i m \alpha} \psi,
	&&\alpha = \frac 1 2 k^2 t - k_i x^i ,
\end{align}
after which $a'_\mu = - \phi d t'$. We see then that the projective transformation (\ref{projTransf}) follows from the diffeomorphism covariant transformation law (\ref{aTransf}) as the result of a combined internal boost plus $U(1)_M$ gauge transformation to preserve the Newtonian form of the gravitational vector potential. The second approach will prove more useful to us since the projective transformation does not readily generalize to curved spacetimes.

Newton-Cartan geometry then does not prove helpful in constructing the Schr\"{o}dinger action, whose invariance is guaranteed rather by a careful cancellation between two separate, non-invariant terms under (\ref{aTransf}). 
There is nothing wrong with this so far as it goes, the action is after all invariant, but we can anticipate this issue will complicate attempts to construct other theories when the need arises. For instance, Greiter, Witten, and Wilczek argue that the Landau-Ginzburg theory for the superconducting order parameter must include higher derivative couplings to properly account for Galilean invariant sound propagation \cite{Greiter:1989qb}. The higher derivative term they introduce is simply the square of (\ref{SchrodL}), but this done we must in principle ask, what else appears at this order? Given our experience so far, answering this question within this approach is likely to be cumbersome. One may check manually that at this order, the combination
\begin{align}\label{higherDeriv}
	 4 m^2 \partial_t  \psi^\dagger \partial_t  \psi - 2 i m \partial^i \psi^\dagger \overset{\leftrightarrow} \partial_t \partial_i \psi + \partial_i \partial_j \psi^\dagger \partial^i \partial^j \psi 
\end{align}
forms another acceptable possibility, but how would we generate this automatically and in a manner that could be easily generalized to even higher derivatives?

A similar problem arises for non-relativistic Goldstones. Greiter, Witten, and Wilczek have demonstrated that the lowest order, Galilean invariant effective action of a non-relativistic Goldstone is given by
\begin{align}\label{goldstoneAction}
	S = \int d^{d+1} x~ p \left( \partial_t \varphi - \frac {\delta^{ij}} {2m} \partial_i \varphi \partial_j \varphi \right) ,
\end{align}
where $p(\mu)$ is an arbitrary function that represents the pressure of the system as a function of chemical potential \cite{Greiter:1989qb} . This is derived by arguing that in a non-relativistic theory, momentum arises purely from the motion matter, and so they set $\tau^{ti} = \frac m e j^i$. Interpreting these currents as the Noether currents for translations and phase rotations, this reads
\begin{align}
	- \frac{\delta \mathcal L}{\delta \partial_0 \varphi} \partial^i \varphi = m \frac{\delta \mathcal L}{\delta \partial_i \varphi}.
\end{align}
The Lagrangian in (\ref{goldstoneAction}) is the solution to this differential equation.
Again, Newton-Cartan geometry is not terribly helpful in reproducing this form. The approach is also specific to systems with excitations that have only a single charge to mass ratio $e / m$ and so does not apply to a wide class of physically relevant systems, for instance, mixtures of superfluid He-3 and He-4. The authors of \cite{Greiter:1989qb} leave generalization of the above action to higher derivative, multi-constituent theories as an open question, one that we hope to address in future work \cite{GeracieForthcoming}.

In chapter \ref{chap:Examples} we present a simple and systematic method to ensure Galilean invariance. The key element is a chronically underused, $(d+2)$-dimensional representation of the Galilean group which we call the extended representation, within which the non-Galilean covariant derivatives $\partial_t \psi$ and $\partial_i \psi$ are collected. The resulting derivative  $D_I$ maps tensor and spinor fields into fields with an additional index valued in an internal space transforming under the extended representation of local Galilean transformations. One can then proceed to construct actions such as (\ref{SchrodL}), (\ref{higherDeriv}), (\ref{goldstoneAction}), and their generalizations along the same lines that one would construct a relativistic action, simply making sure all indices are properly contracted using good geometric objects.

\subsection{The Need for Torsion}

The remaining complication will prove more serious. As mentioned above, there are a number of condensed matter applications in which the presence of torsion is essential. For instance, torsion may be used to mock up the coarse grained effect of lattice defects in a solid \cite{Hughes:2012vg}. The example that will be relevant to us is energy transport, which we shall find requires coupling to torsionful backgrounds. The key issue is that unlike the relativistic case, torsion cannot be chosen independently of the metric data and once torsion is nonzero, the freedom in choosing a connection cannot be fixed in the manner described above. Torsionful Newton-Cartan geometries originally figured prominently in a number of non-relativistic holographic applications \cite{Son:2008ye,Janiszewski:2012nb,Christensen:2013lma,Christensen:2013rfa,Hartong:2014oma,Hartong:2014pma,Bergshoeff:2014uea,Hartong:2015wxa,Hartong:2015zia}. The non-relativistic limits of electrodynamics on torsionful backgrounds were considered in \cite{Festuccia:2016awg,Festuccia:2016caf}.

Perhaps unsurprisingly, and as we shall see in detail in chapter \ref{chap:Ward}, the clock form serves as a source for energy current in non-relativistic theories. 
Defining the energy current and studying it's linear response then requires introducing arbitrary perturbations to $n_\mu$ that bring us outside of the geometry given above. In particular, one finds for arbitrary $n_\mu$, the compatibility condition $\nabla_\mu n_\nu = 0$ implies
\begin{align}\label{torsionConstraint}
	n_\lambda T^\lambda{}_{\mu \nu} = (d \df n)_{\mu \nu} .
\end{align}
For a non-relativistic theory then, torsion cannot be specified independently of metric data, in stark contrast with the relativistic case.

Even defining an energy current then requires understanding how to properly couple a Galilean theory to torsionful backgrounds. Unfortunately, the other key difference with the relativistic case, the ambiguity in the connection, complicates this issue and the manner in which we have fixed this freedom is no longer applicable in the presence of torsion. 

Indeed, after the discussion of the previous section, one might wonder if we have already committed an error. We have fixed the freedom in the connection using the Newtonian potential, but the Newtonian potential depends on a choice of $v^\mu$. Fortunately, one can show the connection does not in fact depend on this choice, but this is no longer true in the presence of torsion. Attempting to do so results in a connection that depends on a choice of $v^\mu$.

For details we refer to \cite{Jensen:2014aia}. In the condensed matter literature this has been dealt with either by continuing to fix the connection using only a Newtonian potential, or by demanding the ``ambiguous part'' of the connection to be zero. In either approach the auxiliary field $v^\mu$ takes on true physical significance and many of the benefits of a covariant treatment are lost. One of the principle advantages of the extended vielbein formalism is that the Galilean invariance of the connection will be automatic even in the presence of torsion, while in its absence it will reduce entirely to Cartan's treatment.

Finally, before we move on, let's take a moment to consider the physical interpretation of spacetimes with temporal torsion. These no longer conform to standard notions of non-relativistic physics.
 In particular, the clock form will no longer be closed and so we lose any notion of absolute time. Time dilation occurs, even non-relativistically, as the elapsed time between events $\Delta t = \int_c \df n$ now depends on the path taken. Such spacetimes are not of direct physical relevance in the non-relativistic regime, but they do enter formally in an intermediate stage of some calculations (another example being the entropy current analysis to appear in chapter \ref{chap:Apps}) and must be considered.

Although such backgrounds do not have any counterpart in real world non-relativistic applications, they are still consistent backgrounds with sensible physics so long as we assume that
\begin{align}\label{causal}
	\df n \wedge d \df n = 0 .
\end{align}
By the Frobenius theorem, this is equivalent to $\df n$ being integrable, that is, that one may locally construct a foliation of codimension one surfaces whose tangent spaces are everywhere orthogonal to $\df n$. When $d \df n \neq 0$ we loose a notion of absolute time, but when $\df n \wedge d \df n \neq 0$ we also loose an absolute notion of space.

This isn't a disaster in and of itself, but spacetimes with $\df n \wedge d \df n \neq 0$ have other features that are far stranger. One can show that for each point $p \in \mathcal M$ where $\df n$ is not integrable, there is a neighborhood of $p$ within which any point may be reached by a null curve (null in the sense that it's tangent is $\df n$ orthogonal) starting at $p$. For a proof, see \cite{Frankel}. It's not difficult to extend this proof to show that indeed any point may be reached by both a future-directed causal curve and a past-directed causal curve. In this neighborhood, all events are both in the future and the past of $p$. We shall then call spacetimes for which (\ref{causal}) holds everywhere causal spacetimes and demand this throughout.

\section{Bargmann Geometry}\label{chap:Formalism}
In this chapter we define the geometries on which the remainder of this thesis will be set. This is a covariant extension of the Newton-Cartan formalism to torsionful backgrounds. For now, we simply state the definition and devote the remainder of the chapter to explanation and discussion.

A Bargmann geometry is a $(d+1)$-dimensional spacetime manifold $\mathcal M$ equipped with an extended coframe $\df e^I$, an extended torsion two-form $\df T^I$, and a Galilean connection one-form $\df \omega \in \mathfrak{gal}(d)$ satisfying the first structure equation
\begin{align}
	D \df e^I = \df T^I .
\end{align}
As we shall see, the extended coframe is a collection of one-form fields $\df e^0 , \df e^a , \df a$, with $a = 1 , \dots, d$ which encode geometry and Newtonian gravity and transform in a $(d+2)$-dimensional representation of the Galilean group. The extended torsion encodes both spacetime torsion $\df T^0, \df T^a$, and a field strength $\df T^M$ that couples to mass as an external Lorentz force.

The spacetime torsion has of course been considered before, but $\df T^M$ was introduced in our initial work on Bargmann geometries \cite{GPR_geometry,GPR_fluids}, as well as in the language of Koszul connections in \cite{BM}.  By the transformation properties of $\df T^I$, this external force cannot be consistently set to zero in a manner that respects Galilean invariance when the spacetime torsion is nonzero and is the missing element in previous approaches to torsionful Newton-Cartan geometries.
Moreover, we will show that when $\df T^I=0$, our formalism reduces entirely to the treatment given by Cartan. 

We introduce the Galilean group and it's representations and establish notation that will be used in the rest of this thesis in section \ref{sec:Galilean}. In section \ref{sec:ExtendedVielbein} we introduce the formalism summarized above and discuss it's interpretation and relationship with previous approaches in section \ref{sec:Discussion}. In appendix \ref{app:Coset}, we discuss how this approach may be obtained in terms of a formal symmetry breaking procedure. 

\subsection{The Galilean Group}\label{sec:Galilean}
In the standard treatment of Galilean relativity, different inertial coordinate systems are related by
\begin{align}\label{Gal}
	t \rightarrow t ,
	&&x^i \rightarrow R^i{}_j x^j - k^i t
\end{align}
where $R^i{}_j \in SO(d)$ relates the relative orientation of coordinate axes and $k^i \in \mathbb R^d$ is their relative velocity. These transformations form a group called the Galilean group $Gal(d)$.

For our purposes, the Galilean group is defined to be the group of $(d+1)$-dimensional matrices $\Lambda^A{}_B$ that leave invariant a co-vector $n_A$ and positive semidefinite symmetric tensor $h^{AB}$
\begin{align}\label{fundRep}
	&n_A =
	\begin{pmatrix}
		1 & 0  
	\end{pmatrix},
	&&h^{AB}=
	\begin{pmatrix}
		0 & 0 \\
		0 & \delta^{ab}
	\end{pmatrix},
	\nonumber \\
	&n_B (\Lambda^{-1})^B{}_A = n_A ,
	&&\Lambda^A{}_C \Lambda^B{}_D h^{CD} = h^{AB} .
\end{align}
These matrices are of the form
\begin{align}\label{GalRep}
	\Lambda^A{}_B =
	\begin{pmatrix}
		1 & 0 \\
		- k^a & R^a{}_b
	\end{pmatrix}
\end{align}
where $R^a{}_b \in SO(d)$. These matrices clearly mimic (\ref{Gal}), but importantly, they act on an internal representation space and are distinct from diffeomorphisms.

There is another, less familiar, $(d+2)$-dimensional representation, given by matrices of the form
\begin{align}\label{extendedRep}
	\Lambda^I{}_J =
	\begin{pmatrix}
		1 & 0 & 0 \\
		- k^a & R^a{}_b & 0 \\
		- \frac 1 2 k^2 & k_c R^c{}_b & 1 
	\end{pmatrix} ,
\end{align}
which we call the extended representation.
Our conventions are that capital Latin indices $A,B,\dots$ transform in the vector/covector representations of $Gal(d)$, while lower case Latin indices $a,b,\dots$ transform under the $SO(d)$ subgroup. Capital Latin indices beginning with $I,J,\dots$ transform in the extended representation. For example,
\begin{align}
	&v^A \rightarrow \Lambda^A{}_B v^B,
	&&w_A \rightarrow w_B ( \Lambda^{-1})^B{}_A, \nonumber \\
	&v^I \rightarrow \Lambda^I{}_J v^J,
	&&w_I \rightarrow w_J ( \Lambda^{-1})^J{}_I.
\end{align}

The extended representation also preserves a set of invariant tensors, one of which is, remarkably, a metric of Lorentzian signature
\begin{align}
	&n_I = 
	\begin{pmatrix}
		1 & 0 & 0
	\end{pmatrix},
	&&
	g^{IJ} =
	\begin{pmatrix}
		0 & 0 & 1 \\
		0 & \delta^{ab} & 0 \\
		1 & 0 & 0
	\end{pmatrix}, \nonumber \\
	&n_J (\Lambda^{-1})^J{}_I = n_I,
	&& \Lambda^I{}_K \Lambda^J{}_L g^{KL} = g^{IJ} .
\end{align}
We will denote the inverse to $g^{IJ}$ as $g_{IJ}$ and use the metric to raise and lower extended indices at will. Importantly, one does not lose information in doing so as one does when raising indices with the degenerate tensor $h^{AB}$. Due to the availability of a metric, manipulations of extended indices will often be easier than (co)vector indices.

The vector and extended representations together also posses a mixed invariant
\begin{align}
	\Pi^A{}_I =
	\begin{pmatrix}
		1 & 0 & 0 \\
		0 & \delta^a{}_b & 0 
	\end{pmatrix},
	&&\Lambda^A{}_B \Pi^B{}_J (\Lambda^{-1})^J{}_I = \Pi^A{}_I ,
\end{align}
with which we may project extended indices to the vector representation or pull back covector indices to the extended representation.
For example
\begin{align}
	n_A \Pi^A{}_I = n_I ,
	&&h^{AB} = \Pi^A{}_I \Pi^B{}_J g^{IJ} .
\end{align}

Finally, since $Gal(d) \leq SL(d+1)$, the Galilean group also preserves the totally antisymmetric tensors $\epsilon^{A_0 \cdots A_d}$ and $\epsilon_{A_0 \cdots A_d}$
\begin{align}\label{epsilons}
	&\epsilon^{A_0 \cdots A_d} = \Lambda^{A_0}{}_{B_0} \cdots \Lambda^{A_d}{}_{B_d} \epsilon^{B_0 \cdots B_d}, \nonumber \\
	&\epsilon_{A_0 \cdots A_d} = \epsilon_{B_0 \cdots B_d} (\Lambda^{-1})^{B_0}{}_{A_0} \cdots (\Lambda^{-1})^{B_d}{}_{A_d} .
\end{align}
where we will take $\epsilon^{01 \cdots d} = \epsilon_{01 \cdots d} = 1$. There is some potential for confusion with this notation since we will occasionally raise Galilean covector indices using the degenerate tensor $h^{AB}$. The epsilon tensor with raised indices is not the ``raised index" version of the epsilon tensor with lowered indices in this sense. Indeed, the fully raised version of $\epsilon_{A_0 \cdots A_d}$ is simply zero.

Let's move on to the Lie algebra $\mathfrak{gal}(d)$ of the Galilean group. The representation induced by the defining representation (\ref{GalRep}) of the Galilean group consists of matrices of the form
\begin{align}\label{lieGalA}
	\Theta^A{}_B = 
	\begin{pmatrix}
		0 & 0 \\
		- k^a & \theta^a{}_b
	\end{pmatrix}
\end{align}
where $\theta^{(ab)} = 0$.
Equivalently, they are the $(d+1)$-dimensional matrices that satisfy
\begin{align}\label{lieFund}
	&n_A \Theta^A{}_B = 0 ,
	&&\Theta^{(A B)} = 0 . 
\end{align}
Similarly, the extended representation of $\mathfrak{gal}(d)$ is given by
\begin{align}\label{lieExt}
	\Theta^I{}_J = 
	\begin{pmatrix}
		0 & 0 & 0\\
		- k^a & \theta^a{}_b & 0 \\
		0 & k_b & 0
	\end{pmatrix}
	&&\text{i.e.}
	&&n_I \Theta^I{}_J = 0,
	&& \Theta^{(IJ)} = 0.
\end{align}
One may also check that the projector $\Pi^A{}_I$ is an intertwiner of the extended and defining representations of $\mathfrak{gal} (d)$
\begin{align}
	\Pi^A{}_J \Theta^J{}_I = \Theta^A{}_B \Pi^B{}_J.
\end{align}

Finally, note that the defining and extended representations of $\mathfrak{gal}(d)$ both contain precisely the same data as a totally antisymmetric matrix with lowered indices 
\begin{align}
	\Theta_{AB} =
	\begin{pmatrix}
		0 & k_b \\
		- k_a & \theta_{ab} 
	\end{pmatrix},
\end{align}
and indeed, this can be used to give us a simple and useful characterization of the two representations given above
\begin{align}\label{loweredTheta}
	\Theta^A{}_B = h^{AC} \Theta_{CB},
	&&\Theta^I{}_J = \Pi^{AI} \Pi^B{}_J \Theta_{AB} .
\end{align}
It will often be easier to write equations in terms of ${\Theta_{AB}}$ rather than $\Theta^A{}_B$ or $\Theta^I{}_J$, despite the fact that it is not a representation of $\mathfrak{gal}(d)$.

\subsection{A First-Order Formalism for Non-Relativistic Geometry}\label{sec:ExtendedVielbein}

The torsionful generalization of Newton-Cartan geometry is far more transparent within a first-order formalism. To motivate the definition of a Bargmann geometry then, let's try constructing non-relativistic geometry in a first-order formalism in as close an analogy as possible with the well understood pseudo-Riemannian case and see what changes need to be made along the way. For an alternate approach to gauging Galilean symmetry, see \cite{Banerjee:2014pya,Banerjee:2014nja,Banerjee:2015tga,Banerjee:2015rca,Banerjee:2016laq}.

To begin, introduce a Galilean coframe $\df e^A$ and torsion $\df T^A$ transforming in the vector representation of $Gal(d)$. As usual, the coframe is equivalent to the ``metric" data of a Newton-Cartan geometry, which may be retrieved using the invariant tensors $n_A$ and $h^{AB}$
\begin{align}\label{NCData}
	n_\mu = n_A e^A_\mu ,
	&&h^{\mu \nu} = e_A^\mu e_B^\nu h^{AB},
\end{align}
where $e^\mu_A$ is the frame field, defined by $e^\mu_A e^A_\nu = \delta^\mu{}_\nu$.
It should be clear that these satisfy the requirements of a Newton-Cartan geometry: $h^{\mu \nu}$ is positive semidefinite with kernel spanned by the non-vanishing form $n_\mu$. We have however generalized somewhat since $\df n$ need not be closed.

One then defines a spin connection $\df \omega^A{}_B$ in the Lie algebra (\ref{lieGalA}) of the Galilean group via the first structure equation
\begin{align}\label{firstTry}
	D \df e^A = \df T^A .
\end{align}
This transforms in the usual manner
\begin{align}\label{connectionTransf}
	\df \omega^A{}_B \rightarrow \Lambda^A{}_C \df \omega^C{}_D (\Lambda^{-1})^D{}_B + \Lambda^A{}_C d ( \Lambda^{-1})^C{}_B .
\end{align}
and defines a spacetime derivative operator $\nabla$ through
\begin{align}\label{nablaDef2}
	\nabla_\mu e^A_\nu + \omega{}_\mu{}^A{}_B e^B_\nu = 0.
\end{align}

Unfortunately, unlike the pseudo-Riemannian case, the first structure equation does {\it not} uniquely specify $\df \omega^A{}_B$ as one of the equations is pure constraint. Contracting (\ref{firstTry}) with $n_A$ we find that the spin connection drops out and we have merely
\begin{align}\label{temporalTorsion}
	d \df n = \df T^0.
\end{align}
What remains of (\ref{firstTry}) is then a collection of only $\frac{ d (d+1)^2}{2} - 1$ equations in $\frac{ d (d+1)^2}{2}$ unknowns and the connection is not completely fixed.

Of course, the constraint is simply that observed previously in equation (\ref{torsionConstraint}) and is the reason we require torsion in the study of energy transport. Moreover, as recounted in section \ref{sec:NewtonCartanGeo}, the fact that the connection is not simply fixed by the torsion and the metric data is the key feature of Newton-Cartan geometry. It is precisely in this freedom that the Newtonian potential sits and allows for the formalism to encode gravity.
What we have done then is not wrong, but we need to figure out how to incorporate the Newtonian potential in a covariant way to complete our description. 

The key to doing so is to consider the transformation law (\ref{aTransf}) for the mass gauge field, which we reproduce here\footnote{The placement of the rotation matrix follows from restoring it in equation (\ref{projTransf}) and can also be justified by the Galilean symmetry breaking procedure discussed in appendix \ref{app:Coset}.}
\begin{align}
	\df a \rightarrow \df a + k_b R^b{}_a \df e^a - \frac 1 2 k^2 \df n .
\end{align}
As it stands, this is pretty inconvenient to work with due to it's nonlinearity. However, we can collect $\df a$ along with the vielbein into a single covariant object using the extended representation (\ref{extendedRep}) of the Galilean group
\begin{align}
	\df e^I = 
	\begin{pmatrix}
		\df n \\
		\df e^a \\
		\df a
	\end{pmatrix},
	&&\df e^I \rightarrow \Lambda^I{}_J \df e^J ,
\end{align} 
which we call the extended vielbein. The coframe $\df e^A$ may always be covariantly isolated from this
\begin{align}
	\df e^A = \Pi^A{}_I \df e^I,
\end{align}
but the final component $\df a$ may not.

The extended vielbein is then an natural object to consider in a non-relativistic gravitational theory. Let's also consider it's covariant exterior derivative. This will be an object in the extended representation, which we call the extended torsion tensor
\begin{align}\label{extendedFirstStructure}
	D \df e^I = \df T^I ,
	&&\df T^I =
	\begin{pmatrix}
		\df T^0 \\
		\df T^a \\
		\df T^M
	\end{pmatrix} .
\end{align}
We can retrieve our original attempt (\ref{firstTry}) from this by simply applying the projector $\Pi^A{}_I$ since $D \Pi^A{}_I = 0$. The key point is then that this equation includes both (\ref{firstTry}) and a single additional equation
\begin{align}
	d \df a - \df \varpi^a{} \wedge \df e_a = \df T^M .
\end{align}
Viewed as an algebraic equation for the connection, (\ref{extendedFirstStructure}) is then a set of $\frac{ d (d+1)^2}{2}$ in $\frac{ d (d+1)^2}{2}$ unknowns and the data in $\df \omega$ and $\df T^I$ is entirely equivalent.

An immediate consequence of these definitions are the Bianchi identities
\begin{align}\label{Bianchi}
	D \df T^I = \df R^I{}_J \wedge \df e^J ,
	&&D \df R^A{}_B = 0 .
\end{align}
where
\begin{align}
	\df R^A{}_B = d \df \omega^A{}_B + \df \omega^A{}_C \wedge \df \omega^C{}_B
\end{align}
is the curvature two-form and 
\begin{align}
	\df R^I{}_J = d \df \omega^I{}_J + \df \omega^I{}_K \wedge \df \omega^K{}_J
\end{align}
is simply $\df R^A{}_B$ considered in the extended representation. From these identities one may obtain the symmetries of the Riemann tensor. These differ in some respects from their pseudo-Riemannian versions and may be found in appendix \ref{app:Riemann}.

Before moving on we note that a Bargmann spacetime admits a natural volume element
\begin{align}\label{vol}
	\df \varepsilon = \frac{1}{(d+1)!} \epsilon_{A_0 \cdots A_d } \df e^{A_0} \wedge \cdots \wedge \df e^{A_d}
\end{align}
which may be used to define integration over spacetime. There is similarly a ``volume element" with raised indices
\begin{align}
	\varepsilon^{\mu_0 \cdots \mu_d} = \epsilon^{A_0 \cdots A_d} e^{\mu_0}_{A_0} \cdots . e^{\mu_d}_{A_d} 
\end{align}
However, as in equation (\ref{epsilons}), $\varepsilon^{\mu_0 \cdots \mu_d}$ is not simply $\varepsilon_{\mu_0 \cdots \mu_d}$ with raised indices. 
In components
\begin{align}
	\varepsilon_{0 \cdots d} = | e | ,
	&&\varepsilon^{0 \cdots d} = | e |^{-1},
	&&\text{where}
	&&|e| = \text{det} (e^A_\mu) .
\end{align}
These are both annihilated by the derivative operator
\begin{align}
	\nabla_\mu \varepsilon_{\mu_0 \cdots \mu_d} = 0,
	&&\nabla_\mu \varepsilon^{\mu_0 \cdots \mu_d} = 0 .
\end{align}

\subsection{Discussion}\label{sec:Discussion}
What is the relationship between this approach and those given previously? Cartan's formalism, recalled at the opening of section \ref{sec:NewtonCartanGeo}, can be retrieved immediately by going to the torsionless case $\df T^I = 0$. Projecting the extended vielbein gives the coframe $\df e^A = \Pi^A{}_I \df e^I$ and spacetime torsion $\df T^A{} = \Pi^A{}_I \df T^I = 0$, from which we may define the clock form and spatial metric via equation (\ref{NCData}).\footnote{It may seem natural at this stage to also consider the ``metric'' $g_{IJ} e^I_\mu e^J_\nu$. This is however not invariant under $U(1)$ gauge transformations $\df a \rightarrow \df a + d f$ and so is not admissible.} Moreover, the clock form is closed by equation (\ref{temporalTorsion}) and the spacetime enjoys a (at least local) definition of absolute time. The derivative operator is defined by $\nabla_\mu e^A_\nu + \omega_\mu{}^A{}_B e^B_\nu = 0$, from which metric compatibility follows
\begin{align}
	\nabla_\mu n_\nu = - n_A \omega_\mu{}^A{}_B e^B_\nu = 0 ,
	&&\nabla_\lambda h^{\mu \nu}  = 2 \omega_\lambda{}^{(AB)} e_A^\mu e_B^\nu = 0 .
\end{align}
Finally, if we specialize to the flat case $\df e^I = ( dt ~ dx^a ~ - \phi dt )^T$ (note we have chosen Newtonian gauge), solving $D \df e^I = 0$ for the connection gives
\begin{align}
	\Gamma^k{}_{tt} = \partial^k \phi ,
 \end{align}
 with all other components zero. We thus see that the connection encodes Newtonian gravity as it should.

Note that as soon as we generalize to torsionful spacetimes, there is no way to discard the final component $\df T^M$ of the extended torsion tensor, since this piece picks up factors of the spacetime torsion under a change of frame
\begin{align}
	\df T^M \rightarrow \df T^M + k_b R^b{}_a \df T^a - \frac 1 2 k^2 \df T^0 .
\end{align}
In previous approaches to torsionful Newton-Cartan geometry in the condensed matter literature, this term was effectively set to zero. The non-invariance of the connection is an artifact of this.

How are we to interpret the mass torsion $\df T^M$ physically? One may view it as an external Lorentz force that couples to mass. We shall see a much more general demonstration of this in section (\ref{sec:Ward}), where this is dictated by Ward identities. However, since our interpretation of $\df a$ in part relies on how it affects the equation of motion for a point particle, it seems appropriate to consider $\df T^M$ in this same light. Let's then consider the simple case of a non-relativistic point particle moving on a background spacetime with nonzero $\df T^I$. In flat space, the action of a point particle coupled to a gravitational potential is
\begin{align}\label{ppActionNonCov}
	S = \int d t \left( \frac 1 2  m \dot x^i (t) \dot x_i (t) - m \phi ( x ( t) )\right),
\end{align}
from which we obtain the equation of motion $\ddot x^i = - \partial^i \phi$. We would like a covariant version of this that will tell us how to couple the particle to the curved backgrounds under consideration. We will then consider how $\df T^M$ appears.

The action that accomplishes this is
\begin{align}
	S = \frac 1 2 m \int d \tau \frac{\xi_I \xi^I}{n_I \xi^I},
	&& \text{where}
	&&\xi^I \equiv e^I_\mu \xi^\mu,
	&&\xi^\mu \equiv  \frac{d x^\mu}{d \tau} ,
\end{align} 
and $x^\mu (\tau)$ is a parameterization of the particle's path through spacetime.\footnote{One is also free to add a term proportional to $\int \df n$. This gives rise to an additional Lorentz force exerted by $\df T^0$ on the particle's energy current and will not affect our discussion.} This is reparameterization invariant, Galilean invariant, and invariant under $U(1)_M$ gauge transformations $\df a \rightarrow \df a + d f$. To see this, note that in terms of the extended coframe a $U(1)_M$ gauge transformation is given by $\delta \df e^I = n^I d f$ and thus
\begin{align}
	\delta S= m \int d f = 0 
\end{align}
for wordlines without endpoints.
Furthermore, one may check that when $\df e^I = ( dt ~ d x^a ~ - \phi dt )$, this action reduces precisely to (\ref{ppActionNonCov}). It also agrees with the coupling to curved backgrounds found in \cite{Kunzle:1972fv,ABGdR,Jensen:2014aia}.

The equations of motion found from varying $S$ is a generalization of the geodesic equation to torsionful backgrounds
\begin{align}\label{ppEOM}
	\xi^\nu \nabla_\nu \xi^\mu = \mathring \xi_I T^{I\mu}{}_\nu \xi^\nu
\end{align}
where for simplicity we have chosen an affine parameterization in the sense that $n_\mu \xi^\mu = 1$. In writing this we have introduced the object $\mathring \xi^I$, which denotes the unique null vector in the extended representation that projects to the particle velocity
\begin{align}
	\mathring \xi^I \mathring \xi_I = 0,
	&&\xi^\mu = \Pi^\mu{}_I \mathring \xi^I .
\end{align}
In components this takes the form $\mathring \xi^I = ( 1 ~ \xi^a ~ - \frac 1 2 \xi^2 )^T$ (this trick will be of great use in our fluid analysis in sections \ref{sec:LLLFluid} and \ref{sec:Fluid} and we will reintroduce it when the time comes).

We see from equation (\ref{ppEOM}) that the extended torsion couples to the point particle as an external force $F^\mu = m \mathring \xi_I T^{I\mu}{}_\nu \xi^\nu$.\footnote{Note the similarity to the relativistic case where we have $\xi^\nu \nabla_\nu \xi^\mu = T_\nu{}^\mu{}_\lambda \xi^\nu \xi^\lambda$.} In particular, the mass component of the torsion $\df T^M$ enters as an external Lorentz force coupling to the mass current
\begin{align}
	F^\mu = m T^{M\mu}{}_\nu \xi^\nu + \cdots,
\end{align}
as we set out to show. This is perhaps rather unphysical since in the real world the only force we know that couples to mass exactly is gravity as dictated by the equivalence principle. However, the physical world is, at least to an excellent approximation, torsionless. If we insist on considering torsionful spacetimes, as our later applications will force upon us, Galilean invariance dictates that we must consider such a force. Since the equivalence principle is broken anyways on torsionful backgrounds, this is not a problem.

\section{Currents and Ward Identities}\label{chap:Ward}

We now begin to present some physical consequences of the formalism introduced above. This chapter concerns the currents present in any non-relativistic theory and the Ward identities they satisfy by virtue of diffeomorphism invariance and internal local Galilean invariance. As such, they are entirely general for theories with underlying Galilean symmetry. 
This chapter will also serve to set up some of the necessary framework for more targeted applications in chapter \ref{chap:Apps}.

In section \ref{sec:Currents} we begin by defining the complete set of currents generated by our background. These include the stress, mass current, momentum current, energy current, and spin current. All of these have been studied in the past and energy current in particular was originally defined in the classic work of Luttinger as response to a fictitious potential (the ``Luttinger potential") that couples to energy \cite{Luttinger:1964zz}. As we shall see, the Luttinger potential lies naturally within the extended vielbein $\df e^I$, which collects the background perturbations available in any Galilean invariant theory into a single covariant object. Specifically, it lies within the clock form $\df n$, which is why it is necessary to couple a theory to torsional geometries when studying energy transport: we must must consider systems in an arbitrary clock form field and hence arbitrary temporal torsion $\df T^0 = d \df n$.

As has been pointed out by several authors, this approach has several advantages over Luttinger's treatment. For one, it allows one to directly define energy current $\varepsilon^i$ in a microscopic theory, whereas Luttinger's treatment merely defines it so as to satisfy local energy conservation. Furthermore, in Luttinger's work, the Luttinger potential was coupled to a theory assuming pre-existing knowledge of the energy density operator. The potential was then not used as a calculational tool to define energy current but rather as an intermediate step to derive Kubo formulas. Given the Luttinger potential's appearance within $\df e^I$ and our knowledge of how to couple microscopic theories to torsionful geometries (see chapter \ref{chap:Examples}), covariance will dictate how the clock form will enter our action so that we may directly derive the energy current operator in an arbitrary microscopic theory. This done, one may generate Kubo formulas for thermal transport, as done in section \ref{sec:Kubo}, of which Luttinger's will be one example, and then compute them in whatever theory is relevant for application. So far as we are aware, this has not yet been done and would be an interesting extension of our work.

To our knowledge, the Luttinger potential was originally identified within the Newton-Cartan data in \cite{Son:2008ye} where the work-energy equation was derived as a Ward identity in flat backgrounds, and discussed further in \cite{Janiszewski:2012nb}. The energy current was later studied in \cite{Geracie:2014nka}, where it’s Ward identity was generalized to curved torsionful backgrounds, though at this stage the extended torsion was still lacking so Galilean invariance was being implicitly broken (see equation (\ref{diff ward}) for the corrected version). Further work in \cite{Gromov:2014vla,Geracie:2014zha,Bradlyn:2014wla} studied energy transport in non-relativistic systems with both preserved and broken Galilean invariance.

The treatment of currents presented here is that of \cite{GPR_fluids}, which differs from previous work insofar as it gives a manifestly Galilean covariant presentation of the Ward identities. This requires collecting the mass, momentum, energy, and stress currents into a single covariant object in the extended representation $\tau^\mu{}_I$, reflecting their transformation properties under $Gal(d)$. The Ward identities of previous treatments mixed under Galilean transformations and we here collect them into a single Galilean invariant Ward identity (\ref{diff ward}). We also present a Ward identity for the stress-mass tensor that follows from local Galilean transformations that to our knowledge is not present in the literature (see equation (\ref{LGTWard}) and discussion below) and makes precise the intuition used by Greiter, Witten and Wilzcek \cite{Greiter:1989qb} that Galilean invariance implies the proportionality of momentum and charge currents and demonstrates how it extends to the multi-component case.

\subsection{Non-Relativistic Currents}\label{sec:Currents}
In this section we define the currents discussed above. A Bargmann geometry defines a set of background data $( \df e^I ,\df \omega_{AB} )$ which may be perturbed to define the stress-energy tensor $\tilde \tau^\mu{}_I$ and spin current $\tilde s^{\mu AB}$
\begin{align}\label{unimprovedCurrentsDef}
	\delta S = \int d^{d+1} x | e | \left(-  { \tilde \tau^\mu}{}_I \delta e^I_\mu + {\tilde s^{\mu AB}} \delta{\omega_{\mu AB} } \right) .
\end{align} 
By $\df \omega_{AB}$ we mean the unique antisymmetric matrix whose first index raises to $\df \omega^A{}_B$ (see equation (\ref{loweredTheta})). From the transformation law for (\ref{connectionTransf}) for $\df \omega^{A}{}_B$ it follows that
\begin{align}
	\df \omega_{AB} \rightarrow (\Lambda^{-1})^C{}_A (\Lambda^{-1})^D{}_B \df \omega_{CD} - d \Theta_{AB}
\end{align}
under a local internal Galilean transformation $\Lambda^A{}_B = (e^\Theta)^A{}_B$.

The tilde's indicate that the components of $\tilde \tau^\mu{}_I$ are not yet the physical currents measured in the lab. For instance, we shall see that in the presence of spinful matter, the stress part $\tilde \tau^a{}_b$ is not necessarily symmetric and so cannot be the Cauchy stress used in condensed matter and engineering applications.  Properly defining the ``true" currents is in some respects more subtle than in the relativistic case, an issue that we will turn to in section \ref{sec:Improvement}. Until then, let's assume that we are considering spinless matter and drop the tildes.

The object $\tau^\mu{}_I$ collects response to the clock form $n_\mu$, spatial vielbein $e^a_\mu$ and mass gauge field $a_\mu$ into a single covariant object. That is, it contains the energy current $\varepsilon^\mu$, stress tensor $T^i{}_a$, momentum current $p_a$, and mass current $\rho^\mu$
\begin{align}\label{stressEnergyComponents}
	\delta S = \int d^{d+1} x | e | \left( - \varepsilon^\mu \delta n_\mu + T^i{}_a\delta e^a_i + p_a \delta e^a_{t}+ \rho^\mu \delta a_\mu + \cdots  \right) .
\end{align}

Though it contains much more information than the name indicates, we shall call $\tau^\mu{}_I$ simply the stress-energy tensor for brevity.
In a Galilean invariant theory, the components listed in (\ref{stressEnergyComponents}) don't have an independent existence, just as stress and energy do not in a relativistic theory.
Separately, they depend on a choice of frame; but in $\tau^\mu{}_I$ they are collected into a single covariant object
\begin{align}\label{stress-energy}
	\tau^\mu{}_I = 
	\begin{pmatrix}
		  \varepsilon^t & - p_{a} & - \rho^t \\
		  \varepsilon^i & - T^i{}_{a} & - \rho^i
	\end{pmatrix} 
\end{align}
whose components transform among each other as indicated by the index structure.

We can however use the invariant tensors introduced in section (\ref{sec:Galilean}) to isolate some of this information in a frame independent way. For example,
we may use $\Pi^A{}_I$ to project out the stress-mass components of $\tau^\mu{}_I$
\begin{align}\label{unimprovedStressMomentum}
	\tau^{\mu \nu} = -\tau^\mu{}_I \Pi^{\nu I} = 
	\begin{pmatrix}
		\rho^t & p^j \\
		\rho^i & T^{ij}
	\end{pmatrix}
\end{align}
where $p^j = e^j_a p^a$ and $T^{ij} = T^{ia} e^j_a$.
We shall refer to $\tau^{\mu \nu}$ as the stress-mass tensor and it is the same as the boost invariant stress tensor of \cite{Jensen:2014aia}. Similarly, the mass current $\rho^\mu $ may be completely separated from the others using $n^I$ (a preferred tensor that points in the ``mass direction" $n^I = ( 0 ~ 0 ~ 1)^T$)
\begin{align}
	\rho^\mu = - \tau^\mu{}_I n^I = \tau^{\mu \nu} n_\nu.
\end{align}

Thus, there is an invariant way to separate the stress-mass tensor from the stress-energy, as well as a way to separate out the mass current from the stress-mass. However, we cannot isolate the stress or momentum from the mass current since different observers will disagree on the measured valued of $T^{i}{}_a$ and $p_a$ by terms proportional to $\rho^\mu$. Similarly, there is no way to isolate the energy current from any of the other components of $\tau^\mu{}_I$ without additional data. We are then forced to discuss the entire object if we wish to consider energy current in a covariant fashion.

Finally, let's conclude this section with a word on the interpretation of these various currents. The interpretation of $\rho^\mu$, $T^{i}{}_a$ and $\varepsilon^\mu$ as the mass current, stress, and energy current should be relatively clear. Referring to $p_a$ as the momentum current may however seem somewhat ad hoc. To justify this, let's compute $p_a$ on flat spacetime explicitly for a theory of a spinless massive field with single derivative couplings $\mathcal L ( \psi , D_t \psi, D_i \psi )$.

Under a variation of the $\delta e^a_t$ components of the coframe about the background $e^A_\mu = \delta^A{}_\mu$, the temporal frame $e^\mu_0$ varies as $\delta e^\mu_0 = - e^\nu_0 e^\mu_A \delta e^A_\nu = -D_a \psi \delta e^a_\mu$. Now the natural covariantization of $D_t \psi$ is $e_0^\mu D_\mu \psi$ which then changes as $\delta( e_0^\mu D_\mu \psi) = - D_a \psi \delta e^a_t$. The Lagrangian then varies as
\begin{align}
	\delta \mathcal L = - \frac{\delta \mathcal L}{\delta D_t \psi} D_a \psi \delta e^a_t - \frac{\delta \mathcal L}{\delta D_t \psi^\dagger} D_a \psi^\dagger \delta e^a_t  
\end{align}
from which we find
\begin{align}
	\tau^{ti} = p^i = - \frac{\delta \mathcal L}{\delta D_0 \psi} D^i \psi   - \frac{\delta \mathcal L}{\delta D_0 \psi^\dagger} D^i \psi^\dagger   , 
\end{align}
which is the canonical Noether current for translations on flat space.\footnote{In this computation we have freely converted between $a$ and $i$ indices using $e^a_i = \delta^a{}_i$.} One can also ascertain this using the formalism of chapter \ref{chap:Examples} via equation (\ref{derivVar}) without the need to argue for natural covariantizations.

\subsubsection{Currents as Measured by Observers}\label{sec:ComovingCurrents}

Though the currents are naturally collected into the single object $\tau^\mu{}_I$, in the presence of an observer with velocity $v^\mu$,\footnote{As always, we are normalizing so that $n_\mu v^\mu = 1$.} one can discuss the stress, momentum, and energy currents as measured by that particular observer. In this section we briefly discuss how to do this and so introduce a number of definitions that will be useful to us in the fluid analysis of section \ref{sec:Fluid}.

To begin with, the mass current measured by $v$ is simply $\rho^\mu = - \tau^\mu{}_I n^I$ and is agreed upon by every observer. Equivalently, one may check directly that the $\rho^\mu$ components of (\ref{stress-energy}) do not transform under internal Galilean transformations. Given $v^\mu$ we may also isolate the momentum current and spatial stress
\begin{align}
	\overset v p{}^\mu = n_\lambda \tau^{\lambda \nu} \overset v P{}^\mu{}_\nu,
	&&\overset v T {}^{\mu \nu} = \overset v P {}^\mu{}_\lambda \tau^{\lambda \rho} \overset v P {}^\nu{}_\rho
\end{align}
where we have introduced the $v^\mu$-orthogonal projector $\overset v P {}^\mu{}_\nu = \delta^\mu{}_\nu - v^\mu n_\nu$, which satisfies
\begin{align}
	\overset v P {}^\mu{}_\lambda \overset v P {}^\lambda{}_\nu = \overset v P {}^\mu{}_\nu,
	&&\overset v P {}^\mu{}_\nu v^\nu = 0,
	&&n_\nu \overset v P {}^\nu{}_\mu = 0 .
\end{align} 
In general, whenever we introduce a frame-dependent quantity, we will denote it with an overset symbol labeling the frame as above, though we shall try to do this as infrequently as possible.
Note that momentum and stress are purely spatial tensors in the sense that $ n_\mu \overset v p{}^\mu = 0$, $\overset v T{}^{\mu \nu} n_\nu = n_\nu \overset v T{}^{\nu \mu} = 0$.

The energy current is contained within the temporal component of the extended stress-energy tensor $\tau^\mu{}_I$ and we need an extended vector that points in a temporal direction to extract it ($n^I$ will not do since it points in the ``mass direction" and moreover has been used to extract $\rho^\mu$). We can always construct such an object given a velocity field. As in our discussion of the point particle equation of motion (\ref{ppEOM}), we need a vector $\mathring v^I$ in the extended representation that projects to $v^\mu$. Of course, there are many such $\mathring v^I$'s, but we can fix this freedom by demanding that it be null with respect to the Lorentzian metric $g_{IJ}$. That is, let $\mathring v^I$ be the unique extended vector satisfying
\begin{align}
	\mathring v^I \mathring v_I = 0,
	&&v^\mu = \Pi^\mu{}_I \mathring v^I .
\end{align}
In components, this is
\begin{align}
	\mathring v^I = 
	\begin{pmatrix}
		1 \\
		v^a \\
		- \frac 1 2 v^2
	\end{pmatrix} .
\end{align}
One may then define
\begin{align}\label{energyCurrent}
	\overset v \varepsilon {}^\mu = \tau^\mu{}_I \mathring v^I
\end{align}
which is the boost invariant energy current of \cite{Jensen:2014ama}, though stated in a different language.

The objects $\rho^\mu$, $\overset v p {}^\mu$, $\overset v T {}^{\mu \nu}$, and $\overset v \varepsilon {}^\mu$ are then the natural objects that can be constructed from the stress-energy tensor $\tau^\mu{}_I$ and an observer's spacetime trajectory and can be interpreted as the mass, momentum, stress, and energy current as measured by an observer with velocity $v^\mu$. To lend additional support to this interpretation, let's consider their relationship with the components of the stress energy tensor given in (\ref{stress-energy}). To make our discussion as transparent as possible we will restrict ourselves to the case of a flat metric and clock form, as in equation (\ref{flatBackground}).

In components then we have
\begin{align}
	v^\mu = 
	\begin{pmatrix}
		1 \\
		v^i 
	\end{pmatrix},
	&&
	\overset v P {}^\mu{}_\nu = 
	\begin{pmatrix}
		0 & 0 \\
		- v^i & \delta^i{}_j
	\end{pmatrix} .
\end{align}
We can then compute the components of the stress-energy tensor in terms of $\overset v p {}^\mu$, $\overset v T {}^{\mu \nu}$, and $\overset v \varepsilon {}^\mu$ and find
\begin{gather}
	p^i = \rho^t v^i + \overset v p{}^i , \nonumber \\
	T^{ij} = v^i \rho^j + \rho^i v^j - \rho^t v^i v^j + \overset v T {}^{ij} , \nonumber \\
	\begin{pmatrix}
		\varepsilon^t \\
		\varepsilon^i 
	\end{pmatrix}
	=
	\begin{pmatrix}
		 v_i \rho^i - \frac 1 2 \rho^t v^2 + \overset v \varepsilon {}^t \\
		\frac 1 2 \rho^i v^2 + v^i \rho^j v_j - \rho^t v^i v^2 + \overset v \varepsilon {}^i + \overset v T{}^{ij} v_j
	\end{pmatrix} .
\end{gather}
In writing this we have used the Ward identity (\ref{LGTWard}) for local Galilean transformations, which we will prove in section \ref{sec:Ward}. In particular, this implies the identity $p^i = \rho^i$.

This will look more familiar if we consider the case of an observer comoving with the mass current. In this case we have $\rho^i = \rho^t v^i$ and the above simplifies to 
\begin{gather}\label{nonCovCur}
	0 = \overset v p{}^i , \nonumber \\
	T^{ij} = \rho^t v^i v^j + \overset v T {}^{ij} , \nonumber \\
	\begin{pmatrix}
		\varepsilon^t \\
		\varepsilon^i 
	\end{pmatrix}
	=
	\begin{pmatrix}
		\frac 1 2 \rho^t v^2 + \overset v \varepsilon {}^t \\
		\frac 1 2 \rho^t v^2 v^i + \overset v \varepsilon {}^i + \overset v T{}^{ij} v_j
	\end{pmatrix} .
\end{gather}
If $\overset v p{}^i$ is the local momentum density measured by comoving observers, then we would expect it to vanish, as it indeed does.
The final two equations are the standard textbook formulas for the stress and energy currents measured in the lab. The lab energy density $\varepsilon^t$ includes both the energy density $\overset v \varepsilon {}^t$ as measured by observers in the rest frame defined by $v^\mu$, plus the kinetic energy $\frac 1 2 \rho^t v^2$, while the energy current includes that measured by the comoving observers, that induced by stress, and the flow of kinetic energy in the direction $v^i$. Similarly, the lab frame stress $T^{ij}$ includes both the stress measured by comoving observers and the standard kinetic contribution $\rho^t v^i v^j$ familiar from non-relativistic continuum dynamics. We see that all of these kinetic contributions are in fact fixed by Galilean invariance.

\subsection{Physical Currents}\label{sec:Improvement}
In relativistic theories with spinful matter there are two possible definitions of the stress-energy. One can vary the action at fixed connection, as we have done above
\begin{align}
	\delta S = \int d^{d+1}x | e| \left( - \tilde T^\mu{}_a \delta e^a_\mu + \tilde s^{\mu ab} \delta \omega_{\mu ab}\right),
\end{align}
or at fixed torsion
\begin{align}
	\delta S = \int d^{d+1}x | e| \left( - T^\mu{}_a \delta e^a_\mu + s_a{}^{\mu \nu} \delta T^a{}_{\mu \nu}\right) .
\end{align}

While both are good operators, the later is the more physically relevant of the two since it is $T^{\mu \nu} = T^{\mu a} e^\nu_a$ that appears in the Einstein equations. It is also more natural from a condensed matter/engineering perspective for several reasons. For one, when local rotational invariance is unbroken, conservation of angular momentum ensures that the Cauchy stress tensor is symmetric \cite{irgens2008continuum}, which is true of $T^{\mu \nu}$ but not $\tilde T^{\mu \nu}$. More directly, stresses in the lab are induced by a shear with the background torsion held fixed. We shall then refer to $T^{\mu \nu}$ as the Cauchy, or true, stress tensor.

Though not of direct physical relevance, $\tilde T^{\mu \nu}$ is often useful as an intermediate step since it somewhat easier to compute. However one must remember to always then obtain the true stress, which in the torsionless case is related to $\tilde T^{\mu \nu}$ by
\begin{align}
	&T^{\mu \nu} = \tilde T^{\mu \nu} - \nabla_\lambda (\tilde s^{\lambda \mu \nu} + \tilde s^{\mu \nu \lambda} - \tilde s ^{\nu \lambda \mu }) , 
\end{align}
a process we will refer to as ``improvement''.

Though well understood, we dedicate some space here to considering the non-relativistic case for completeness, but especially because because the energy current involves some subtleties not present in the relativistic example. The reason is the same that has troubled us before: it is impossible to introduce a perturbation to the clock form without introducing torsion. For this reason it will prove impossible to define improvements to $\varepsilon^\mu$ (and thus the full stress-energy tensor $\tilde \tau^\mu{}_I$, though we can improve $\tilde \tau^{\mu \nu}$). We will argue below that the unimproved energy current defined above already corresponds to the true internal energy of the system, and so should not be improved in any case.

\subsubsection{The Cauchy Stress-Mass Tensor}
Before we turn to the energy current however, let's consider the stress-mass tensor, whose story is straightforward. The physical currents are to be defined at fixed torsion, so we will require that the variation of $\df e^I$ does not involve the clock form. This in turn implies that the variation may be written as the pullback of some form $\delta \df e_A$ to the extended representation 
\begin{align}
	n_I \delta \df e^I = 0,
	&&\implies
	&& \delta \df e^I = \Pi^{AI} \delta \df e_A .
\end{align}
By design, we then also have
\begin{align}
	n_I \delta \df T^I = 0,
	&&\implies
	&&\delta \df T^I  = \Pi^{A I} \delta \df T_{A}
\end{align} 
for some $(\delta \df T_{A})_{ \mu \nu}$. We then define the Cauchy stress-mass $\tau^{\mu \nu}$ and spin current $s^{A\mu \nu}$ by the variation
\begin{align}\label{improvedCurrentsDef}
	\delta S = \int  d^{d+1} x | e | \left( \tau^{\mu A} (\delta \df e_A)_{ \mu} + s^{A \mu \nu} \delta (\df T_{A})_{ \mu \nu}\right) .
\end{align}

To perform the translation between (\ref{unimprovedCurrentsDef}) and (\ref{improvedCurrentsDef}) we shall need the variation of the spin connection $\delta \omega_{\mu AB}$ in terms of $(\delta \df e_A)_\mu$ and $(\delta \df T_A)_{ \mu \nu}$. We can retrieve this from the first structure equation $\df T^I = D \df e^I$, which gives
\begin{align}
	\delta\df  T^I = D \delta \df e^I + \delta \df \omega^I{}_J \wedge \df e^J .
\end{align}
Using $\delta \df e^I = \Pi^{AI} \delta \df e_A$, $\delta \df T^I = \Pi^{AI} \delta \df T_A $, and $\delta \df \omega_{IJ} = \Pi^A{}_I \Pi^B{}_J \delta\df  \omega_{AB}$, this reads
\begin{align}
	\delta\df  \omega_{AB} \wedge \df e^B = \delta \df T_A  - D \delta\df  e_A ,
\end{align}
which after some algebraic rearrangement gives
\begin{align}
	\delta \omega_{CAB} &= \delta \omega_{[CA]B} + \delta \omega_{[BC]A}-\delta \omega_{[AB]C} \nonumber \\
		&= \frac 1 2 \bigg(  ( D \delta \df e_A )_{BC} + (D \delta \df e_B)_{CA} - ( D \delta \df e_C )_{AB} \nonumber \\
		& \qquad \qquad \qquad - ( \delta \df T_A)_{BC} - (\delta \df T_B)_{CA} + (\delta \df T_C)_{AB} \bigg) .
\end{align}

Finally, using $\delta \df e^I = \Pi^{AI} \delta \df e_A$ and the definition (\ref{unimprovedStressMomentum}) of the unimproved stress-mass tensor $\tilde \tau^{\mu \nu}$, we find that
\begin{align}
	\delta S &= \int d^{d+1} x | e | \left( \tilde \tau^{\mu A} \delta e_{A \mu} + \tilde s^{\mu AB} \delta \omega_{\mu AB}\right) \nonumber \\
	&= \int d^{d+1} x | e | \left( \tau^{\mu A} (\delta \df e_A)_{ \mu} + s^{A \mu \nu} (\delta \df T_A)_{ \mu \nu}\right) ,
\end{align}
which implies
\begin{align}\label{stressMomentumImprovement}
	&  s^{\lambda \mu \nu} =  \frac 1 2 \left( \tilde s^{\lambda \mu \nu}  - \tilde s^{\mu \nu \lambda} - \tilde s^{\nu \lambda \mu} \right), \nonumber \\
	&\tau^{\mu \nu} = \tilde \tau^{\mu \nu} - 2 ( \nabla_\lambda - T^\rho{}_{\rho \lambda} ) s^{\nu \mu \lambda} - T^\mu{}_{\lambda \rho} s^{\nu \lambda \rho} .
\end{align}
In particular, this gives a physical mass current $\rho^\mu = \tau^{\mu \nu} n_\nu$
\begin{align}\label{rhoImprovement}
	\rho^\mu = \tilde \rho^\mu -2 ( \nabla_\lambda - T^\rho{}_{\rho \lambda} ) s^{\nu \mu \lambda}n_\nu - T^\mu{}_{\lambda \rho} s^{\nu \lambda \rho} n_\nu .
\end{align}
This is the physical mass current for non-scalar matter since the Newtonian gravitational potential is contained within the connection $\nabla$. That is, it is $\rho^\mu$ rather than $\tilde \rho^\mu$ that flows in response to a gravitational perturbation at fixed torsion.

\subsubsection{The Stress-Energy Tensor}
Finally, let's turn to the energy current and thus, by necessity, the full stress-energy tensor $\tilde \tau^\mu{}_I$. A perturbation to the the clock form necessarily introduces temporal torsion and so we cannot perform a variation at fixed torsion as done above. One might hope there is a way around this, for instance, by performing the variation at fixed spatial torsion $T^a{}_{\mu \nu}$. However, since $\df T^a$ is merely the spatial component of the Galilean vector $\df T^A$, there is no invariant way to do this. That is, since $\df T^a \rightarrow \df T^a - k^a \df T^0$ under a Galilean boost, once temporal torsion is present, there is no way to set spatial torsion to zero in a manner that all observers can agree upon.

Let's see what would go wrong if we tried to naively define the stress-energy at fixed extended torsion $\df T^I$
\begin{align}
	\delta S = \int d^{d+1} x | e |\left( - \tau^\mu{}_I \delta e^I_\mu + s_I{}^{\mu \nu} (\delta \df T^I)_{\mu \nu} \right).
\end{align}
Because $\df T^0 = d\df n$, this is a constrained variation of the vielbein $n_I d \delta \df e^I = 0$ and we are free to add a term $n_I H^{\mu \nu} ( d  \delta \df  e^I )_{\mu \nu}$, where $H^{\mu \nu}$ is an arbitrary antisymmetric tensor. This shifts the stress-energy tensor
\begin{align}
	\tau^\mu{}_I \rightarrow \tau^\mu{}_I - \left(2 (\nabla_\nu - T^\lambda{}_{\lambda \nu} )H^{\mu \nu} + T^\mu{}_{\nu \lambda} H^{\nu \lambda}\right) n_I
\end{align}
which, comparing to equation (\ref{stress-energy}), alters the energy current while leaving the stress, momentum, and mass components fixed. Thus we can unambiguously improve the $\tilde \tau^{\mu \nu}$, but not $\tilde \tau^\mu{}_I$.

Of course, this is unphysical. One considers the Cauchy stress since this is the stress measured in the lab, induced by shearing the system at fixed torsion. The energy measured in the lab of course has no such ambiguity. Moreover, the energy density $\varepsilon^t = \tilde \tau^t{}_0$ already corresponds to the Hamiltonian density of the system and so truly is the physical energy density of the system.

This can been seen in simple case of the spinful Schr\"odinger equation, whose action on the background $n_\mu = ( n_t ~ 0),~ h^{\mu \nu} = \begin{pmatrix} 0 & 0 \\ 0 & \delta^{ij} \end{pmatrix}$ is
\begin{gather}
	S = \int d^{d+1} x \sqrt{h} n_t \left( \frac{1}{n_t} \frac i 2 \psi^\dagger \overset{\leftrightarrow} D{}_t \psi - \frac{\delta^{ij}}{2m} D_i \psi^\dagger D_j \psi \right),
	\nonumber \\
	\text{where}
	\qquad \qquad
	D_\mu \psi = \left( \partial_\mu - i q A_\mu - i m a_\mu - \frac{i}{2} J^{ab} \omega_{ab} \right) \psi
\end{gather}
and $J^{ab}$ are the spin representation matrices. One then finds
\begin{align}
	\varepsilon^t = - \frac{\delta S}{\delta n_t} = \frac{\delta^{ij}}{2m} D_i \psi^\dagger D_j \psi  ,
\end{align}
whereas the Hamiltonian density for this system is
\begin{align}
	\mathcal H = \frac{\delta^{ij}}{2m} D_i \psi^\dagger D_j \psi - q A_t \psi^\dagger \psi - m a_t \psi^\dagger \psi - \frac 1 2 \omega_{t ab} \psi^\dagger J^{ab} \psi .
\end{align}
We see that $\varepsilon^t = \tilde \tau^t{}_0$ is the Hamiltonian density minus coupling to external potentials and so corresponds to the {\it internal} kinetic and interaction energy of a system. 

While we have motivated this in the specific case of the Schr\"odinger theory, it is actually quite general. The proper, coordinate invariant definition of the Hamiltonian is given by Iyer and Wald in the appendix to \cite{Iyer:1994ys}. Carrying out their analysis on a Bargmann geometry yields the same result.

\subsection{Ward Identities}\label{sec:Ward}
Finally, we turn to the principle results of this chapter, the Ward identities satisfied by the currents
\begin{align}\label{unimprovedCurrentsDef2}
	\delta S = \int d^{d+1} x | e | \left(-  { \tilde \tau^\mu}{}_I \delta e^I_\mu + {\tilde s^{\mu AB}} \delta{\omega_{\mu AB} } + j^\mu \delta A_\mu \right) .
\end{align}
as a result of diffeomorphism invariance, $U(1)$ gauge invariance, and internal Galilean invariance. Note we have now included an external electromagnetic gauge field $A_\mu$ and its current $j^\mu$ for later applications. There is no essential complication in considering general non-abelian gauge fields and the generalization of our results to this case should be clear throughout. However, since we have an eye to quantum Hall applications, we shall restrict to $U(1)$ for simplicity. In what follows we will refer to the mass and charge $U(1)$'s as $U(1)_M$ and $U(1)_Q$ respectively.

We will first consider the consequences of $U(1)$ invariance since this is the simplest case, and perform the calculation in detail. Under a $U(1)_Q$ transformation $\delta A_\mu = \nabla_\mu \alpha_Q$, the action varies as
\begin{align}
	\delta S = \int d^{d+1} x | e | j^\mu \nabla_\mu \alpha_Q = - \int d^{d+1} x | e | \alpha_Q \left( \nabla_\mu - T^\nu{}_{\nu \mu} \right) j^\mu
\end{align}
where we have integrated by parts in the second equality.\footnote{Note that in the torsionful case, integration by parts also introduces a torsional term. This is because upon integration by parts, one generates  the term $\partial_\mu | e | = \Gamma^\nu{}_{\nu \mu} - T^\nu{}_{\nu \mu}$.} Since we assume the action is gauge invariant, this must be zero for all choices of $\alpha_Q$ and thus
\begin{align}
	( \nabla_\mu - T^\nu{}_{\nu \mu} ) j^\mu = 0 .
\end{align}
Identities generated in this fashion are called Ward identities and are true on shell (since the variation of dynamical fields is proportional to the equations of motion, which we have set to zero).
The Ward identity for $U(1)_M$ invariance follows in exactly the same manner. After a variation $\delta e^I_\mu = n^I \nabla_\mu \alpha_M$, we generate the Ward identity
\begin{align}\label{unimprovedMassCons}
	( \nabla_\mu - T^\nu{}_{\nu \mu} ) \tilde \rho^\mu = 0 .
\end{align}

The next easiest Ward identity follows from local Galilean invariance
\begin{align}
	\df e^I \rightarrow \Lambda^I{}_J \df e^J,
	&& \df \omega_{AB} \rightarrow (\Lambda^{-1})^C{}_A (\Lambda^{-1})^D{}_B \df \omega_{CD} - d \Theta_{AB} .
\end{align}
In their infinitesimal forms $\Lambda^A{}_B = \delta ^A{}_B + \Theta^A{}_B + \cdots$, we have
\begin{align}\label{infinitesimalGal}
	\delta \df e^I = \Theta^I{}_J \df e^J=  \Pi^{AI} \Theta_{AB} \df e^B,
	&&\delta\df  \omega_{AB} = - D \Theta_{AB} .
\end{align}
The action then transforms as
\begin{align}
	\delta S &= \int d^{d+1} x | e | \left( - \tilde \tau^\mu{}_I \Pi^{AI} \Theta_{AB} e^B_\mu - \tilde s^{\mu AB} D_\mu \Theta_{AB} \right) \nonumber \\
		&= \int d^{d+1} x | e | \Theta_{AB} \left( - \tilde \tau^{AB} + (D_\mu - T^\nu{}_{\nu \mu} ) \tilde s^{\mu AB}\right) ,
\end{align}
from which we find
\begin{align}
	\tilde \tau^{[\mu \nu]} = ( \nabla_\lambda - T^\rho{}_{\rho \lambda} ) \tilde s^{\lambda \mu \nu} .
\end{align}

Now we turn to the Ward identities for diffeomorphisms. Under an infinitesimal coordinate transformation $\xi^\mu$, the background fields transform as
\begin{align}\label{diffeos}
	&\delta \df  e^I = \pounds_\xi \df  e^I = \xi \cdot ( d \df  e^I ) + d (\xi \cdot \df e^I ) = D ( \xi \cdot \df  e^I )+ \xi \cdot \df T^I - ( \xi \cdot \df  \omega^I{}_J ) \wedge \df e^J ,\nonumber \\
	&\delta \df  \omega_{AB} = \pounds_\xi \df  \omega_{AB} = \xi \cdot ( d \df  \omega_{AB} ) + d ( \xi \cdot \df \omega_{AB} ) = \xi \cdot \df R_{AB} + D (\xi \cdot \df  \omega_{AB}) , \nonumber \\
	&\delta \df  A = \pounds_\xi \df  A = \xi \cdot ( d \df A )+ d ( \xi \cdot \df A )= \xi \cdot \df F + d (\xi \cdot \df A ).
\end{align}
These transformations aren't terribly convenient since they refer to non-covariant objects such as $\df A$ and $\df \omega_{AB}$. However, they only differ from covariant expressions by a $U(1)_Q$ gauge transformation $\alpha_Q = - \xi \cdot \df  A$ and an infinitesimal local Galilean transformation $\Theta_{AB} = \xi  \cdot \df \omega_{AB}$. Since our theory is assumed to be gauge and local Galilean invariant we may perform a simultaneous diffeomorphism, $U(1)_Q$, and $Gal(d)$ transformation, after which we have
\begin{align}
	\delta \df e^I = D ( \xi \cdot \df e^I ) + \xi \cdot \df T^I
	&&\delta \df \omega_{AB} = \xi \cdot \df R_{AB}
	&&\delta \df A = \xi \cdot \df F.
\end{align}

Variation of the action then yields the Ward identity
\begin{align}\label{diff ward}
	- e^I_\mu ( D_\nu - {T^\lambda}_{\lambda \nu}) {\tilde \tau^\nu}{}_I = F_{\mu \nu} j^\nu + R_{AB\mu \nu} {\tilde s^{\nu AB}} - {T^I}_{\mu \nu} {\tilde \tau^\nu}{}_I.
\end{align}
The explicit appearance of $e^I_\mu$ may be concerning since it includes the $U(1)_M$ connection $a_\mu$, which is not gauge invariant. This is however not a problem since $a_\mu$ multiplies precisely the Ward identity (\ref{unimprovedMassCons}) and so drops out of the equation. To see this, perform a $U(1)_M$ transformation $e^I_\mu \rightarrow e^I_\mu + n^I \nabla_\mu \alpha_M$, after which we pick up a term proportional to $(\nabla_\mu - T^\nu{}_{\nu \mu} ) \tilde \rho^\mu$.

This identity may seem somewhat novel due to it's unfamiliar presentation. It is however well known in the case of flat spacetimes. Setting $\df e^I = ( dt ~ d x^a ~ - \phi dt )^T$ and $\df T^I = 0$ and churning through (\ref{diff ward}) in components (\ref{stress-energy}), we find the $\mu = t$ and $\mu = i$ components of the Ward identity are respectively\footnote{Let's consider spinless matter for now so that we can ignore the issues discussed in section \ref{sec:Improvement}.}
\begin{align}\label{flatSpaceFluid}
	\dot \varepsilon^t + \partial_i \varepsilon^i = E_i j^i + g_i \rho^i,
	&&\dot \rho^i + \partial_j T^{ij} = j^t E^i + F^{ij} j_j + \rho^t g^i ,
\end{align}
where $E_i = F_{i0}$ is the electric field, $F_{ij}$ the magnetic field, and $g_i = - \partial_i \phi$ the gravitational field.

These should be quite familiar. The first is simply the work-energy equation expressing the work done by external background fields on  the system. The second is the continuum version of Newton's second law, also known as the Cauchy momentum equation, expressing the local flow of momentum density subject to internal stresses $T^{ij}$ and external electromagnetic and gravitational forces. Once supplied with constitutive relations in terms of local thermodynamic degrees of freedom, this is the Navier-Stokes equation. These equations, supplemented by local mass and charge conservation, collectively serve as fluid equations of motion \cite{landau1987fluid}.
The Ward identity (\ref{diff ward}) then collects the work-energy and Cauchy momentum equations into a single covariant package and generalizes them to the curved, torsionful geometries we are considering. It will serve as the starting point for our fluid analyses in chapter \ref{chap:Apps}.

It will prove convenient later on to isolate the stress-mass part of equation (\ref{diff ward}). To do this, we raise the index using the spatial metric $h^{\mu \nu}$. Using the identity $e^{I \mu} = \Pi^{\mu I } + (a^\mu - e^\mu_0 ) n^I$ and (\ref{unimprovedMassCons}), we find this gives
\begin{align}
	( \nabla_\nu - {T^\lambda}_{\lambda \nu}) {\tilde \tau^{\nu \mu}} = F^\mu{}_{\nu} j^\nu + R_{AB}{}^\mu{}_\nu {\tilde s^{\nu AB}} - {T^{I\mu}}_{\nu} {\tilde \tau^\nu}{}_I.
\end{align} 
This also includes the local conservation of mass (\ref{unimprovedMassCons}), which can be obtained by contracting with $n_\mu$.

\subsubsection{Ward Identities for the Cauchy Stress-Mass}
In summary, the Ward identities for diffeomorphism invariance, gauge invariance and local internal Galilean invariance are
\begin{gather}
	- e^I_\mu ( D_\nu - {T^\lambda}_{\lambda \nu}) {\tilde \tau^\nu}{}_I = F_{\mu \nu} j^\nu + R_{AB\mu \nu} {\tilde s^{\nu AB}} - {T^I}_{\mu \nu} {\tilde \tau^\nu}{}_I , \label{diff ward1} \\
	(\nabla_\mu - T^\nu{}_{\nu \mu}) \tilde \rho^\mu = 0, \\
	(\nabla_\mu - T^\nu{}_{\nu \mu}) j^\mu = 0, \\
	\tilde \tau^{[\mu \nu]} = ( \nabla_\lambda - T^\rho{}_{\rho \lambda} ) \tilde s^{\lambda \mu \nu} \label{LGTWard1},
\end{gather}
which in particular include
\begin{align}\label{cauchy1}
	( \nabla_\nu - {T^\lambda}_{\lambda \nu}) {\tilde \tau^{\nu \mu}} = F^\mu{}_{\nu} j^\nu + R_{AB}{}^\mu{}_\nu {\tilde s^{\nu AB}} - {T^{I\mu}}_{\nu} {\tilde \tau^\nu}{}_I.
\end{align}
However, these are for the currents defined at fixed connection and we would like to state them for the physical currents.

As discussed in section \ref{sec:Improvement}, there is simply nothing we can do about $\tilde \tau^\mu{}_I$ as a whole and so (\ref{diff ward1}) must remain as it is.
We can however find the Ward identities satisfied by the Cauchy stress-mass $\tau^{\mu \nu}$. Plugging in the expressions (\ref{stressMomentumImprovement}) into (\ref{LGTWard1}) and using the identity (\ref{RicciSymmetry}), we find the antisymmetric part of the stress-mass tensor is fixed by the torsion and spin current
\begin{align}\label{LGTWard}
	\tau^{[\mu \nu]} = - T^{[\mu}{}_{\lambda \rho} s^{\nu] \lambda \rho } . 
\end{align}
In particular, on torsionless backgrounds, $\tau^{\mu \nu}$ is totally symmetric, which is indeed required of the Cauchy stress tensor $T^{ij}$ when local rotational invariance is unbroken \cite{irgens2008continuum}. Referring to equation (\ref{unimprovedStressMomentum}), we see that this also implies the equality of the momentum and mass currents
\begin{align}\label{coordLGTWard}
	p^i = \tau^{ti} = \rho^i ,
\end{align}
generalizing the relation (\ref{wittenGalInv}) used by Greiter, Witten, Wilczek and many subsequent authors to impose Galilean invariance to the multi-constituent case. Satisfyingly, this identity is precisely the Ward identity for internal Galilean boosts.

Similarly, in terms of the Cauchy stress-mass, equation (\ref{cauchy1}) reads
\begin{align}
	(\nabla_\nu - T^\lambda{}_{\lambda \nu}) \tau^{\nu \mu} = F^\mu{}_\nu j^\nu  +(  2 R_{ \rho \nu \lambda}{}^\mu - R^\mu{}_{\rho \nu \lambda}  ) s^{\rho \nu \lambda} - T^{I \mu}{}_\nu \tilde \tau^\nu{}_I ,
\end{align}
where we have made use of the identity (\ref{RicciSymmetry}).
We can remove the explicit dependence on the curvature by using the symmetry of the Riemann tensor under exchange of the first and second pairs of indices (\ref{indexSym2}). One then finds
\begin{align}\label{cauchyEquation}
	(\nabla_\nu - &T^\lambda{}_{\lambda \nu}) \tau^{\nu \mu} = F^\mu{}_\nu j^\nu  + (D \df T_A )^\mu{}_{\nu \lambda} s^{A \nu \lambda} - T^{I \mu}{}_\nu \tilde \tau^\nu{}_I .
\end{align}
In appendix \ref{app:Riemann} we have defined $D \df T_A$ by the equation $D \df T_I = \Pi^A{}_I D \df T_A$. Note that as such it is not necessarily the exterior derivative of a two-form (we use the notation since $D \df T^A = h^{AB} D \df T_B$). Contracting with $n_\mu$ we find that the physical mass current is also conserved
\begin{align}
	(\nabla_\mu - T^\nu{}_{\nu \mu} ) \rho^\mu = 0.
\end{align}

Equation (\ref{cauchyEquation}) is the covariant generalization of the Cauchy momentum equation to general Bargmann spacetimes. In particular we see that there are external forces exerted by extended torsion in addition to the usual electromagnetic force. Also note that, as promised in section (\ref{sec:ExtendedVielbein}), the mass component of the extended torsion $\df T^M$ exerts a Lorentz force on the (unimproved) mass current. This is then a general feature of physics on Bargmann spacetimes and not merely of the non-relativistic point particle.
While it may seem awkward to include the unimproved tensor $\tilde \tau^\mu{}_I$ in (\ref{cauchyEquation}), having converted everything else to physical currents, this is something we must simply accept as we have shown there is no unambiguous way to improve the energy current. We simply observe that the external force exerted by extended torsion couples to the unimproved stress-energy.

In summary then, we have the following (somewhat redundant) list of Ward identities for the physical currents
\begin{gather}
	- e^I_\mu ( D_\nu - {T^\lambda}_{\lambda \nu}) {\tilde \tau^\nu}{}_I = F_{\mu \nu} j^\nu + R_{AB\mu \nu} {\tilde s^{\nu AB}} - {T^I}_{\mu \nu} {\tilde \tau^\nu}{}_I ,\nonumber \\
	(\nabla_\nu - T^\lambda{}_{\lambda \nu}) \tau^{\nu \mu} = F^\mu{}_\nu j^\nu  + (D \df T_A )^\mu{}_{\nu \lambda} s^{A \nu \lambda} - T^{I \mu}{}_\nu \tilde \tau^\nu{}_I , \nonumber \\
	(\nabla_\mu - T^\nu{}_{\nu \mu}) \rho^\mu = 0, \qquad \qquad
	(\nabla_\mu - T^\nu{}_{\nu \mu}) j^\mu = 0, \nonumber \\
	\tau^{[\mu \nu]} = - T^{[\mu}{}_{\lambda \rho} s^{\nu] \lambda \rho } .
\end{gather}
In the torsionless case relevant for most applications these simplify to
\begin{gather}
	- e^I_\mu  D_\nu{\tilde \tau^\nu}{}_I = F_{\mu \nu} j^\nu + R_{AB\mu \nu} {\tilde s^{\nu AB}} ,\nonumber \\
	\nabla_\nu \tau^{\mu \nu} = F^\mu{}_\nu j^\nu   , \qquad \qquad
	\nabla_\mu  \rho^\mu = 0, \qquad \qquad
	\nabla_\mu  j^\mu = 0, \qquad \qquad
	\tau^{[\mu \nu]} =0 .
\end{gather}

\section{Galilean Invariant Theories}\label{chap:Examples}

In chapter \ref{chap:Ward}, we considered very general consequences of Galilean and diffeomorphism invariance, valid in any theory that realizes them. Now we turn to the question of how to actually write down theories that do so. As overviewed in the introduction, the metric formalism of Newton-Cartan geometry isn't terribly helpful in this regard, as both the Schr\"odinger and superfluid Lagrangians cannot be written with only the data $( n_\mu , h^{\mu \nu} , \nabla)$. The Schr\"odinger action requires a careful cancellation between the kinetic and time derivative terms of (\ref{CovSchrodL}), while the action (\ref{goldstoneAction}) for superfluid Goldstones is instead solved for from a differential equation in the dependence of the Lagrangian on fields and their derivatives. It is lowest order in derivatives and specific to single-constituent superfluids.
These are among the simplest and most useful actions we know of, so one might justifiably be concerned that writing down more complicated Galilean field theories will quickly become extremely cumbersome.

One approach, common in the literature, is to construct such theories by writing down a ``parent" relativistic theory in a Lorentzian spacetime of one higher dimension. One then demands a null isometry and chooses to keep modes of only a certain momentum in that direction, which will then be the mass of the resulting non-relativistic field. The resulting theory is guaranteed to realize Galilean symmetry. This method requires a fair amount of excess conceptual baggage though and can be computationally cumbersome. 
It is however a systematic means to churn out theories of the Schr\"odinger type. 

In this chapter we introduce a simple and efficient means to realize Galilean symmetry in Wilsonian effective actions. The key ingredient is a derivative operator $D_I$, introduced in section \ref{sec:Derivative}, that is valued in the extended representation of the Galilean group. We claim this is the natural derivative operator that can be defined on the $U(1)_M$ bundle encoding the kinematics of massive non-relativistic fields. Only when operating on massless fields can $D_I$ be restricted to an invariant spacetime derivative operator $\nabla$.

One can then systematically generate actions with manifest Galilean symmetry by writing them in terms of fields, their extended derivatives, and Galilean invariant tensors as we will do in section \ref{sec:Schrodinger}. For massive fields, this generates the Schr\"odinger action at lowest order and easily produces higher derivative terms to any order one desires. We shall find that this is formally equivalent to the procedure of null reduction, but has the benefit of being intrinsic to the non-relativistic spacetime under consideration and not requiring any computation. In future work we will show how this method may also be used to reproduce Greiter, Witten, and Wilczek's theory of superfluids in a systematic way that generalizes to the multi-constituent case \cite{GeracieForthcoming}.

We then consider theories of a different type, the non-relativistic analog of the Dirac equation. Such actions were originally considered by Levy-Leblond in \cite{LevyLeblond:1967zz}
and generalized to a larger class of representations by de Montigny, Niederle, and Nikitn in \cite{Montigny,Niederle:2007xp}.
Recently, Levy-Leblond's action has been coupled to Newton-Cartan geometries \cite{Fuini:2015yva}, though their method should also work for the more general Dirac-like representations of de Montigny et al. In section \ref{sec:Dirac} we review their constructions. While we are not aware of any physical systems that realize these possibilities (other than the non-relativistic electron with $g$-factor 2, which may also be discussed in the Schr\"odinger approach), they are interesting to consider and one might hope nature will be kind enough to furnish us with examples some day. The content of section \ref{sec:Dirac} is then entirely the product of other researchers. We include it here out of interest and a sense of completeness as another example of a collection of actions one may write down with manifest Galilean symmetry.

\subsection{The Extended Derivative Operator}\label{sec:Derivative}
To recap from section \ref{sec:ActionPrinciples}, generalizing the Schr\"odinger action to curved spacetimes requires referencing an auxiliary timelike vector field $v^\mu$, which one uses to write down the time derivative term. Identifying this field with the timelike portion of the frame $e^\mu_0$,\footnote{At this point, these are conceptually distinct. However, the actions we will write down in this section will naturally churn out theories with $v^\mu= e^\mu_0$.} we have
\begin{gather}\label{schrod}
	S = \int d^{d+1} x | e | \left( \frac i 2 e^\mu_0 \psi^\dagger \overset{\leftrightarrow} D{}_\mu \psi - \frac{h^{\mu \nu}}{2m} D_\mu \psi^\dagger D_\nu \psi \right).
\end{gather}
Here we are considering massive matter with arbitrary charge and spin. The covariant derivative is then
\begin{align}
	D_\mu = \partial_\mu - i A_\mu Q - i a_\mu M - \frac i 2 \omega_{\mu AB} J^{AB} ,
\end{align}
where $Q$ is the charge, $M$ is the mass, and $J^{AB}$ the rotation (and boost) operators acting on $\psi$ (see equation (\ref{standardBasis})). For example, for a spin $1/2$ particle, $J^{ab} =  \frac i 4 [ \gamma^a , \gamma^b ]$ where the $\gamma^a$'s satisfy the Clifford algebra $\{ \gamma^a , \gamma^b  \} =  2 \delta^{ab}$, and $J^{a0} = 0$.\footnote{An important ingredient for Galilean invariance is that $\psi$ is boost trivial, that is, $J^{AB} n_B = 0$. As in the relativistic case, boosts cannot be realized unitarily and so the action would not be invariant under a non-trivially realized boost transformation. For further discussion see section \ref{sec:Dirac}.}

Of course, $e^\mu_0$ is not boost invariant $e^\mu_0 \rightarrow e^\mu_0 + e_a^\mu (R^{-1})^a{}_b k^b $.
Fortunately, the second term is not either, owing to the transformation of the mass gauge field $a_\mu \rightarrow a_\mu + k_b R^b{}_a e^a_\mu - \frac 1 2 k^2 n_\mu$. 
Galilean invariance of (\ref{schrod}) then relies on a careful cancellation between the kinetic and time derivative terms as discussed in section \ref{sec:ActionPrinciples}.

In other words, the ``covariant derivative" $D_\mu$ is not covariant at all, but transforms under Galilean boosts as
\begin{align}
	D_\mu \rightarrow D_\mu - i k_b R^b{}_a e^a_\mu M + \frac 1 2 i k^2 n_\mu M .
\end{align}
The key observation of this section is that it can however be collected into a larger derivative operator valued in the extended representation of the Galilean group
\begin{align}
	D_I = ( D_A ~ i M ),
	&&\text{where}
	&&D_A = e^\mu_A D_\mu .
\end{align}
One may then check by hand that if $\psi$ transforms in some representation $\psi \rightarrow \Lambda_R \psi$ of $Gal(d)$, it's extended derivative also transforms covariantly
\begin{align}
	D_I \psi \rightarrow \Lambda_R (\Lambda^{-1})^J{}_I D_J \psi .
\end{align} 
It is then $D_I$ rather than $D_\mu$ that is the natural derivative operator on massive non-relativistic fields.

Note that for massless fields $f$, we have $n^I D_I f = 0$. In this case the derivative operator may be written as the pullback of the usual derivative operator to the extended representation
\begin{align}
	D_I f = \Pi^\mu{}_I D_\mu f.
\end{align}
In particular we may retrieve in this way the Newton-Cartan derivative operator $\nabla$ which we know to be Galilean invariant, but this derivative operator does not act on massive fields.

\subsection{Schr\"{o}dinger Fields}\label{sec:Schrodinger}
Now let's use this to start writing down theories. The first obvious thing to try is the ``Klein-Gordon"-type action
\begin{align}\label{covariantSchrodinger}
	S = - \frac{1}{2m } \int d^{d+1} x | e | D_I \psi^\dagger D^I \psi ,
\end{align}
which upon expanding yields precisely (\ref{schrod}). We have thus found a manifestly invariant formulation of the Schr\"odinger equation. Moreover, it's now a trivial matter to consider more general theories without breaking Galilean invariance.

As an example, the first higher derivative term that appears is
\begin{align}
	D_I D_J \psi^\dagger D^I D^J \psi .
\end{align}
The get a sense for this term, let's expand it in flat space
\begin{align}
	D_I D_J \psi^\dagger D^I D^J \psi &= \text{Tr}
	\begin{pmatrix}
		D_0 D_0  \psi^\dagger & D_0 D_c \psi^\dagger & -i m D_0 \psi^\dagger \\
		D_a D_0 \psi^\dagger & D_a D_c \psi^\dagger & -i m D_a  \psi^\dagger \\
		-i m D_0 \psi^\dagger & -im D_c \psi^\dagger & - m^2 \psi^\dagger
	\end{pmatrix}	
	\begin{pmatrix}
		- m^2 \psi & i m D^b \psi & i m D_ 0 \psi \\
		i m D^c \psi & D^c D^b \psi &  D^c D_0  \psi \\
		 i m  D_0 \psi &  D_0 D^b \psi & D_0 D_0  \psi
	\end{pmatrix}\nonumber \\
	&=
	m^2 \left( - D_t D_t \psi^\dagger \psi + 2 D_t \psi^\dagger D_t \psi - \psi^\dagger D_t D_t \psi \right) \nonumber \\
	&\qquad + i m \left( \{ D_t , D_i \} \psi^\dagger D^i \psi  - D^i \psi^\dagger \{ D_t , D_i \} \psi  \right) + D_i D_j \psi^\dagger D^i D^j \psi .
\end{align}
After integration by parts, this is the term (\ref{higherDeriv}) given in the introduction to this thesis.
One may check by explicit computation that this is in fact invariant under the projectively realized Galilean transformation
that one uses to check boost invariance in the standard, textbook approach.

One can now extend this method to any order in derivatives one wishes
\begin{align}
	S [ \psi , D_I \psi , \cdots , D_{I_1} \cdots D_{I_k} \psi] .
\end{align}
So long as one properly contracts indices using the invariant metric $g_{IJ}$, the resulting theory will be Galilean invariant.\footnote{One can also contract with $n^I$, but this is rather trivial since $n^I D_I \psi= i m \psi$.}

As promised, this method of generating non-relativistic theories is clearly equivalent to the procedure of null reduction. The extended derivative $D_I$ is simply the covariant derivative in the higher-dimensional spacetime, which of course should be contracted with a $(d+2)$-dimensional Lorentzian metric. Selection of a particular momentum in the null direction is simply the fact that $n^I D_I \psi = i m \psi$. We hope however that our formulation has certain conceptual advantages since it is intrinsic to the non-relativistic spacetime. It is certainly more streamlined from a calculational point of view, as one may write down theories directly, without the intervening steps of postulating a parent unphysical theory and performing reduction, which can be cumbersome in the presence of interacting modes.

\subsubsection{Currents}

As an example, let's compute the currents defined in chapter \ref{chap:Ward} for the Schr\"odinger field (\ref{covariantSchrodinger}). For this we will need the variation of the extended derivative operator acting on $\psi$. 
The non-covariant derivative $D_\mu$ simply varies with the mass gauge field $\delta D_\mu \psi = - i \delta a_\mu M \psi$, from which we find
\begin{gather}
	\delta (e^\mu_A D_\mu \psi) = - \delta e^B_\nu e^\nu_A e^\mu_B D_\mu \psi - i \delta a_\nu e^\nu_A M \psi = - \delta e^I_\nu e^\nu_A D_I \psi  .
\end{gather}
Including the variation of the spin connection and electromagnetic gauge field then gives
\begin{align}\label{derivVar}
	\delta D_I \psi = - \Pi^\mu{}_I \left( \delta e^J_\mu D_J \psi	+ \frac i 2 J^{AB} \psi \delta \omega_{\mu AB} + i q \psi \delta A_\mu \right).
\end{align}
Similarly, the variation of the volume element is
\begin{align}
	\delta | e | = | e | e^\mu_A \delta e^A_\mu = | e | \Pi^\mu{}_I \delta e^I_\mu .
\end{align}

Using these formulas, it's then not difficult to find the currents
\begin{gather}
	\tilde s^{\mu AB} = - \frac{i}{4m} \psi^\dagger J^{AB} \overset \leftrightarrow D {}^\mu \psi , \nonumber \\
	\tilde \tau^\mu{}_I = - \frac{1}{2m} \left( D^\mu \psi^\dagger D_I \psi + D_I \psi^\dagger D^\mu \psi \right) + \frac 1 {2m} D_J \psi^\dagger D^J \psi \Pi^\mu{}_I , \nonumber \\
	j^\mu = - \frac{iq}{2m} \psi^\dagger \overset \leftrightarrow D {}^\mu \psi  ,
\end{gather}
where importantly, by $D^\mu \psi$ we do not mean the raised index version of the non-covariant derivative $D_\mu \psi$, but rather $e^\mu_A$ contracted with
\begin{align}
	D^A \psi = \Pi^{A I} D_I \psi = 
	\begin{pmatrix}
		im  \psi \\
		D^a \psi
	\end{pmatrix} .
\end{align}

Before writing the Cauchy currents, we note that the spin current for the Schr\"odinger field is conserved on-shell \(D_\mu \tilde s^{\mu AB} = 0\), as one would expect in a theory with no spin-orbit coupling
\begin{align}\label{spinConservation}
	D_\mu \tilde s^{\mu AB} & = - \frac{i}{4m} \Pi^{\mu I} D_\mu \left( \psi^\dagger J^{AB} \overset \leftrightarrow D_I \psi \right) \nonumber \\
		& = - \frac{i}{4m} D^I \left( \psi^\dagger J^{AB} \overset \leftrightarrow D_I \psi \right) \nonumber \\
		& = - \frac{i}{4m} \left( \psi^\dagger J^{AB} D^I D_I \psi \right) + c.c. = 0
\end{align}
where in the second equality we have used the fact that \(\psi^\dagger J^{AB} \overset \leftrightarrow D_I \psi \) is a massless ($U(1)_M$ invariant) field which implies that \(\Pi^{\mu I}D_\mu = D^I\). Finally, the last line vanishes by the covariant form of Schr\"odinger's equation (on torsionless backgrounds) \(D^I D_I \psi = 0\).

The physical currents are then
\begin{gather}
	s^{\lambda \mu \nu} = - \frac{i}{8m} \left( \psi^\dagger J^{\mu \nu} \overset \leftrightarrow D {}^\lambda \psi  -  \psi^\dagger J^{\nu \lambda} \overset \leftrightarrow D {}^\mu \psi  -  \psi^\dagger J^{\lambda \mu} \overset \leftrightarrow D {}^\nu \psi \right) , \nonumber \\
	\tau^{\mu \nu} =  \frac{1}{m}  D^{(\mu} \psi^\dagger D^{\nu)} \psi  - \frac 1 {2m} D_I \psi^\dagger D^I \psi h^{\mu \nu} - \frac{i}{2m} \nabla_\lambda \left( \psi^\dagger  J^{ \lambda (\mu} \overset \leftrightarrow D {}^{\nu )}  \psi \right) + \frac{i}{4m} \nabla_\lambda \left( \psi^\dagger J^{\mu \nu} \overset \leftrightarrow D {}^\lambda \psi\right), \nonumber \\
	\rho^\mu = - \frac i 2 \psi^\dagger \overset \leftrightarrow D {}^\mu \psi - \frac 1 2 \nabla_\nu \left( \psi^\dagger J^{\mu \nu} \psi \right) , \nonumber \\
	j^\mu = - \frac {iq}{2m} \psi^\dagger \overset \leftrightarrow D {}^\mu \psi ,
\end{gather}
where we have restricted to torsionless backgrounds for simplicity. In the above we have used (\ref{spinConservation}) to write the stress-mass tensor in a manifestly symmetric form on shell, something which we know is guaranteed by the Ward identity for local Galilean transformations (\ref{LGTWard}).

Since this likely all looks rather unfamiliar, we also give component expressions of the above equations for a spin $1/2$ particle in flat, three-dimensional space. On shell, these are
\begin{gather}
	s^{\mu}_j = 
	\begin{pmatrix}
		\psi^\dagger S_j \psi \\
		- \frac{i}{2m} \psi^\dagger S_j \overset \leftrightarrow D{}^i \psi
	\end{pmatrix} , \nonumber \\
	\varepsilon^\mu =
	\begin{pmatrix}
		\frac{1}{2m} D_i \psi^\dagger D^i \psi \\
		- \frac 1 {2m} \left( D^i \psi^\dagger D_t \psi + D_t \psi^\dagger D^i \psi \right)
	\end{pmatrix}, 
	\nonumber \\
	\rho^\mu =
	\begin{pmatrix}
		m \psi^\dagger \psi \\
		- \frac{i}{2} \psi^\dagger \overset \leftrightarrow D{}^i \psi + \frac 1 2 \epsilon^{ijk} \partial_j \left( \psi^\dagger S_k \psi \right)
	\end{pmatrix} , \qquad \qquad 
	j^\mu =
	\begin{pmatrix}
		q \psi^\dagger \psi \\
		- \frac{iq}{2m} \psi^\dagger \overset \leftrightarrow D{}^i \psi
	\end{pmatrix} ,
	\nonumber \\
	T^{ij} = \frac{1}{m} D^{(i} \psi^\dagger D^{j )} \psi + \left( \frac{i}{2} \psi^\dagger \overset \leftrightarrow  D_t \psi - \frac{1}{2m} D_i \psi^\dagger D^i \psi \right) \delta^{ij} - \frac i {2m} \epsilon^{kl(i} \partial_k \left( \psi^\dagger S_l \overset \leftrightarrow D{}^{j )} \psi \right) ,
\end{gather}
where $S^i = \frac 1 2 \sigma^i$ are the Pauli spin operators. Note the energy current is the anticommutator of the energy $- i D_t $ and velocity $- \frac{i} m D^i$ and represents kinetic energy being transported with the velocity of the particle.


For the sake of familiarity, we have defined $s^\mu_j = - \epsilon_{jkl} s^{kl \mu}=- \frac{i}{2m} \psi^\dagger S_j \overset \leftrightarrow D {}^\mu \psi$,\footnote{The formula $\tilde s^{\lambda ij} = - 2 s^{[ij] \lambda}$ is useful here.} whose density is the spin density of standard quantum mechanics and whose current may also be interpreted along the lines of the energy current as it is half the anticommutator of the spin density and velocity. By virtue of (\ref{spinConservation}), this is conserved on shell
\begin{align}
	\partial_\mu s^\mu_i = 0 .
\end{align}

The mass current is simply $m$ times the probability current, plus a magnetization term $ \frac 1 2 \nabla \times \overset \rightarrow S$ that can be very roughly interpreted as mass flow due to spinning matter whose rate of precession is not uniform. As usual, one should not take this too interpretation too seriously. Note in particular, variation in the spin density does not give rise to an electromagnetic magnetization current and the charge and mass currents are distinct even in the single-constituent case in the presence of spinful matter.

Finally, the stress tensor is the familiar stress tensor for spinless Schr\"odinger fields, plus a contribution $\epsilon^{kl ( i} \partial_k s^{j)}_l$ arising from non-uniformities in the spin current. Of course, we do not display the momentum current since it is equal to the mass current on shell.

\subsection{Non-Relativistic Dirac Fields}\label{sec:Dirac}
There are some options for non-relativistic actions which we have not considered. If a massive field $\psi$ transforms under boosts, the action (\ref{covariantSchrodinger}) is not invariant since boosts cannot be realized unitarily. 
Importantly, we are not referring to the projective boost transformation of equation (\ref{projTransf}), which we long ago absorbed into the the mass gauge field; rather, we are considering internal boosts
\begin{align}
	\Psi \rightarrow e^{- \frac i 2 J^{AB} \Theta_{AB}} \Psi,
	&&\text{where}
	&&
	\Theta_{AB}
	=
	\begin{pmatrix}
		0 & k_b \\
		- k_a & \theta_{ab}
	\end{pmatrix} 
\end{align}
parametrizes an arbitrary $Gal(d)$ transformation and the generators $J^{0a} =K^a$ are nontrivial. 
To our knowledge, a general theory of the finite-dimensional representations of the Galilean group has not been developed and finding them is a surprisingly difficult task, but in four spacetime dimensions, a complete list with spin 1 and lower has been obtained and may be found in equation (11) and Table 1 of \cite{Montigny}. To convert their results to our conventions, one should take $J^{ab} = - \epsilon^{abc} S_c$, $K^a = -\eta^a$.

Of those representations with $K^a \neq 0$, there is one with spin $1/2$ and eight with spin 1. The spin $1/2$ representation for instance contains two spin $1/2$ fields 
\begin{align}
	\Psi =
	\begin{pmatrix}
		\psi \\
		\chi
	\end{pmatrix}
\end{align}
that transform into each other under boosts
\begin{align}\label{eq:NR_dirac_rep}
	J^{ab} = - \epsilon^{abc}
	\begin{pmatrix}
		\frac 1 2 \sigma_c & 0 \\
		0 & \frac 1 2 \sigma_c
	\end{pmatrix},
	&&J^{a0} =
	\begin{pmatrix} 
		0 & 0 \\
		\frac i 2 \sigma^a & 0
	\end{pmatrix},
\end{align}
and was originally discovered by Levy-Leblond in \cite{LevyLeblond:1967zz}. The non-relativistic Dirac equation (\ref{NRDirac}) for this field is precisely the $c \rightarrow \infty$ limit of the relativistic 4-component Dirac equation: the second spinor is auxiliary and integrating it out yields the Schr\"odinger equation with a $g$-factor of 2. We are unfortunately not aware of physical realizations of the other boost-nontrivial representations listed in \cite{Montigny}.

However, given a field in one of these representations $\psi \rightarrow \Lambda_R \psi$, one may follow the standard procedure to construct a Dirac equation for it. In a follow up to \cite{Montigny}, the authors find a collection of matrices $\beta^I$ for each $R$ such that \cite{Niederle:2007xp}
\begin{align}
	\Lambda_R^\dagger \beta^I \Lambda_R = \Lambda^I{}_J \beta^J .
\end{align}
The non-relativistic Dirac action is then
\begin{align}\label{NRDirac}
	S = \int d^4x | e | \frac i 2 \Psi^\dagger \beta^I \overset \leftrightarrow D_I \Psi,
\end{align}
where we have taken the liberty of minimally coupling their action to a Bargmann spacetime. The $\beta$-matrices they obtain may be found in equation (10) and equations (42-45) of \cite{Niederle:2007xp}.\footnote{In comparing to \cite{Niederle:2007xp}, we have $\beta^I_{ours} = ( \beta_{0} , ~ \beta_a , - \beta_4)_{theirs}$.} For the Leby-Leblond representation they are
\begin{align}\label{betaMatrices}
	\beta^0 = 
	\begin{pmatrix}
		I & 0 \\
		0 & 0 
	\end{pmatrix},
	&&\beta^a = 
	\begin{pmatrix}
		0 & \sigma^a \\
		\sigma^a & 0 
	\end{pmatrix},
	&&\beta^{M} =
	\begin{pmatrix}
		0 & 0 \\
		0 & - 2 I 
	\end{pmatrix}.
\end{align}

The non-relativistic Dirac equation obtained from \eqref{NRDirac} is then precisely the $c \rightarrow \infty$ limit of the relativistic 4-component Dirac equation (see also \cite{Fuini:2015yva}). Expanding this out we have
\begin{align}
	 S = \int d^4 x | e | \left( \frac i 2 \psi^\dagger \overset \leftrightarrow D_0 \psi + \frac i 2 \psi^\dagger \sigma^a \overset \leftrightarrow D_a \chi + \frac i 2 \chi^\dagger \sigma^a \overset \leftrightarrow D_a \psi + 2 m \chi^\dagger \chi \right)
\end{align}
The bottom component \(\chi\) is auxiliary, satisfying the constraint (we are now restricting to flat spacetimes)
\begin{align}\label{auxiliarySol}
	\chi = - \frac{i}{2m} \sigma^a D_a \psi .
\end{align}

Plugging this in on torsionless backgrounds then gives, after integration by parts, the Schr\"odinger action  with a $g$-factor of 2 for the top spinor \(\psi\)
\begin{align}\label{gFactor}
	S = \int d^4 x | e |  \left( \frac i 2 \psi^\dagger \overset \leftrightarrow D_0 \psi - \frac{\delta^{ab}}{2m} D_a \psi^\dagger D_b \psi + \frac q m B_a \psi^\dagger S^a\psi - \frac{R}{8m} \psi^\dagger \psi\right)
\end{align}
where $B^a =  \frac 1 2 \epsilon^{abc} F_{bc}$ is the magnetic field and $S^a = \frac 1 2 \sigma^a$ is the spin operator. Note that the $g$-factor also induces a non-minimal coupling to the Ricci scalar $R$. This follows from the same commutator of derivatives giving rise to the magnetic moment.

\subsubsection{Currents}

From the action \eqref{NRDirac} we find the stress-energy and spin current to be
\begin{align}
	\tilde \tau^\mu{}_I = \frac i 2 \Psi^\dagger \beta ^\mu \overset \leftrightarrow D_I \Psi - \frac i 2 \Psi^\dagger \beta ^J \overset \leftrightarrow D_J \Psi \Pi^\mu{}_I,
	&&\tilde s^{\mu AB} = \frac 1 4 \Psi^\dagger M^{\mu AB} \Psi ,
\end{align}
where we have defined $M^{C AB} = \beta^C J^{AB} + (J^{AB})^\dagger \beta ^C $ and as usual $\beta^C = \Pi^C{}_I \beta^I$. Computing the components of $M^{ABC}$ from \eqref{eq:NR_dirac_rep} and \eqref{betaMatrices}, we find that $M^{ABC}$ is in fact totally antisymmetric in its indices. Given this total anti-symmetry, it's independent components are
\begin{align}
	M^{0ab} = - \epsilon^{abc}
	\begin{pmatrix}
		\sigma_c & 0 \\
		0 & 0 
	\end{pmatrix},
	&&M^{abc} = - \epsilon^{abc}
	\begin{pmatrix}
		0 & 1 \\
		1 & 0
	\end{pmatrix} .
\end{align}

The Cauchy stress-mass in torsionless backgrounds is then
\begin{gather}
	s^{\lambda \mu \nu} =- \frac 1 8 \Psi^\dagger M^{  \lambda \mu \nu } \Psi , \nonumber \\
	T^{\mu \nu} = - \frac i 2 \Psi^\dagger \beta ^{( \mu} \overset \leftrightarrow { D} {}^{\nu ) } \Psi + \frac i 2 \Psi^\dagger \beta ^I \overset \leftrightarrow D_I \Psi h^{\mu \nu}, \nonumber \\ 
	\rho^\mu =  \frac m 2  \Psi^\dagger \beta^\mu \Psi - \frac i 4 \Psi^\dagger \beta^0 \overset \leftrightarrow  { D} {}^\mu \Psi, \qquad \qquad  
	j^\mu =  q  \Psi^\dagger \beta^\mu \Psi .
\end{gather}
In simplifying this we have used the total antisymmetry of $M^{CAB}$ and used the Ward identity $T^{\mu \nu} = T^{( \mu \nu )}$ since we are considering the on shell currents.

This form has the benefit of being manifestly Galilean covariant, but is perhaps somewhat mysterious. To get a feel for this let's restrict to the Levy-Leblond representation and simplify using the constraint \eqref{auxiliarySol}. In components we then have
\begin{gather}
	\varepsilon^\mu =
	\begin{pmatrix}
		- \frac{1}{4m} \psi^\dagger D_k D^k \psi - \frac{1}{4m} D_k D^k \psi^\dagger \psi - \frac q m B_i \psi^\dagger S^i \psi \\
		- \frac{1}{4m} D^i \psi^\dagger \overset \leftrightarrow D_t \psi + \frac{i}{2m} \epsilon^{ijk} D_j \psi^\dagger S_k \overset \leftrightarrow D_t \psi + c.c.
	\end{pmatrix}
	\nonumber \\
	\rho^\mu =
	\begin{pmatrix}
		m \psi^\dagger \psi \\
		- \frac{i}{2} \psi^\dagger \overset \leftrightarrow D {}^i \psi + \frac 1 2\epsilon^{ijk} \partial_j \left( \psi^\dagger S_k \psi \right)
	\end{pmatrix} ,
	\qquad \qquad
	j^\mu =
	\begin{pmatrix}
		q \psi^\dagger \psi \\
		- \frac{iq}{2m} \psi^\dagger \overset \leftrightarrow D {}^i \psi + \frac q m \epsilon^{ijk} \partial_j \left( \psi^\dagger S_k \psi \right)
	\end{pmatrix} ,
	\nonumber \\
	T^{ij} = \frac{1}{2m} D^{(i} \psi^\dagger D^{j)} \psi - \frac 1{4m} \psi^\dagger D^{(i} D^{j)} \psi - \frac 1 {4m} D^{(i} D^{j)} \psi^\dagger \psi + \frac q m B^{(i } \psi^\dagger S^{j)} \psi \nonumber \\
		\qquad \qquad  \qquad + \left( \frac i 2 \psi^\dagger \overset \leftrightarrow D_t \psi + \frac 1 {4m} \psi^\dagger D_k D^k \psi + \frac{1}{4m} D_k D^k \psi^\dagger \psi\right) \delta^{ij} - \frac i{2m} \epsilon^{kl(i} \partial_k \left( \psi^\dagger S_l \overset \leftrightarrow D{}^{j)} \psi\right) .
\end{gather}
In particular, we have the standard charge and mass currents, plus magnetization currents arising from the magnetic moments that the $g$-factor attaches to particles. Note that the mass magnetization enters as though it had $g$-factor 1. In comparison to the Schr\"odinger case, these currents also have additional terms in the energy current and stress arising from the non-minimal coupling terms
\begin{align}
	\frac{q}{m} B_i \psi^\dagger S^i \psi - \frac{R}{8m} \psi^\dagger \psi .
\end{align}

 \subsection{Quantum Hall EFT}\label{sec:effectiveAction}
 
We conclude with an application to the quantum Hall effect, writing down most general effective action (to our order in power counting) consistent with Galilean invariance for systems gapped by a non-vanishing magnetic field $B$. This problem has been considered by numerous authors, beginning with \cite{FROHLICH1991517,BlokWen,Wen:1992uk,Wen:1992ej}, which characterized the long distance physics of quantum Hall systems by the quantized coefficients of the Chern-Simons and Wen-Zee terms (we shall specialize to the case of the integer quantum Hall states in this section and not consider the heirarchical construction). 

Additional restrictions owing to Galilean invariance were considered in \cite{Hoyos:2011ez}, which also imposed an additional symmetry
\begin{align}\label{singleConstituent}
	\df A \rightarrow \df A + \df \xi,
	&&\df a \rightarrow \df a - \frac q m \df \xi
\end{align}
specific to single particle systems (see \cite{Geracie:2016inm}). In this section we enlarge the the backgrounds considered by previous authors to general Bargmann spacetimes and impose Galilean invariance without resort to \eqref{singleConstituent}, so finding the most general effective action to our order in power counting, applicable to systems composed of arbitrary numbers of fundamental constituents. The added generality leads us to introduce three new Chern-Simons-type terms with quantized coefficients $\nu_M, \kappa_M$, and $\zeta_M$. These terms attach mass to vorticity, charge to vorticity, and mass to curvature respectively.

All results will be stated on torsionless backgrounds. Hence though torsion will be essential to carry out the variations correctly, it will not appear in the final formulas for the variations themselves.

\subsubsection{Setup}

As is usual, we shall assume slowly varying fields and so adopt a power counting scheme in which derivatives are small
\begin{align}
	\partial_t \sim \epsilon^2,
	&&\partial_i \sim \epsilon .
\end{align}
We are concerned with backgrounds in which there is a large magnetic field $B$ gapping the system and so
\begin{align}
	A_t \sim 1,
	&&A_i \sim \epsilon^{-1} .
\end{align}
The power counting is chosen so that $B \sim 1$, but electric fields are small $E_i \sim \epsilon$. Although in experimental setups, the other background curvatures $d \df a$, $d \df \omega$, and $d \df n$ are small or vanishing, it will be convenient to consider all potentials on the same level as the electromagnetic one
\begin{align}
	a_t \sim 1,
	&&a_i \sim \epsilon^{-1} ,
	&&\omega_t \sim 1,
	&&\omega_i \sim \epsilon^{-1} ,
	&&n_t \sim 1,
	&&n_i \sim \epsilon^{-1} .	
\end{align}

The most general effective action to order $\epsilon^0$ is then
\begin{align}
	\label{QHEAction}
	 S &= \int \bigg(   \frac{\nu}{4 \pi} \df A \wedge d \df A + \frac{\kappa}{2 \pi} \df \omega \wedge d \df A + \frac{\nu_M}{4\pi} \overset u {\df a}  \wedge d \overset u {\df a}  + \frac{\zeta_M}{2 \pi} \overset u {\df a}  \wedge d \df A + \frac{\kappa_M}{2 \pi} \df \omega \wedge d \overset u {\df a} - \frac{c}{4 8 \pi} \df \omega \wedge d \df \omega \\
		&\qquad \qquad - \epsilon ( B , R , \overset u \Omega , G ) \df\varepsilon  \bigg)  ,
\end{align}
where we have made the following definitions. First of all, since we have a nowhere vanishing magnetic field, we may define the drift velocity 3-vector
\begin{align}
	u^A = \frac{1}{2 B} \epsilon^{ABC} F_{BC} = 
	\begin{pmatrix}
		1 \\
		\frac{\epsilon^{ab} E_b}{B}
	\end{pmatrix} ,
\end{align}
where $E_a = F_{a0}$.

The gravitational vector potential $\df a$ is not boost invariant, however in the presence of a preferred frame such as $u^A$ we may define $\overset u {\df a}$ to be $\df a$ measured in that frame. One convenient way to do this is
\begin{align}
	\overset u {\df a} = \mathring u_I \df e^I = \df a + u_a \df e^a - \frac 1 2 u^2 \df n .
\end{align}
$\overset u {\df a}$ is then simply $\df a$ in the frame where $u^a = 0$. The one-form $\df \omega$ is the spatial part of the spin connection ${\df \omega}_{AB}$
\begin{align}
	\df \omega = \frac 1 2 {\df \omega}_{AB} \epsilon^{AB} = \frac 1 2 { \df \omega}_{ab} \epsilon^{ab} .
\end{align}
It is not possible to involve the boost part of the connection $\df \varpi^a$ to this order.

Finally, the equation of state $\epsilon$ is a function of the curvatures
\begin{align}
	B = \frac 1 2 \varepsilon^{\mu \nu} F_{\mu \nu},
	&&R = \frac 1 2 \varepsilon^{\mu \nu} R_{\mu \nu},
	&&\overset u \Omega = \frac 1 2 \varepsilon^{\mu \nu} \overset u f_{\mu \nu},
	&&G = \frac 1 2 \varepsilon^{\mu \nu} G_{\mu \nu}.
\end{align}
where
\begin{align}
	\df F = d \df A,
	&& \df R = d \df \omega,
	&&\overset u {\df f} = d \overset u{\df a},
	&&\df G = d \df n .
\end{align}
$B$ is of course the magnetic field and $R$ the curvature of a spatial slice. The other two curvatures may require some comment on interpretation however. In flat space, the third may be calculated to be
\begin{align}
	\overset u \Omega = \epsilon^{ij} \partial_i u_j + \epsilon^{ij} \partial_i a_j 
\end{align}
and so is the vorticity of the quantum Hall fluid. The term $\epsilon^{ij} \partial_i a_j$ ensures the boost invariance of $\overset u \Omega$ and can be thought of as the Coriolis contribution to the vorticity in a rotating frame.
The curvature $G$ is proportional to $\df n \wedge d \df n$ and so is nonzero only in spacetimes that are not causal. While not physical, considering non-zero $G$ is a useful intermediate step to derive energy currents and has been used for this purpose in a number of works \cite{Gromov:2014vla,Bradlyn:2014wla}.

In what follows, it will be convenient to introduce the definitions
\begin{align}
	\overset u R_\mu = R_{\mu \nu} u^\nu , 
	&&	\overset u g_\mu = \overset u f_{\mu \nu} u^\nu , 
	&&\overset u G_\mu = G_{\mu \nu} u^\nu  .
\end{align}
These are the ``electric'' components of their associated field strengths in the frame where $u^a = 0$. Note in particular that in the case of pure Newtonian gravity $\df a = - \phi \df n$\footnote{There is always a Galilean frame in which this is true.}
\begin{align}
	\overset u g {}^A =
	\begin{pmatrix}
		0 \\
		- \partial^a \phi
	\end{pmatrix}
\end{align}
so $\overset u g_\mu$ represents the Newtonian gravitational force. Of course, there is no comoving electric field in the frame defined by the drift velocity: $F_{\mu \nu} u^\nu = 0$. The field strengths then have the decomposition
\begin{align}
\label{fieldStrengthDecomposition}
	F_{\mu \nu} &= B \overset u \varepsilon_{\mu \nu} , \\
	R_{\mu \nu} &= \overset u R_\mu n_\nu - n_\mu \overset u R_\nu + R \overset u \varepsilon_{\mu \nu}  ,  \\
	\overset u f_{\mu \nu} &= \overset u g_\mu n_\nu - n_\mu \overset u g_\nu + \overset u \Omega \overset u \varepsilon_{\mu \nu} , \\
	G_{\mu \nu} &= \overset u G_\mu n_\nu - n_\mu \overset u G_\nu + G \overset u \varepsilon_{\mu \nu}
\end{align}
where
\begin{align}
	\overset u \varepsilon_{\mu \nu} = \varepsilon_{\mu \nu \lambda} u^\lambda
\end{align}
is the $u^\mu$-orthogonal spatial volume form. One may check that $\varepsilon^{\mu \lambda} \overset u \varepsilon_{\nu \lambda}$ is $n_\mu$ orthogonal in it's first index, $u^\nu$-orthogonal in it's second index, and squares to itself, so that
\begin{align}
	\varepsilon^{\mu \lambda} \overset u \varepsilon_{\nu \lambda} = \overset u P {}^\mu{}_\nu.
\end{align} 

\subsubsection{The Stress-Energy Tensor}
Let's begin by calculating the stress energy tensor $\tilde \tau^\mu{}_I$ for the action \eqref{QHEAction}. We'll begin with the (non-gravitational) field strengths $B$, $R$, and $G$. The covariant vorticity $\overset u \Omega$ is more involved due to $\overset u {\df a}$'s dependence on the extended tetrad and we return to it in a moment. For this we need the variation of the ``spatial volume element'' $\varepsilon^{\mu \nu} = e^\mu_A e^\nu_B \epsilon^{AB}$. Using $\delta e^\mu_A = - e^\mu_B e^\nu_A \delta e^B_\nu = - \Pi^\mu{}_I e^\nu_A \delta e^I_\nu$ we find
\begin{align}
	\delta \varepsilon^{\mu \nu} &= 2 \delta e^{[\mu}_A e^{\nu ]}_B \epsilon^{AB}
		= 2 \Pi^{[\mu}{}_I  \varepsilon^{\nu ] \lambda} \delta e^I_\lambda
\end{align}
so that
\begin{align}
	\delta B = \frac 1 2 \delta \varepsilon^{\mu \nu} F_{\mu \nu} = \Pi^\mu{}_I \varepsilon^{\nu \lambda} F_{\mu \nu} \delta e^I_\lambda = - B \overset u P {}^\mu{}_I \delta e^I_\mu,
\end{align}
where in the final step we have plugged in \eqref{fieldStrengthDecomposition}. The same steps also lead to
\begin{align}
	&\delta R =  \Pi^\mu{}_I \varepsilon^{\nu \lambda} R_{\mu \nu} \delta e^I_\lambda  = - R \overset u P {}^\mu{}_I \delta e^I_\mu + n_I \varepsilon^{\mu \nu} \overset u R_\nu \delta e^I_\mu ,
\end{align}
the additional term arising from the fact that $R_{\mu \nu} u^\nu \neq 0$. Similarly
\begin{align}
	\delta G &= \frac 1 2 \delta \varepsilon^{\mu \nu} G_{\mu \nu}+ \frac 1 2 \varepsilon^{\mu \nu} (d \delta \df n)_{\mu \nu} =  \varepsilon^{\mu \nu} \nabla_\mu (n_I \delta e^I_\nu).
\end{align}
Note the terms resulting from $\delta \varepsilon^{\mu \nu}$ vanish since we are varying about torsionless backgrounds.

Since $\overset u {\df a}$ and $\overset u \Omega$ both depend on the drift velocity $u^A = \frac 1 {2B} \epsilon^{ABC} F_{BC}$, we must find how this changes under a geometric perturbation $\delta e^I_\mu$ to continue. From the expression for $\delta B$, we find
\begin{align}
 \delta u^A = u^A \overset u P {}^\mu{}_I \delta e^I_\mu .
\end{align}
Now the variation of the null extended frame $\mathring u^I$ is determined by $n_I \delta \mathring u^I = \mathring u_I \delta \mathring u^I = 0$ and $\Pi^A{}_I \delta \mathring u^I = \delta u^A$, which are solved by the formula
\begin{align}
	\delta \mathring u^I = u^\mu \overset u P {}^I{}_J \delta e^J_\mu,
		&&\text{where}
		&&\overset u P {}^I{}_J = \delta^I{}_J - \mathring u^I n_J - n^I \mathring u_J
\end{align}
is the $n_I$ and $\mathring u_I$-orthogonal extended projector, which one may check satisfies $\Pi^A{}_I \overset u P {}^I{}_J = \overset u P {}^A{}_J$.

Finally, we turn to  $\overset u {\df a}$
\begin{align}
	\delta \overset u a_\mu = \mathring u_I \delta e^I_\mu +  \delta \mathring u_I e^I_\mu = \mathring u_I \delta e^I_\mu + e^I_\mu u^\nu \overset u P_{IJ} \delta e^J_\nu 
	= ( \delta^\nu{}_\mu \mathring u_I + u^\nu \overset u h_{\mu I} ) \delta e^I_\nu
\end{align}
where in the final step we have noted that $\overset u P_{IJ} e^J_\mu = \overset u h_{\mu I}$, which may be checked by noting that both sides of this equation are $n^I$ and $\mathring u^I$ orthogonal in their extended indices, and that both sides project to $\overset u P {}^\nu{}_\mu$ upon application of $\Pi^{\nu I}$. From this, we obtain the variation of the covariant vorticity
\begin{align}
	\delta \overset u \Omega &= \frac 12 \delta \varepsilon^{\mu \nu} \overset u f_{\mu \nu} + \varepsilon^{\mu \nu} \nabla_\nu \delta \overset u a_\nu \nonumber \\
		&= - \overset u \Omega \overset u P {}^\mu{}_I \delta e^I_\mu + n_I \varepsilon^{\mu \nu} \overset u g_\nu \delta e^I_\mu + \varepsilon^{\mu \nu} \nabla_\mu \left( ( \delta^\lambda{}_\nu \mathring u_I + u^\lambda \overset u h_{\nu I}) \delta e^I_\lambda \right) .
\end{align}

The variations in hand, it is now a straightforward exercise to compute the stress-energy tensor. Let's begin with the unimproved mass current, which is
\begin{align}
	\tilde \rho^\mu &= \frac{\nu_M}{4 \pi} \varepsilon^{\mu \nu \lambda} \overset u f_{\nu \lambda} + \frac{\zeta_M }{4 \pi} \varepsilon^{\mu \nu \lambda} F_{\nu \lambda} + \frac{\kappa_M}{4 \pi} \varepsilon^{\mu \nu \lambda} R_{\nu \lambda}  - \varepsilon^{\mu \nu} \nabla_\nu \epsilon_R \nonumber \\
	&= \left( \frac{\nu_M \overset u \Omega}{2 \pi} + \frac{\zeta_M B}{2 \pi}+ \frac{\kappa_M R}{2 \pi} \right) u^\mu + \frac{\nu_M}{2 \pi} \varepsilon^{\mu \nu} \overset u g_\nu + \frac{\kappa_M}{2 \pi} \varepsilon^{\mu \nu} \overset u R_\nu - \varepsilon^{\mu \nu} \nabla_\nu \epsilon_\Omega .
\end{align}
The unimproved stress-mass is then simply a pressure term, plus contributions determined by the mass current (here $\tilde \rho = n_\mu \tilde \rho^\mu$)
\begin{gather}
	\tilde T^{\mu \nu} = \tilde \rho^\mu u^\nu + u^\mu \tilde \rho^\nu -  \tilde \rho u^\mu u^\nu + p h^{\mu \nu} ,
	\qquad \qquad
	\text{where}
	\nonumber \\
	p = B \epsilon_B + R \epsilon_R + \overset u \Omega \epsilon_\Omega + G \epsilon_G - \epsilon.
\end{gather}
The full stress-energy contains these contributions, plus energy transport
\begin{align}
	\tilde \tau^\mu{}_I = \overset u{  \varepsilon}{}^\mu n_I - \tilde T^{\mu \nu} \overset u h_{\nu I} - \tilde \rho^\mu \mathring u_I
\end{align}
where $\overset u{ \varepsilon}{}^\mu = \tilde \tau^\mu{}_I \mathring u^I$ is the comoving energy current
\begin{align}
	\overset u{ \varepsilon}{}^\mu = \epsilon u^\mu + \epsilon_\Omega \varepsilon^{\mu \nu} \overset u g_\nu + \epsilon_R \varepsilon^{\mu \nu} \overset u R_\nu + \varepsilon^{\mu \nu} \nabla_\nu \epsilon_G .
\end{align}
We postpone a discussion of the physical content of these formulas until we pass to the Cauchy stress-mass.

\subsubsection{The Cauchy Stress-Mass Tensor}
To pass to the Cauchy stress-mass, we need to first calculate the spin-current $\tilde s^{\mu AB}$. This is rather easier than finding $\tilde \tau^\mu{}_I$, the only necessary variations being
\begin{align}
	\delta R = \varepsilon^{\mu \nu} \nabla_\mu \delta \omega_\nu,
	&& \delta \omega_{\mu} = \frac 1 2 \epsilon^{AB} \delta \omega_{\mu AB} .
\end{align}
The spin current is then $\tilde s^{\mu AB} = \frac 1 2 s^\mu \epsilon^{AB}$ with
\begin{align}\label{QHESpin}
	s^{\mu} =  \frac{\kappa}{4 \pi} \varepsilon^{\mu \nu \lambda} F_{\nu \lambda} + \frac{\kappa_M}{4 \pi} \varepsilon^{\mu \nu \lambda} \overset u f_{\nu \lambda} - \frac{c}{48 \pi} \varepsilon^{\mu \nu \lambda} R_{\nu \lambda} - \varepsilon^{\mu \nu} \nabla_\nu \epsilon_R  .
\end{align}
It will prove convenient to write this as
\begin{align}
	s^\mu = s u^\mu + \varepsilon^{\mu \nu} \zeta_\nu
\end{align}
where $s = n_\mu s^\mu$ is the spin density
\begin{align}
	s = \frac{\kappa B}{2 \pi} + \frac{\kappa_M \overset u \Omega}{2 \pi} - \frac{c R}{24 \pi} 
\end{align}
and
\begin{align}
	\zeta_\mu = \frac{\kappa_M}{2 \pi} \overset u g_\mu - \frac{c}{24 \pi} \overset u R_\mu - \nabla_\mu \epsilon_R .
\end{align}

In terms of $s^\mu$, the Cauchy stress-mass is
\begin{align}\label{improveSimple}
	T^{\mu \nu} = \tilde T^{\mu \nu} + \varepsilon^{\lambda ( \mu} \nabla_\lambda s^{\nu)} .
\end{align}
where we have used \eqref{stressMomentumImprovement} and $\nabla_\mu s^\mu = 0$, which follows from \eqref{QHESpin}.
An easy way of computing this is to decompose it into its $u^\mu$-parallel and $n_\mu$-orthogonal parts
\begin{align}
	T^{\mu \nu} = \rho^\mu u^\nu + u^\mu \rho^\mu - \rho u^\mu u^\nu + t^{\mu \nu} ,
\end{align}
where $t^{\mu \nu} n_\nu = 0$
and compute these separately. Explicitly, $\rho^\mu = T^{\mu \nu} n_\nu$ and $t^{\mu \nu} = \overset u P {}^\mu{}_\lambda \overset u P {}^\nu{}_\rho T^{\lambda \rho}$. From \eqref{improveSimple}, the improved mass current is
\begin{align}
	\rho^\mu = \tilde \rho^\mu - \frac 1 2 \varepsilon^{\mu \nu} \nabla_\nu s,
\end{align}
Finally, the spin contribution to $t^{\mu \nu}$ is
\begin{align}
	\overset u P {}^{(\mu}{}_\alpha \overset u P {}^{\nu )}{}_\beta \varepsilon^{\lambda \alpha} \nabla_\lambda s^\beta &= \overset u P {}^{(\mu}{}_\alpha \overset u P {}^{\nu )}{}_\beta \varepsilon^{\lambda \alpha} \nabla_\lambda \left( s u^\beta + \varepsilon^{\beta \gamma}\zeta_\gamma \right) \nonumber \\
		&= \frac 1 2 s \tilde \sigma^{\mu \nu} + \varepsilon^{\lambda (\mu} \varepsilon^{\nu ) \rho} \nabla_\lambda \zeta_\rho \nonumber \\
		&= \frac 1 2 s \tilde \sigma^{\mu \nu} - \nabla_\lambda \zeta^\lambda h^{\mu \nu} + \nabla^{(\mu} \zeta^{\nu )} .
\end{align}
We have used the identity $\epsilon^{AB} \epsilon^{CD} = h^{AC} h^{BD} - h^{AD} h^{BC}$ to simplify the final term in the second line. Moreover we have introduced the shear tensor
\begin{align}
	\sigma^{\mu \nu} = \nabla^\mu u^\nu + \nabla^\nu u^\mu - \nabla_\lambda u^\lambda h^{\mu \nu}
\end{align}
and it's dual
\begin{align}
	\tilde \sigma^{\mu \nu} = \varepsilon^{\lambda ( \mu} \overset u h_{\lambda \rho} \sigma^{\nu) \rho} .
\end{align}
Perturbing around flat space, it's coefficient is the Hall viscosity.

\subsubsection{Summary}

The currents encoded in the action \eqref{QHEAction} are given bellow. We see that the new quantized gravitational coefficients $\nu_M$, $\kappa_M$, and $\zeta_M$ lead to the following effects. $\nu_M$ attaches mass to vorticity and leads to mass transport transverse to the gravitational field. $\kappa_M$ attaches mass to curvature in the same manner $\kappa$ attaches charge. It also leads to a Hall viscosity proportional to vorticity and transverse response to time dependent metric perturbations encoded in $\overset u R_\mu$. This is all summarized in the formulas below. Though we have not carried out the computation of $j^\mu$ here, it proceeds along lines entirely similar to those given above.

Let's begin with the charge current
\begin{gather}
	j^\mu = \left( \frac{\nu B}{2 \pi} +  \frac{\zeta_M \overset u \Omega}{2 \pi} + \frac{\kappa R}{2 \pi} - \nabla_\nu \xi^\nu  \right)u^\mu + \frac{\zeta_M}{2 \pi} \varepsilon^{\mu \nu} g_\nu + \frac{\kappa}{2 \pi} \varepsilon^{\mu \nu} R_\nu , \nonumber \\
	 \qquad \qquad  + \dot \xi^\mu - \frac 1 2 \left( \sigma^{\mu \nu} - \theta h^{\mu \nu} - \overset u \Omega \varepsilon^{\mu \nu} \right) \xi_\nu - \varepsilon^{\mu \nu} \nabla_\nu  \epsilon_B     \nonumber \\
	\text{with} 
	\qquad \qquad \xi^\mu = \frac{\nu_M}{2 \pi B} g^\mu + \frac{\kappa_M}{2 \pi B} R^\mu - \frac 1 B \nabla^\mu  \epsilon_\Omega 
\end{gather}
and $\theta = \nabla_\mu u^\mu$ is the expansion.
The Cauchy stress-mass is
\begin{align}
	T^{\mu \nu}& = \rho^\mu u^\nu + u^\mu \rho^\nu - \rho u^\mu u^\nu \nonumber \\
		& \quad + \left( p - \nabla_\lambda \zeta^\lambda \right) h^{\mu \nu} + \frac 1 2 s \tilde \sigma^{\mu \nu} + \nabla^{(\mu} \zeta^{\nu )} 
\end{align}
where $\rho^\mu$ is the Cauchy mass current
\begin{align}
	\rho^\mu &=  \left( \frac{\nu_M \overset u \Omega}{2 \pi} + \frac{\zeta_M B}{2 \pi}+ \frac{\kappa_M R}{2 \pi} \right) u^\mu + \frac{\nu_M}{2 \pi} \varepsilon^{\mu \nu} \overset u g_\nu + \frac{\kappa_M}{2 \pi} \varepsilon^{\mu \nu} \overset u R_\nu  - \varepsilon^{\mu \nu} \nabla_\nu \left(  \epsilon_R + \frac 1 2 s\right).
\end{align}
and
\begin{align}
	&s = \frac{\kappa B}{2 \pi} + \frac{\kappa_M \overset u \Omega}{2 \pi} - \frac{c R}{24 \pi}, 
	&&\zeta_\mu = \frac{\kappa_M}{2 \pi} \overset u g_\mu - \frac{c}{24 \pi} \overset u R_\mu - \nabla_\mu \epsilon_R .
\end{align}
$s$ is the spin density and we see receives contributions not only from the Wen-Zee coefficient $\kappa$, but also the ``gravitational'' Wen-Zee coefficient $\kappa_M$, which leads to a Hall viscosity contribution proportional to the vorticity and the tying of mass to curvature.

The full stress-energy on the other hand is
\begin{align}
	\tilde \tau^\mu{}_I = \overset u{ \varepsilon}{}^\mu n_I - \tilde T^{\mu \nu} \overset u h_{\nu I} - \tilde \rho^\mu \mathring u_I ,
\end{align}
where $\overset u{\tilde \varepsilon}{}^\mu = \tilde \tau^\mu{}_I \mathring u^I$ is the comoving energy current
\begin{align}
	\overset u{ \varepsilon}{}^\mu = \epsilon u^\mu + \epsilon_\Omega \varepsilon^{\mu \nu} \overset u g_\nu + \epsilon_R \varepsilon^{\mu \nu} \overset u R_\nu + \varepsilon^{\mu \nu} \nabla_\nu \epsilon_G  ,
\end{align}
and $\tilde T^{\mu \nu}$ and $\tilde \rho^\mu$ are the unimproved stress-mass and mass currents
\begin{align}
	\tilde T^{\mu \nu} &= \tilde \rho^\mu u^\nu + u^\mu \tilde \rho^\nu -  \tilde \rho u^\mu u^\nu + p h^{\mu \nu} , \\
	\tilde \rho^\mu & = \frac{\nu_M}{4 \pi} \varepsilon^{\mu \nu \lambda} \overset u f_{\nu \lambda} + \frac{\zeta_M }{4 \pi} \varepsilon^{\mu \nu \lambda} F_{\nu \lambda} + \frac{\kappa_M}{4 \pi} \varepsilon^{\mu \nu \lambda} R_{\nu \lambda}  - \varepsilon^{\mu \nu} \nabla_\nu \epsilon_R.
\end{align}

\section{Non-Relativistic Fluid Dynamics}\label{chap:Apps}

In this chapter we apply the formalism established in the previous chapters to non-relativistic fluid dynamics and linear response. As a phenomenological theory given in a derivative expansion, this is a particularly easy setting within which to enforce symmetry principles and indeed, the fluid equations of motion are themselves the Ward identities of spacetime symmetries. We will review these aspects and develop the theory of non-relativistic fluids on Bargmann spacetimes in section (\ref{FluidGeneralities}).

Fluid mechanics is a theory of nearly-thermal systems when fluctuations occur on length and time scales much larger than any microscopic scales characterizing it. As such, it is an extremely powerful tool for probing the near-equilibrium physics of a variety of systems whose microscopic descriptions are resistant to traditional field-theoretic techniques. One such system we will be interested in is the fractional quantum Hall states, which are known to form an incompressible quantum fluid at zero temperature \cite{Laughlin:1983}.

In section \ref{sec:LLLFluid}, we consider the fluid dynamics of a system constrained by energetics to lie in the lowest Landau level, allowing us to extract universal predictions for quantum Hall states at finite temperature with $\nu \leq 1$, so long as the temperature lies far below the cyclotron gap. We find that the results are strongly constraining: to first order in a derivative expansion charge transport is determined entirely by the equation of state while energy response includes only an independent thermal conductivity, Righi-Leduc or thermal Hall conductivity, and magnetization. Many of these constraints follow from consistency of the fluid description on {\it torsionful} backgrounds and so demand the treatment of chapter \ref{chap:Formalism} to be consistent with Galilean invariance. In section \ref{sec:Fluid} we turn to the standard $(2+1)$-dimensional fluid in a perturbing electromagnetic field. Kubo formulas for all response coefficients are given in section \ref{sec:Kubo}.

\subsection{Generalities}\label{FluidGeneralities}

In this section we discuss the generalities that will be necessary for our later studies of quantum Hall fluids and standard multi-component fluids. The general form of the Ward identities on curved torsionful spacetimes has already been given in section \ref{sec:Ward}, but in section \ref{sec:ComovingWard} we rephrase these in terms of the currents measured by comoving fluid observers. In section \ref{sec:PerfectFluid} we supply constitutive relations to leading order in a derivative expansion and obtain the curved space perfect fluid equations of motion. Technical aspects of fluid frame transformations are recalled in sections \ref{sec:FluidFrames} and \ref{sec:CanonicalEntropy}.

The program of fluid dynamics is to describe the fluctuations of thermodynamic variables in slightly out of equilibrium media. These variables include the temperature and velocity $u^\mu$ as well as a chemical potential for each conserved charge. For us, there are two such quantities, the electric charge and mass, whose associated chemical potentials we will denote $\mu_Q$ and $\mu_M$. In all we have $d+3$ degrees of freedom
\begin{align}\label{therm var}
	T , 
	&&\mu_Q , 
	&&\mu_M , 
	&&u^\mu,
\end{align}
where the velocity has been normalized so that $n_\mu u^\mu = 1$ and thus represents $d$ degrees of freedom.
The equilibrium properties of the system are then completely characterized by an equation of state $p(T, \mu_Q , \mu_M )$, which expresses the pressure as a function of the state variables. The entropy, charge, mass, and energy densities are then defined by
\begin{align}\label{thermo}
	dp = s dT + q d \mu_Q + \rho d \mu_M,
	&&\epsilon + p = T s + q \mu_Q + \rho \mu_M .
\end{align}

Just out of equilibrium, the thermodynamic variables are allowed to vary slowly in space and time, where slow is compared to the mean free path and mean free time, so that local equilibrium is always a good approximation. The Ward identities 
\begin{align}\label{fluid eom}
	&(\nabla_\mu - {T^\nu}_{\nu \mu} ) j^\mu = 0 , \qquad \qquad
	(\nabla_\mu - {T^\nu}_{\nu \mu} ) \rho^\mu = 0 , \nonumber \\
	&- e^I_\mu ( D_\nu - {T^\lambda}_{\lambda \nu}) {\tau^\nu}_I = F_{\mu \nu} j^\nu - {T^I}_{\mu \nu} {\tau^\nu}_I   ,
\end{align}
are then sufficient to serve as equations of motion since they are the same in number as the thermodynamic variables. We need only constitutive relations for the currents in terms of these degrees of freedom and their derivatives. In (\ref{fluid eom}) and for most of this chapter we will principally restrict our attention to spinless fluids, except for the quantum Hall fluid which we take to be spin polarized and minimally coupled to curvature. It would be interesting to generalize our results beyond this case.

Since we are perturbing away from equilibrium, these constitutive relations naturally organize themselves in a gradient expansion where higher derivatives of (\ref{therm var}) take on diminishing importance. To complete our power counting scheme, one needs also specify the backgrounds to be perturbed around. In this paper we shall assume a trivial spacetime background background
\begin{align}
	\df e^I =
	\begin{pmatrix}
		dt \\
		d x^a \\
		0
	\end{pmatrix},
	&&\df T^I = 0
\end{align}
in equilibrium so that 
\begin{align} 
	&\nabla,
	&&\df T^I
\end{align}
are first order in derivatives.

This is not a unique selection and corresponds to a choice of regime in which we expect our results to be applicable. The power counting for $F_{\mu \nu}$ will depend on the problem at hand. The quantum Hall fluid is gapped by a large magnetic field and fluctuations are assumed to be small enough to not significantly alter the size of the gap. We will then take the magnetic field to be order zero in derivatives in our study of the LLL fluid in section \ref{sec:LLLFluid}, but order one in the standard fluid in section \ref{sec:Fluid}.

The constitutive relations so obtained are not in general sensible and could lead to an on shell decrease in entropy, the canonical example being that of a negative shear viscosity \cite{landau1987fluid}. As such we must also impose the second law of thermodynamics as an additional constraint on fluid flows, diminishing the freedom present in the gradient expansion and returning a reduced set of transport coefficients. This procedure is known as an entropy current analysis and has been carried out in many systems including  $(2+1)$-dimensional relativistic normal fluids \cite{Jensen:2011xb} and $(3+1)$-dimensional relativistic superfluids \cite{Bhattacharya:2011tra}.
In this thesis we will work to first order in derivatives. It's well known that even at first order, the dynamics is very rich in the parity odd sector when $d=2$ and so we shall eventually restrict to two-dimensional fluids.

\subsubsection{Comoving Currents}
As we have seen, the stress-energy tensor $\tau^\mu{}_I$ cannot in general be separated into it's constituent currents in a manner that is consistent with Galilean invariance. Notions of energy, stress, and momentum are then observer dependent and can be retrieved from knowledge of their spacetime trajectory as discussed in section \ref{sec:ComovingCurrents}.
In the hydrodynamic setting however, Galilean invariance is broken by the boost invariant fluid velocity $u^\mu$, which defines a preferred rest frame: the frame of observers comoving with the fluid. 

It will thus prove natural to organize our calculation in terms of the comoving currents
\begin{align}
	\varepsilon^\mu = \tau^\mu{}_I \mathring u^I,
	&&T^{\mu \nu} = \overset u P{}^\mu{}_\lambda \tau^{\lambda \rho} \overset u P{}^\nu{}_\rho,
	&&p^\mu = n_\nu \tau^{\nu \lambda} \overset u P{}^\mu{}_\lambda,
	&&\rho^\mu = \tau^{\mu \nu} n_\nu ,
\end{align}
in terms of which the stress-energy and stress-mass tensors decompose as
\begin{gather}
	\tau^\mu{}_I = \varepsilon^\mu n_I - \rho^\mu \mathring u_I - ( u^\mu p^\nu + T^{\mu \nu}) \overset u \Pi_{\nu I}, \nonumber \\
	\tau^{\mu \nu} = \rho^\mu u^\nu + u^\mu p^\nu + T^{\mu \nu} .
\end{gather}
In the above we have defined a $u^\mu$ orthogonal ``inverse" $\overset u h_{\mu \nu}$ to $h^{\mu \nu}$, with which we have lowered the index on the projector
\begin{align}
	h^{\mu \lambda} \overset u h_{\lambda \nu} = \overset u P{}^\mu{}_\nu,
	&&\overset u h_{\mu \nu} u^\nu = 0,
	&& \overset u \Pi_{\mu I} = \overset u h_{\mu \nu} \Pi^\nu{}_I .
\end{align}
Of course, since the stress-mass tensor is symmetric in the spinless case (equation (\ref{LGTWard})), the spatial stress is symmetric and the comoving momentum is simply the $u^\mu$ orthogonal part of the mass current
\begin{align}
	T^{\mu \nu} = T^{( \mu \nu)},
	&&p^\mu = \overset u P{}^\mu{}_\nu \rho^\nu .
\end{align}
Note that for notational simplicity we have dropped the over-script $u$'s of section \ref{sec:ComovingCurrents} since these are the only currents with which we shall have to deal with for most of this analysis. One should always keep in mind however that all of these quantities are as measured in by comoving observers. To distinguish them from the components of $\tau^\mu{}_I$, we will refer to these as the ``lab" currents
\begin{align}
	\tau^\mu{}_I = 
	\begin{pmatrix}
		\varepsilon^t_\text{lab} & - p_{\text{lab} a} & - \rho^t \\
		\varepsilon^i_\text{lab} & - T_\text{lab}^i{}_a & - \rho^i 
	\end{pmatrix}.
\end{align}

\subsubsection{Incorporating Torsion}
As we have seen, knowing how to covariantly couple to torsion is necessary for a sensible derivation of the energy current of a given theory. It will thus prove necessary in any microscopic calculation of the transport coefficients to be discussed (see section \ref{sec:Kubo} for a collection of Kubo formulas). In our fluid analysis it will however play a different role.

The key idea, introduced in \cite{Jensen:2011xb}, is that the more general a background a system is placed on, the more constraining the entropy current analysis is. For example, in this work the authors demonstrate entropy flow induced by acceleration $s^\mu =  \nu_1 u^\nu \nabla_\nu u^\mu + \cdots$ sources entropy density
\begin{align}
	\nabla_\mu s^\mu = - \nu_1 u^\mu u^\nu R_{\mu \nu} + \cdots .
\end{align}
Since this term is sign indefinite, one concludes that the coefficient $\nu_1$ must vanish. Thus, even if one is studying fluids in flat spacetime, considering them on a curved space is a useful theoretical experiment. Time evolution should still be consistent with the second law of thermodynamics and imposing this yields additional relations not apparent in flat space.

We will also get a lot of mileage out this reasoning and further find it useful to extended the analysis to torsionful spacetimes.
Spatial torsion may be interpreted in the condensed matter setting as a coarse-grained Burgers vector density and so is necessary in the study of media with dislocation defects  \cite{Hughes:2012vg}. However, we will be mainly concerned with temporal torsion $\df T^0 = d \df n$, since it may be used to mock up the effect of steady state temperature gradients.

To see why, consider a system in equilibrium subjected to a static but spatially varying clock form
\begin{align}\label{GTCclockForm}
	n_\mu = ( n_t ~ 0 ) 
\end{align}
and consider the flow induced in the energy current with constitutive relations
\begin{align}
	\varepsilon^i = \Sigma_T \partial^i T - \Sigma_G \partial^i n_t.
\end{align}
Since $n_t$ couples to energy density, under a perturbation $n_t = 1 + \delta n_t$, energy will flow from higher $n_t$ to lower $n_t$ until equilibrium is regained. This occurs when \cite{Luttinger:1964zz}
\begin{align}\label{TEquilibrium}
	\delta T = - T \delta n_t .
\end{align}
We thus see that in thermodynamic equilibrium, non-relativistic fluids on backgrounds with temporal torsion do {\it not} have uniform temperature, similar to the Tolman-Ehrenfest effect \cite{Tolman} for fluids in a gravitational background in general relativity.

Once equilibrium is reached, energy must cease to flow $\varepsilon^i = 0$.\footnote{Or at the very least, longitudinal flow vanishes. Persistent currents proportional to the curl of $T$ (so-called ``magnetization" currents) may remain since they do not lead to a buildup of energy density.} This constrains the constitutive relations so that
\begin{align}
	\Sigma_T = - \frac 1 T \Sigma_G ,
\end{align}
a torsional version of the Einstein relation $\sigma = - \frac{1}{T} D$ relating the longitudinal conductivity and diffusion coefficients $j^i = \sigma E^i + D \partial^i ( \mu / T)$. As reasoned by Luttinger, even if we cannot turn on $n_t$ in the lab, this is an extremely useful relation since it allows us to derive Kubo formulas for thermal transport since $n_t$ may be viewed as a background field perturbing a microscopic system.

This relation and many others will follow from the entropy current analysis on torsionful spacetimes $d \df n \neq 0$, so the additional complications torsion introduces are worth the effort. We will however restrict to causal spacetimes $\df n \wedge d \df n = 0$ since there is otherwise no consistent notion of future or past, making the second law of thermodynamics meaningless (see the concluding paragraph of chapter \ref{chap:Intro}).\footnote{The entropy current analysis may still be carried out formally in this case. However, we find that no new relations amongst the physical coefficients are imposed upon doing so since the proliferation of new terms proportional to $\df n \wedge d \df n$ out-paces the number of additional constraints.} 

To simplify the analysis, we shall also take spatial torsion to vanish, though generalization to the fully torsionful case would be an interesting extension of our work. However, as discussed in the paragraph below equation \ref{rhoImprovement}, there is no Galilean invariant notion of spatial torsion once we allow temporal torsion to be nonzero. To settle this, we agree that space will be torsionless for comoving fluid observers
\begin{align}
	\overset u h_{\lambda \rho} T^\rho{}_{\mu \nu} = 0,
	&&\text{which implies}
	&&T^\lambda{}_{\mu \nu} = u^\lambda (d \df n)_{\mu \nu} .
\end{align}

The full extended torsion is then of the form
\begin{align}\label{ComovingTorsion}
	\df T^I = \mathring u^I d \df n + n^I \df f
\end{align}
where $\df f = \mathring u_I \df T^I$ is the mass torsion measured in the comoving frame. As we have seen, this acts as an external Lorentz force coupling to $\rho^\mu$. Including $\df f$ is natural in the presence of timelike torsion (it must be nonzero for {\it some} observers and generically will be for fluid observers in particular) and does not really complicate the analysis. Unfortunately it doesn't lead to any interesting relations either.

\subsubsection{The Navier-Stokes and Work-Energy Equations}\label{sec:ComovingWard}

In this section we restate the Ward identity for diffeomorphisms in terms of the comoving currents. Plugging the extended torsion (\ref{ComovingTorsion}) into stress-energy conservation (\ref{fluid eom}), we find
\begin{align}
	- e^I_\mu ( D_\nu - G_\nu) {\tau^\nu}_I = F_{\mu \nu} j^\nu + f_{\mu \nu} \rho^\nu - G_{\mu \nu} \varepsilon^\nu 
\end{align}
where we have introduced the shorthand $G_{\mu \nu} = (d \df n)_{\mu \nu}$ and $G_\mu = - G_{\mu \nu} u^\nu$. As we saw in equation (\ref{cauchyEquation}), the spatial components of this identity, obtained by raising the index with $h^{\mu \nu}$ give the Cauchy equation
\begin{align}
	(\nabla_\nu - G_{\nu}) \tau^{\mu \nu} = F^\mu{}_\nu j^\nu  + f^\mu{}_\nu \rho^\nu - G^\mu{}_\nu \varepsilon^\nu ,
\end{align}
which will serve as the Navier-Stokes equation once we provide constitutive relations.

We saw previously (equation (\ref{flatSpaceFluid})) that the temporal component of the diffeomorphism Ward identity contains the work-energy equation. We can obtain the covariant version of this by contracting the Ward identity with $u^\mu$. In doing so, the following identity is useful
\begin{align}\label{extendedContract}
	e^I_\mu u^\mu = \mathring u^I + \mathring u_J e^J_\mu u^\mu n^I.
\end{align}
Using this equation and mass conservation, one finds
\begin{align}
	\mathring u^I (D_\mu - G_\mu ) \tau^\mu{}_I &= E_\mu j^\mu + e_\mu \rho^\mu + G_\mu \varepsilon^\mu, \nonumber \\
	\implies \qquad \qquad
	( \nabla_\mu - G_\mu ) \varepsilon^\mu &= E_\mu j^\mu + e_\mu \rho^\mu + G_\mu \varepsilon^\mu + \tau^\mu{}_I D_\mu \mathring u^I ,
\end{align}
where $E_\mu = F_{\mu \nu} u^\nu$ and $e_\mu = f_{\mu \nu} u^\nu$ are the comoving electric fields for charge and mass.

To simplify the final term, we note that $n_I D_\mu \mathring u^I = \mathring u_I D_\mu \mathring u^I = 0$, so there is a $u^\nu$ orthogonal tensor $t_{\mu \nu}$ such that $D_\mu \mathring u^I = \Pi^{\lambda I} t_{\mu \lambda}$. Contracting this equation with $\Pi^\nu{}_I$ we find that $t_{\mu}{}^\nu = \nabla_\mu u^\nu$ and so
\begin{align}\label{extendedShear}
	D_\mu \mathring u^I = \overset u \Pi_\nu{}^I \nabla_\mu u^\nu .
\end{align}
Plugging this in, one finds the work-energy equation for the comoving energy current is
\begin{align}
	( \nabla_\mu - G_\mu ) \varepsilon^\mu &= E_\mu j^\mu + e_\mu \rho^\mu + G_\mu \varepsilon^\mu - \tau^\mu{}_\nu \nabla_\mu u^\nu 
\end{align}
where we have lowered the index on the stress-mass tensor with $\overset u h_{\mu \nu}$. In summary, in addition to the work done by external fields, the comoving energy density is dissipated by fluid shear coupling to the stress-mass tensor.\footnote{That energy is dissipated rather than sourced will follow from the constitutive relations and upcoming entropy current analysis.}

In summary, we may restate the Ward identities in terms of the comoving currents as the work-energy equation and Cauchy momentum equation, plus conservation laws for charge and mass
\begin{gather}\label{parsedWard}
	 ( \nabla_\mu - G_\mu  ) j^\mu  = 0 , \nonumber \\
	 ( \nabla_\mu - G_\mu ) \rho^\mu = 0 , \nonumber \\
	(\nabla_\nu - G_{\nu}) \tau^{\mu \nu} = F^\mu{}_\nu j^\nu  + f^\mu{}_\nu \rho^\mu - G^\mu{}_\nu \varepsilon^\nu , \nonumber \\
	( \nabla_\mu - G_\mu ) \varepsilon^\mu = E_\mu j^\mu + e_\mu \rho^\mu + G_\mu \varepsilon^\mu - \tau^\mu{}_\nu \nabla_\mu u^\nu  .
\end{gather}

\subsubsection{Perfect Fluids}\label{sec:PerfectFluid}
These Ward identities serve as dynamical equations once constitutive relations have been supplied, specifying the currents $j^\mu$, $\rho^\mu$, $\varepsilon^{\mu}$ and $T^{\mu \nu}$ in terms of the thermodynamic degrees of freedom
\begin{align}\label{HydroDOF}
	T,
	&&\mu_Q,
	&&\mu_M,
	&&u^\mu.
\end{align}
At zeroth order in derivatives, the most general tensors we can construct using these variables and the Newton-Cartan structure are
\begin{align}
	j^\mu = q u^\mu,
	&&\rho^\mu = \rho u^\mu,
	&&\varepsilon^\mu = \epsilon u^\mu,
	&&T^{\mu \nu} = p h^{\mu \nu} .
\end{align}
Here $q$, $\rho$, $\epsilon$ and $p$ are functions of $(T, \mu_Q , \mu_M)$. They are the thermodynamic charge density, mass density, energy density and pressure and satisfy the relations (\ref{thermo}).

Feeding these into the Ward identities (\ref{parsedWard}), we obtain the perfect fluid equations of motion
\begin{align}\label{Navier}
	&\dot q + q \theta = 0 ,
	\qquad \qquad
	 \dot \rho + \rho \theta = 0 ,
	\qquad \qquad
	\dot \epsilon + ( \epsilon + p ) \theta = 0 , \nonumber \\
	& \rho \alpha^\mu = \rho e^\mu + q E^\mu + ( \epsilon + p ) G^\mu - \nabla^\mu p 
\end{align}
where $\theta = \nabla_\mu u^\mu$ is the fluid expansion and $\alpha^\mu = u^\nu \nabla_\nu u^\mu$ the acceleration.
In these equations and those that follow, dotted objects indicate the material derivative, $\dot f = u^\mu \nabla_\mu f$. 
The final equation is simply Newton's second law and an obvious covariant generalization of Euler's equation \cite{landau1987fluid}.

\subsubsection{Fluid Frames}\label{sec:FluidFrames}
To go beyond perfect fluids one needs to expand the currents to first order in derivatives. The constitutive relations are then the perfect fluid ones plus $\mathcal O (\partial^1 )$ corrections
\begin{align}\label{corrections}
	&j^\mu = ( q + \mathcal Q ) u^\mu + \nu^\mu, 
	&& \rho^\mu = ( \rho + \varrho ) u^\mu + \mu^\mu , \nonumber \\
	& \varepsilon^\mu = ( \epsilon + \mathcal E ) u^\mu + \xi^\mu , 
	&&T^{\mu \nu} = ( p + \mathcal P ) h^{\mu \nu} + \pi^{\mu \nu}.
\end{align}
In the above the, vector corrections are defined to be transverse
\begin{align}
	n_\mu \nu^\mu = n_\mu \mu^\mu = n_\mu \xi^\mu = 0 
\end{align}
while the tensor correction is traceless
\begin{align}
	h^{\mu \nu} \pi_{\mu \nu}
\end{align}
($\pi^{\mu \nu}$ is of course already transverse since $T^{\mu \nu}$ is). This is convenient as it separates the first order corrections into irreducible representations of $SO(d)$.

The decomposition (\ref{corrections}) is subject to a well known ambiguity stemming from the need to define $T$, $\mu_Q$, $\mu_M$, and $u^\mu$ out of equilibrium. Any such definition is admissible so long as it reduces to the equilibrium values at order zero and so is subject to a $d+3$ parameter $\mathcal O( \partial^1 )$ field redefinition
\begin{align}
	T \rightarrow T + \delta T ,
	&& \mu_Q \rightarrow \mu_Q + \delta \mu_Q,
	&& \mu_M \rightarrow \mu_M + \delta \mu_M,
	&&u^\mu \rightarrow u^\mu + \delta u^\mu
\end{align}
called a fluid frame transformation (not to be confused with a Galilean frame transformation). To deal with this ambiguity we may either fix the frame by imposing extra conditions, or work in a manifestly frame invariant manner. 

Frame transformations are worked out in detail in \cite{Jensen:2014ama} (see \cite{Bhattacharya:2011tra} for the relativistic case) and in this section and the next, we refer the reader to this treatment for the details. For our purposes we only note that (besides those related to the entropy, which we have saved for the next section) the complete set of first order frame invariants is
\begin{align}\label{frame invariants}
	&\mathcal S  = \mathcal P - \partial_\epsilon p \mathcal E - \partial_q p \mathcal Q - \partial_ \rho p \varrho,
	\qquad \qquad
	\mathcal T^{\mu \nu} = \pi^{\mu \nu} ,\nonumber \\
	&\mathcal J^\mu = \nu^\mu - \frac{q}{\rho} \mu^\mu, \qquad \qquad
	\mathcal E^\mu = \xi^\mu - \frac{\epsilon + p}{\rho} \mu^\mu .
\end{align}
 Although we shall usually take $p$ to be a function of temperature and the chemical potentials, here we have taken $p = p(\epsilon , q , \rho )$ and the partial derivatives $\partial_\epsilon$, $\partial_q$ and $\partial_\rho$ are defined accordingly.
Note we have an additional vector frame invariant $\mathcal J^\mu$ compared to either the relativistic case or non-relativistic single-component case due to the independence of the mass and charge currents.

\subsubsection{The Canonical Entropy Current}\label{sec:CanonicalEntropy}
There is an additional frame invariant involving the entropy, which we treat separately due to it's importance in the ensuing calculation. This is the ``canonical entropy current"
\begin{gather}
	 T s^\mu_\text{can} = p u^\mu + \varepsilon^\mu - \mu_Q j^\mu - \mu_M \rho^\mu \nonumber \\
	\text{i.e.} \qquad s^\mu_\text{can} = s u^\mu - \frac{\mu_Q}{T} ( \mathcal Q u^\mu + \nu^\mu ) - \frac{\mu_M}{T} ( \varrho u^\mu + \mu^\mu ) + \frac{1}{T} ( \mathcal E u^\mu + \xi^\mu ) .
\end{gather}
Intuitively, this is designed to match the combination of currents found in the equilibrium identity (\ref{thermo}), but it's true importance stems from it's manifest frame invariance.
Out of equilibrium the entropy flow will in general deviate from the canonical part
\begin{align}
	s^\mu = s^\mu_\text{can} + \zeta^\mu .
\end{align}
Of course, since $s^\mu_\text{can}$ is a frame invariant, $\zeta^\mu$ is as well. It's divergence is a quadratic form in first order data 
\begin{align}\label{quadraticForm}
	( \nabla_\mu - G_\mu ) s^\mu_\text{can} &=  - \frac{1}{T} \mathcal S \theta  - \frac{1}{2 T} \sigma_{\mu \nu} \mathcal T^{\mu \nu}
	+ \frac{1}{T} \mathcal J^\mu \left( E_\mu - T \nabla_\mu \left( \frac{\mu_Q}{T} \right)  \right) - \frac{1}{T^2} \mathcal E^\mu  ( \nabla_\mu T - T G_\mu )  .
\end{align}

Here we have performed a decomposition of $\nabla_\mu u^\nu$ into $u^\mu$ and $n_\mu$ orthogonal tensors
\begin{align}\label{DECOMP}
	\nabla_\mu u^\nu = n_\mu \alpha^\nu + \frac 1 2 \sigma_\mu{}^\nu + \frac 1 d\theta \overset u P {}^\nu{}_\mu + \frac 1 2 \omega_\mu{}^\nu
\end{align}
where
\begin{align}
	&\alpha^\mu = u^\nu \nabla_\nu u^\mu ,
	&&\sigma^{\mu \nu} = 2 \nabla^{( \mu }u^{\nu )} - \frac{2}{d} h^{\mu \nu} \theta, 
	&&\theta = \nabla_\mu u^\mu,
	&&\omega^{\mu \nu} = 2 \nabla^{[\mu} u^{\nu ]} .
\end{align}
Since these are all $n_\mu$ orthogonal tensors, there is no information lost in lowering their indices with $\overset u h_{\mu \nu}$, as we have done above and shall continue to do throughout.

\subsection{Hydrodynamics on the Lowest Landau Level}\label{sec:LLLFluid}

In this section we turn to the quantum Hall effect, whose study inspired much of the initial interest in Newton-Cartan geometry within the condensed matter literature \cite{Son:2013}. Since its discovery, the fractional quantum Hall (FQH) 
effect~\cite{Tsui:1982yy} has been subjected to intense 
study and proven a fruitful playground for new concepts in both 
condensed matter and high energy physics. Beyond its quantized Hall 
conductance, FQH states exhibit a number of interesting 
features including anyonic 
excitations~\cite{Halperin:1984fn,Arovas:1984qr}, edge 
states~\cite{Wen:1992vi}, and are prime examples of topological phases 
of matter~\cite{Wen-book}.  Beginning with the Laughlin 
wave function~\cite{Laughlin:1983}, it has been attacked with numerous 
approaches including Chern-Simons field 
theory~\cite{Zhang:1988wy,Halperin:1992mh} and the composite-fermion 
approach~\cite{Jain:1989}.

One reason it has proven such fruitful ground for theoretical studies is that it is a famously difficult problem exhibiting a variety of physics in different regimes. The integer quantum Hall effect arises from the Landau level quantization (with gap of order of the cyclotron frequency) of free electrons combined with Anderson localization \cite{AndersonLoc}, while the fractional quantum Hall effect relies on intra-Landau level strong coupling physics (with gap of order the Coulomb energy). For systems with temperature on the order of than the Coulomb gap, little can be said.

However, the zero temperature integer and fractional quantum Hall states both form an incompressible quantum fluid \cite{Laughlin:1983}. At finite temperature and slightly out of equilibrium, one might ask what are the dynamics of this fluid? In this section we will answer this question by supplementing the formalism introduced in section \ref{sec:Fluid} with a lowest Landau level projection.

The results are applicable whenever the temperature lies far below the cyclotron gap. We are thus interested in the universal properties of systems constrained to the lowest Landau level. In this regime, response is very simple and highly constrained: we can have an equation of state $p$, the usual suite of viscosities $\zeta, \eta, \tilde \eta$, a thermal conductivity $\Sigma_T$, magnetization currents $M$ and $M_\epsilon$ for the charge and energy currents, and a thermal Hall conductivity $c_{RL}$
\begin{align}\label{current summary}
	j^t &= n, \qquad 
	\varepsilon^t_\text{lab} = \epsilon ,\nonumber \\
	\rho^i &= \frac 1 2 \mathfrak s \epsilon^{ij} \partial_j q , \nonumber \\
	j^i &= \epsilon^{ij} \left( \frac{n}{B} \left(E_j - 
\d_j \mu \right) - \frac{s}{B}\d_j T + \d_j  M \right), \nonumber  \\
	\varepsilon^i_\text{lab} &= 
	\Sigma_T \partial^i  T  + \epsilon^{ij} \left(  \frac{\epsilon 
+ p}{B} \left( E_j - \d_j   \mu  \right) - M \d_j \mu 
	-  T c_{RL} \partial_j T 
	+ \partial_j  M_E  \right), \nonumber \\
	T_\text{lab}^{i j} &= \left( p_\text{int} - \zeta \theta + \frac 1 2 \mathfrak  s q \omega + \frac{\mathfrak s}{B} E^k \partial_k q \right) \delta^{i j} - \frac{\mathfrak s}{B} E^{(i} \partial^{j)} q - \eta \sigma^{i j} - \tilde \eta \tilde \sigma^{i j} .
\end{align}
Note we also have a mass magnetization current and vorticity induced pressure tied to the spin density. Here $\mathfrak s = 1/2$ denotes the spin of the approximately spin polarized electrons that make up the quantum Hall fluid. $\sigma^{ij}$, $\theta$ and $\omega$ are the traceless shear, expansion and vorticity of the drift velocity $u^i = \epsilon^{ij} E_j / B$, and $\tilde \sigma^{ij} = \epsilon_k{}^{(i} \sigma^{j) k}$. For instance, in the case of flat space and uniform magnetic field, they read
\begin{align}
	&\sigma^{ij} = - \frac{2}{B} \epsilon^{k (i} \partial^{j)} E_k + \frac{\dot B}{B} \delta^{ij},
	&&\tilde \sigma^{ij} = \frac{2}{B} \partial^{(i} E^{j)} - \frac{1}{B} \partial_k E^k \delta^{ij} , \nonumber \\
	&\theta = - \frac{\dot B}{B},
	&&\omega = - \frac{\partial_i E^i}{B}.
\end{align}

Finally, we present a set of generalized St\v reda formulas. We find a fractional quantum Hall fluid in thermodynamic equilibrium 
has persistent mass, electric, and energy currents,
\begin{align}
	\rho^i &=  \epsilon^{ij} \big( 
\lambda^{\text{eq}}_H E_j  +  \lambda^{B\text{eq}}_H \d_j B + 
\lambda^{G\text{eq}}_H G_j \big) , \nonumber \\
	j^i &=  \epsilon^{ij} \big( \sigma^{\text{eq}}_H 
E_j  +  \sigma^{B\text{eq}}_H \d_j B + \sigma^{G\text{eq}}_H G_j \big),
\nonumber \\
	\varepsilon^i_\text{lab} &=  \epsilon^{ij} \big( 
\kappa^{\text{eq}}_H E_j  +  \kappa^{B\text{eq}}_H \d_j B + 
\kappa^{G\text{eq}}_H G_j \big),
\end{align}
where
\begin{align}
	\lambda^\text{eq}_H &= \frac 1 2 \mathfrak s \left( \frac{\partial q}{\partial \mu} \right)_{T,B},
	\qquad \qquad \qquad \lambda^{B\text{eq}}_H = \frac 1 2 \mathfrak  s \left( \frac{\partial q}{\partial B} \right)_{T,\mu} , \nonumber \\
	\lambda^{G\text{eq}}_H &= \frac 1 2 \mathfrak s \left( T \left( \frac{\partial q}{\partial T} \right)_{\mu,B} +  \mu \left( \frac{\partial q}{\partial \mu} \right)_{T,B} - q \right) , \nonumber \\
	\sigma^\text{eq}_H &= \left( \frac{\d q}{\d B} \right)_{T,\mu},
\qquad \qquad \qquad ~~
	\sigma^{B\text{eq}}_H = \left( \frac{\d M}{\d B} 
\right)_{T,\mu}, \nonumber \\
	\sigma^{G\text{eq}}_H &= T \left( \frac{\d s}{\d B} 
\right)_{T,\mu} + \mu \left( \frac{\d q}{\d B}\right)_{T , \mu} - M,
\nonumber \\
	\kappa^\text{eq}_H &= \left( \frac{\d M_E}{\d \mu} 
\right)_{T,B} - M, \qquad \qquad
	\kappa^{B\text{eq}}_H = \left( \frac{\d M_E}{\d B} 
\right)_{T,\mu}, \nonumber \\
	\kappa^{G\text{eq}}_H &= T\left( \frac{\d M_E}{\d T} 
\right)_{\mu , B}  + \mu \left( \frac{\d M_E}{\d \mu} \right)_{T , B} 
- 2 M_E ,
\end{align}
of which $\sigma^\text{eq}_H$ may be recognized as the familiar St\v reda
formula of \cite{streda1982theory}. The remaining formulas generalize this result to persistent currents driven by non-uniform magnetic fields and a Luttinger potential (here $G_i = - \partial_i n_t$). We note that these St\v eda formula do not agree with those given in \cite{Gromov:2014vla}, we believe due to differing factors of the Luttinger potential. Of course, throughout this section we will be specializing to the case $d=2$.

\subsubsection{The LLL Projection}\label{sec:LLLProjection}
We begin by addressing how to perform the lowest Landau level projection within the hydrodynamic setting. The idea is to take the massless limit $m \rightarrow 0$, or more precisely,
\begin{align}
	m \ll \frac B T ,\frac B \mu
\end{align}
in which the temperature and chemical potential are much less than the Landau level splitting $\omega_c = \frac B m $.
Electrons are then overwhelmingly likely to be found in the lowest Landau level. Unfortunately, this limit is typically singular due the ground state energy of the lowest Landau level, which also goes as $B/m$.

This is however not the case for electrons whose $g$-factor is precisely two. Consider the Schr\"odinger action with arbitrary $g$-factor $\frac{g B}{4m} \psi^\dagger \psi$. The curved space generalization of this action is
\begin{align}\label{gFactorAction}
S = \int d^3 x \sqrt{h} e^{-\Phi} \left( \frac{i}{2} e^{\Phi} \psi^\dagger
\overset{\leftrightarrow}{D}_0 \psi - \frac{1}{2m} \big( h^{ij} + 
\frac{i
g}{2} \varepsilon^{ij} \big) D_i \psi^\dagger D_j \psi \right) .
\end{align}
In writing this down we have assumed spinless electrons and specialized to causal spacetimes $\df n \wedge d \df n =0$, where we can always choose coordinates in which $n_i = 0$, and parameterized the Luttinger potential as $n_t = e^{- \Phi}$. We have also defined a spatial epsilon tensor $\varepsilon^{\mu \nu} = \varepsilon^{\mu \nu \lambda} n_\lambda$ whose spatial components are $\varepsilon^{ij}$.

The matrix $\varepsilon^{ij}$ has eigenvalues $\pm i$ and so the value 
$g =2$ is distinguished, for the matrix $g^{ij} + i 
\varepsilon^{ij}$
is degenerate. In terms of the zweibein $e^a_i$, we have
\begin{equation}
\varepsilon^{ij} = \epsilon_{ab} e^{ai} e^{bj} .
\end{equation}
The eigenvectors of $\varepsilon^{ij}$ are the chiral basis vectors
\begin{equation}
e_i = \frac1{\sqrt2} (e_i^1 + ie_i^2), \qquad \bar e_i = 
\frac1{\sqrt2}
(e_i^1 - ie_i^2) ,
\end{equation}
in terms of which we have the convenient formulas
\begin{align}
h^{ij} &= e^i \bar e^j + \bar e^i e^j ,\qquad \varepsilon^{ij} = i(e^i 
\bar
e^j - \bar e^i e^j) ,\qquad h^{ij} + i\varepsilon^{ij} = 2\bar e^i e^j 
.
\end{align}

Hence the $g=2$ action may be written as
\begin{equation}
S = \int d^3 x \sqrt h e^{-\Phi} \left( \frac i2 e^\Phi \psi^\dagger 
\overset{
\leftrightarrow}{D_0} \psi - \frac1{m} (\bar e^i D_i\psi^\dagger)(e^j 
D_j\psi)  \right) .
\end{equation}
In flat space, $e^i D_i \psi = D_{\bar z} \psi$ and we see the 
degeneracy
direction corresponds precisely to electrons in the LLL. Using a
Hubbard-Stratonovich transformation, we write this as
\begin{equation}
S = \int d^3 x \sqrt g e^{-\Phi} \left( \frac i2 e^\Phi \psi^\dagger 
\overset{
\leftrightarrow}{D_0}\psi - \chi (\bar e^i D_i\psi^\dagger) -\bar\chi (e^i
D_i\psi) + m \bar\chi\chi  \right) .
\end{equation}
The $m \rightarrow 0$ limit is then manifestly regular and the higher 
Landau
levels are now completely trivial to integrate out as $\chi$ and $\bar 
\chi$
simply become Lagrange multipliers enforcing the constraint
\begin{align}  \label{holomorphic}
e^i D_i \psi = 0 ,
\end{align}
which is the curved space equation for the LLL wave function. The 
many-body
problem of electrons confined to the LLL can then be understood as a 
system
of interacting particles with no kinetic energy
\begin{equation}\label{S-LLL}
S = \int d^3 x \sqrt h \left( \frac i2 \psi^\dagger 
\overset{\leftrightarrow}{D_0}
\psi \right)
\end{equation}
for which path integration is carried out subject to the 
holomorphic
constraint (\ref{holomorphic}).

Of course, the $g$-factor is generally not equal to 2 (for GaAs, it is in fact nearly zero: $g \approx - 0.03$ \cite{Weisbuch:1977rz}), but we can can consider a fictitious, placeholder system with $g=2$ by absorbing the $g$-factor term into the electromagnetic vector potential. That is, perform a redefinition of fields
\begin{align}
	A'_\mu = A_\mu - \frac{g'-g}{4m} B n_\mu.
\end{align}
If we denote the action (\ref{gFactorAction}) by $S_g[A_\mu]$, then we have
\begin{align}
	S_g [A_\mu] = S_{g'} [A'_\mu] ,
\end{align}
and one may simply consider $S_{g'}$ as the theory of interest. In this section we will take $g'=2$ so as to render the massless limit finite.

This doesn't actually do anything physical, the two theories are the same by construction. One can think of this procedure simply as a redefinition of currents to separate out the divergent parts. That is, if $\tau^\mu{}_I$ and $j^\mu$ are the currents of the original theory, the currents $\tau'^\mu{}_I$ and $j'^\mu$ defined by taking $e^I_\mu$ and $A'_\mu$ as the independent variables are related to the physical ones by
\begin{gather}\label{gFactorTransf}
	\tau'^\mu{}_I = \tau^\mu{}_I - \frac{g'-g}{4m} B j^\mu n_I - \frac{g'-g}{4m} j \varepsilon^{\mu \nu } F_{\nu \lambda} \Pi^\lambda{}_I, \nonumber \\
	j'^\mu = j^\mu + \frac{g' - g}{4m} \varepsilon^{\mu \nu} ( \nabla_\nu - T^\lambda{}_{\lambda \nu}) j + \frac{g'-g}{8m} j T^\mu{}_{\nu \lambda} \varepsilon^{\nu \lambda} ,
\end{gather}
where $j = n_\mu j^\mu$ as usual, which one may derive by varying $S_g[A_\mu] = S_{g'} [A'_\mu]$. When $g' = 2$, the primed currents are finite in the massless limit, while the unprimed currents are not. In what follows we will always work with the finite $g'=2$ currents and drop the subscripts, translating back to the physical $g$-factor when all is said and done. Note this procedure is unique to single constituent fluids, which we will be implicitly assuming throughout.\footnote{Also note we have not included a $g$-factor for $a_\mu$. This may be done in a Galilean invariant manner, but takes some care and is not particularly needed here.}

To this point, we have also assumed spinless electrons, while in reality electrons in quantum Hall samples are nearly spin polarized. If we assume only minimal coupling to curvature
\begin{align}
	D_\mu = \partial_\mu - i e A_\mu - i m a_\mu + i \mathfrak s \omega_\mu,
	&&\text{where}
	&&\omega_\mu = \frac{1}{2} \omega_{\mu AB} \epsilon^{AB}
\end{align}
(the physical value being $\mathfrak s = \frac 1 2$), then we may similarly perform a redefinition
\begin{align}
	A'_\mu = A_\mu + (\mathfrak s'-\mathfrak s) \omega_\mu
\end{align}
to change the spin from $\mathfrak s$ to $\mathfrak s'$. That is,
\begin{align}
	S_{g\mathfrak s} [A_\mu] = S_{g'\mathfrak s'} [A'_\mu] .
\end{align}
For convenience we will use this freedom to set $\mathfrak s'=0$ throughout this section. This alters the definition of the spin current in the manner $\tilde s'^{\mu AB} = \tilde s^{\mu AB} - \frac{1}{2} (\mathfrak  s'-\mathfrak  s) j^\mu \epsilon^{AB}$. Using the improvement formula (\ref{stressMomentumImprovement}) we then find
\begin{align}\label{sTransf}
	\tau'^{\mu \nu} = \tau^{\mu \nu} - (\mathfrak s' - \mathfrak s ) \varepsilon^{\lambda ( \mu} ( \nabla_\lambda - T^\rho{}_{\rho \lambda} ) j^{\nu)}+ (\mathfrak s'-\mathfrak s) T^\mu{}_{\lambda \rho} \left( j^\lambda \varepsilon^{\nu \rho} + \frac 1 2 j^\nu \varepsilon^{\lambda \rho} \right) 
\end{align}
while the charge and energy currents are not altered.
Performing this redefinition, we may assume a vanishing spin current throughout, which greatly simplifies the fluid analysis. 
Note that while taking $\mathfrak s' = 0$ is simply a matter of convenience, the choice $g'=0$ is essential to our logic. 

\subsubsection{The LLL Fluid}
Returning to the hydrodynamic regime, we may then perform the lowest Landau level limit by taking $m \rightarrow 0$, in which case the mass current vanishes\footnote{Of course, it is the unimproved mass current $\tilde \rho^\mu$ that vanishes, but these are the same since we have set $\mathfrak s'=0$.}
\begin{align}
	\rho^\mu \rightarrow 0.
\end{align}
The hydrodynamic equations of motion (\ref{parsedWard}) for the comoving currents then reduce to
\begin{gather}\label{LLLEOM}
	( \nabla_\mu - G_\mu ) j^\mu = 0, \nonumber \\
	( \nabla_\nu - G_\nu ) T^{\mu \nu } = F^\mu{}_\nu j^\nu - G^\mu{}_\nu \varepsilon^\nu , \nonumber \\
	( \nabla_\mu - G_\mu ) \varepsilon^\mu = E_\mu j^\mu + G_\mu \varepsilon^\mu - T_{\mu \nu} \nabla^\mu u^\nu . 
\end{gather}

We must now specify constitutive relations for the currents in terms of background fields and the hydrodynamic degrees of freedom (\ref{HydroDOF}). However, in the massless limit we lose three of these. The mass chemical potential $\mu_M$ drops out since there is no longer a conserved mass. Furthermore, note that the stress-mass tensor reduces to simply the stress tensor $\tau^{\mu \nu} \rightarrow T^{\mu \nu}$, which only has spatial components. 
A more radical consequence of the massless limit is then that stress-mass conservation, the continuum analog of Newton's second law,  loses it's time derivatives and becomes a force-free constraint. That is, since the fluid is massless it is obliged to flow in a manner such that the applied and internal forces exactly cancel.  We may use this constraint to eliminate any two dynamical degrees of freedom, which we will take to be the fluid velocity $u^\mu$.

\subsubsection{Constitutive Relations}\label{sec:constitutive relations}

Having done so, the currents are to be specified in terms of background fields, the temperature, and the charge chemical potential, which we will denote simply as $\mu$
\begin{align}
	T,
	&&\mu.
\end{align}
We will do so about equilibrium in flat spacetime with a large constant magnetic field. That is, constitutive relations are organized into a derivative expansion in which
\begin{align}
	T,
	&&\mu,
	&&F_{\mu \nu},
	&&e^A_\mu,
\end{align}
are specified to be order $\mathcal O( \partial^0 )$ and their derivatives order $\mathcal O( \partial^1 )$, etc. Note in particular the magnetic field
\begin{align}
	B = \frac 1 2 \varepsilon^{\mu \nu} F_{\mu \nu}
\end{align}
is large and nowhere vanishing.

As before, to zeroth order we have
\begin{align}
	j^\mu = q u^\mu , 
	&& \varepsilon^\mu = \epsilon u^\mu ,
	&&T^{\mu \nu} = p_\text{int} h^{\mu \nu} .
\end{align}
where these are arbitrary functions of our zeroth order scalars $T, \mu$, and $B$, so that in particular
\begin{align}\label{magnetoThermo}
	dp = s dT + q d \mu + M dB,
	&&\epsilon + p = T s + q \mu ,
\end{align}
where $M$ is the magnetization density.
Importantly, the pressure that appears in the stress tensor is not the thermodynamic pressure $p$ but the so-called internal pressure $p_\text{int} = p - M B$.
To see this, note that the variation of the grand canonical partition function $\Omega = \int d^3 x | e | p (T , \mu , B)$ picks up a term from the variation of $B$ with respect to the metric. 
In what follows, the analysis is simplified if we exchange $B$ for $M$ and take all functions of thermodynamic variables to be functions of $T$, $\mu$, and $M$.

Now, to zeroth order the equations of motion then read simply
\begin{align}
	F^\mu{}_\nu j^\nu = 0 ,
\end{align}
which we use to solve for $u^\mu$
\begin{align}
	u^\mu = \frac 1 {2B} \varepsilon^{\mu \nu \lambda} F_{\nu \lambda} = e^\Phi 
	\begin{pmatrix}
		1 \\
		\frac{\varepsilon^{ij} E_j}{B}
	\end{pmatrix},
\end{align}
which one may check is the unique zero eigenvector of $F_{\mu \nu}$. That is, unsurprisingly, the LLL projection constrains the fluid to move at the drift velocity $u^i = e^\Phi \frac{\varepsilon^{ij} E_j}{B}$ in equilibrium.
Note that if we define $\overset u \varepsilon_{\mu \nu} = \varepsilon_{\mu \nu \lambda} u^\lambda$, we have the useful identities
\begin{align}
	F_{\mu \nu} = B \overset u \varepsilon_{\mu \nu} ,
	&&\varepsilon^{\lambda \mu} \overset u \varepsilon_{\lambda \nu} = \overset u P{}^\mu{}_\nu ,
\end{align}
which we will often use in computations that follow.

Let's continue our expansion.
Organized into representations of $SO(2)$, the complete set of first order data is 
\begin{center}
	\begin{tabular}{ l | c c c}
			& Independent Data  \\
		Scalar & $\theta \qquad  ( \dot  T ) 
\qquad ( \dot \mu )$   \\
		Vector & $\nabla^\mu T \qquad \nabla^\mu \mu\qquad 
 \nabla^\mu M \qquad G^\mu$\\
		Traceless Symmetric Tensor & $\sigma^{\mu \nu}$ .
	\end{tabular}
\end{center}
The material derivative of all three 
thermodynamic variables are not included since one may always be eliminated by the 
constraint
\begin{align}
	\nabla_\mu ( B u^\mu ) = 0 , && \implies &&
	\dot B = - B \theta ,
\end{align}
where the magnetic field is viewed as a function $B= B(T , \mu , M )$. This follows from the Bianchi identity $d \df F = 0$.
Note we have used the decomposition (\ref{DECOMP}) of $\nabla_\mu u^\nu$, which we repeat here
\begin{align}
	\nabla_\mu u^\nu = n_\mu \alpha^\nu + \frac 1 2 \sigma_\mu{}^\nu + \frac 1 2\theta \overset u P {}^\nu{}_\mu + \frac 1 2 \omega_\mu{}^\nu.
\end{align}
Of course, the comoving electric field $E_\mu = F_{\mu \nu} u^\nu$ is zero by definition and does not appear.

Not all of this data is independent on shell and we may choose to 
eliminate some in favor of the others by solving the equations of 
motion. In our case, there are two scalar equations: the continuity 
equation and the work-energy equation. We use these to eliminate the 
material derivatives of $T$ and $\mu$, as indicated by parentheses.

We have not included the acceleration and vorticity amongst the allowed data since these necessarily involve the mass vector potential $a_\mu$. We can show this quite easily in the simple case of a flat metric and clock form, in which case $\Gamma^i{}_{t t} = - \partial^i a_t + \partial_t a^i$ and $\Gamma^j{}_{ti} = \frac 1 2 ( d \df a )_i{}^j$ are the only independent non-zero Christoffel components. Then
\begin{align}
	\alpha^i = u^\mu \partial_\mu u^i - (d \df a)^i{}_j u^j + \partial_t a^i - \partial^i a_t,
	&&\omega^{ij} = 2 \partial^{[i} ( u^{j]} + a^{j]} ).
\end{align}
The first equation exhibits the expected appearance of the gravitational potential in the acceleration which is the hallmark of Newton-Cartan geometry, while the appearance of $a_i$ in the vorticity encodes the Coriolis effect. In either case, a massless, spinless theory cannot couple to $\df a$ and so $\omega$ and $\alpha^\mu$ are off limits.

Finally, a few words are due on the fluid frame invariants (\ref{frame invariants}) before we expand them in first order data. Typically, under a frame transformations, the transverse corrections $j^\mu = (q + \mathcal Q) u^\mu + \nu^\mu$, $\varepsilon^\mu = (\epsilon + \mathcal E ) u^\mu + \xi^\mu$ vary as
\begin{align}
	\delta \nu^\mu = q \delta u^\mu,
	&&\delta \xi^\mu = ( \epsilon + p_\text{int} ) \delta u^\mu
\end{align}
(see section 4.2 of \cite{Jensen:2014ama}). However, since in our case the fluid velocity is no longer a dynamical variable but rather a constrained function of background fields, there is no freedom to redefine it. Hence $\nu^\mu$ and $\xi^\mu$ are themselves frame invariant. Indeed, since the mass current vanishes, we have
\begin{align}
	\mathcal J^\mu = \nu^\mu ,
	&&\mathcal E^\mu = \xi^\mu .
\end{align}

Now let's expand the frame invariants in our data
\begin{align}\label{constitutiveRelations}
	\mathcal S &= - \zeta \theta ,
	\qquad \qquad 
	\mathcal T^{\mu \nu} = - \eta 
\sigma^{\mu \nu} - \tilde \eta \tilde \sigma^{\mu \nu}\nonumber \\
	\mathcal J^\mu &= \chi_T \nabla^\mu T + \chi_{\mu} \nabla^\mu \mu 
+ \tilde \chi_M \nabla^\mu M + \chi_G G^\mu \nonumber \\ 
	&\qquad\qquad+ \tilde \chi_T \varepsilon^{\mu \nu} \nabla_\nu T + 
\tilde \chi_{\mu} \varepsilon^{\mu \nu} \nabla_\nu \mu + \chi_M 
\varepsilon^{\mu \nu} \nabla_\nu M + \tilde \chi_G  \varepsilon^{\mu 
\nu} G_\nu , \nonumber \\
	\mathcal E^\mu &= \Sigma_T \nabla^\mu T + \Sigma_{\mu} \nabla^\mu \mu 
+ \tilde \Sigma_m \nabla^\mu m + \Sigma_G G^\mu \nonumber \\ 
	&\qquad\qquad+ \tilde \Sigma_T \varepsilon^{\mu \nu} \nabla_\nu T + 
\tilde \Sigma_{\mu} \varepsilon^{\mu \nu} \nabla_\nu \mu + \Sigma_M 
\varepsilon^{\mu \nu} \nabla_\nu M + \tilde \Sigma_G  \varepsilon^{\mu 
\nu} G_\nu , \nonumber \\
	\zeta^\mu &= \zeta_\theta \theta u^\mu + \zeta_T \nabla^\mu T 
+ \zeta_{\mu} \nabla^\mu \mu + \tilde \zeta_M \nabla^\mu M + \zeta_G 
G^\mu \nonumber \\ 
	&\qquad \qquad + \tilde \zeta_T \varepsilon^{\mu \nu} \nabla_\nu T + \tilde 
\zeta_{\mu} \varepsilon^{\mu \nu} \nabla_\nu \mu + \zeta_M 
\varepsilon^{\mu \nu} \nabla_\nu M + \tilde \zeta_G \varepsilon^{\mu 
\nu} G_\nu .
\end{align}
To save on notation, we have used a tilde to denote oddness under parity.

\subsubsection{Force-Free Flows}
We first turn to consistency with the force-free constraint.
As mentioned previously, we may use the force balance constraint to 
solve for the charge current $j^\mu = (q + \mathcal Q) u^\mu + \nu^\mu$
\begin{align}\label{restrictions1}
	\nabla^\mu p_\text{int} &= B \varepsilon^{\mu \nu} 
\nu_\nu 
+ ( \epsilon + p_\text{int} ) G^\mu  \nonumber \\
	\tilde \chi_T &= - \frac{s}{B}\,, \qquad
	\tilde \chi_\mu = - \frac{q}{B}\,,  \qquad
	\chi_M = 1, \qquad
	\tilde \chi_G =  \frac{\epsilon + p_\text{int}}{B}\, .
\end{align}
All charge transport coefficients are thus determined by the equation 
of 
state. Also note that all longitudinal responses are zero. This is 
because the Lorentz force must cancel forces from pressure gradients 
and 
the magnetic field always produces a force perpendicular to the 
current.
Hence the current must be perpendicular to pressure gradients.

\subsubsection{Entropy Current Analysis}\label{sec:entropy analysis}
The constitutive relations (\ref{constitutiveRelations}) and (\ref{restrictions1}) are the most general possible
that are consistent with the equations of motion and
constraint. However, it is still possible to generate flows that
violate the second law of thermodynamics. For example, it is well
known that a negative shear viscosity allows one to remove entropy
from an isolated system and so we should have $\eta \geq 0$
\cite{landau1987fluid}. To derive all such restrictions, we perform an
entropy current analysis along the lines of
\cite{Jensen:2011xb}.
This involves imposing the second law of thermodynamics locally. For non-negative entropy production 
between all spatial slices, we must have
\begin{align}\label{second law}
	(\nabla_\mu - G_\mu ) s^\mu \geq 0 .
\end{align}

We begin by considering the terms in $(\nabla_\mu - G_\mu ) s^\mu$ that involve genuine second order data. That is, terms that are not simply products and contractions of the first-order data given above. Since the divergence of the canonical entropy current is a quadratic form in first order data (see \ref{quadraticForm}), this gives merely
\begin{align}
	\nabla_\mu \zeta^\mu \Big|_{2 - \partial} = \zeta_\theta \dot \theta  + \zeta_T \nabla^2 T + \zeta_{Q} \nabla^2 \mu_Q + 
\tilde \zeta_M \nabla^2 M + \zeta_G  \nabla_\mu G^\mu ,
\end{align}
where we have used the Newton-Cartan identities $\varepsilon^{\mu \nu} 
G_{\mu \nu} = 0$ and $\varepsilon^{\mu \nu} \nabla_\mu G_\nu = 0$ (which follow from the causality of spacetime and the Bianchi identity $d \df G = 0$ respectively). 
Since each term is sign indefinite and independent, these coefficients must all vanish for entropy production to be strictly non-negative.

The remaining first order data is then
\begin{align}\label{entropyCurrentAnalysis}
	(\nabla_\mu - G_\mu ) s^\mu 
	&= \frac{1}{T} \zeta \Theta^2 + \frac{1}{2T} \eta \sigma_{\mu 
\nu} \sigma^{\mu \nu} + \frac{1}{T} \Sigma_G G_\mu G^\mu \nonumber \\
	&+ \frac{1}{T} \Big( \Sigma_T - \frac{1}{T} \Sigma_G \Big) 
G^\mu \nabla_\mu T + \frac{1}{T} \Sigma_\mu G^\mu \nabla_\mu \mu + 
\frac{1}{T} \tilde \Sigma_M G^\mu \nabla_\mu M \nonumber \\
	&- \Big( \partial_T \tilde \zeta_G + \tilde \zeta_T - \frac{1}
{T}\tilde \Sigma_T - \frac{1}{T^2} \tilde \Sigma_G + \frac{\mu}{T^2} 
\tilde \chi_G \Big)  \varepsilon^{\mu \nu} G_\mu \nabla_\nu  T 
\nonumber \\
	&- \Big( \partial_\mu \tilde \zeta_G + \tilde \zeta_\mu - 
\frac{1}{T}\tilde \Sigma_\mu - \frac{1}{T} \tilde \chi_G \Big) 
\varepsilon^{\mu \nu} G_\mu \nabla_\nu \mu
	- \Big( \partial_M \tilde \zeta_G + \zeta_M - \frac{1}{T} 
\Sigma_M \Big)  \varepsilon^{\mu \nu} G_\mu \nabla_\nu  M \nonumber \\
	&- \frac{1}{T^2} \Sigma_T \nabla_\mu T \nabla^\mu T - \frac{1}
{T^2} \Sigma_\mu \nabla_\mu T \nabla^\mu \mu - \frac{1}{T^2} \tilde 
\Sigma_M \nabla_\mu T \nabla^\mu M \nonumber \\
	& +\Big( \partial_T \tilde \zeta_\mu - \partial_\mu \tilde 
\zeta_T + \frac{1}{T} \tilde \chi_T + \frac{\mu}{T^2} \tilde \chi_\mu 
- \frac{1}{T^2} \tilde \Sigma_\mu \Big) \varepsilon^{\mu \nu} 
\nabla_\mu T \nabla_\nu \mu \nonumber \\
	& +\Big( \partial_T \zeta_M - \partial_M \tilde \zeta_T + 
\frac{\mu}{T^2} \chi_M - \frac{1}{T^2} \Sigma_M \Big) \varepsilon^{\mu 
\nu} \nabla_\mu T \nabla_\nu M \nonumber \\
	&+\Big( \partial_\mu \zeta_M - \partial_M \tilde \zeta_\mu - 
\frac{1}{T} \chi_M \Big) \varepsilon^{\mu \nu} \nabla_\mu \mu 
\nabla_\nu M \nonumber \\
	&\geq 0.
\end{align}
Note that by $\partial_\mu$ we mean the partial derivative with 
respect to the chemical potential, not a spacetime derivative. For 
clarity we will always use $\nabla_\mu$ for the spacetime derivative.

The $\nabla_\mu T \nabla^\mu T$, $G^\mu \nabla_\mu T$, and $G_\mu 
G^\mu$ terms need not be separately constrained. We obtain a less 
stringent condition by setting $\Sigma_T = - \frac{1}{T} \Sigma_G$, in 
which case they arrange into a perfect square
\begin{align}
	- \frac{1}{T^2} \Sigma_T (\nabla_\mu T - T G_\mu) (\nabla^\mu 
T - T G^\mu) .
\end{align}
We note in passing that in thermal equilibrium there can be no entropy 
production. This implies
\begin{align}\label{equilibrium T}
	\nabla^\mu T = T G^\mu ,
\end{align}
or $\partial_i T = T \partial_i \Phi$ in coordinates where $n_\mu = ( e^{-\Phi} ~0)$. The physics of 
this was discussed below equation (\ref{GTCclockForm}). Heat will 
tend to flow from regions of higher $-\Phi$ to lower $-\Phi$. 
Equilibrium is reached once the temperature profile is such that 
(\ref{equilibrium T}) is satisfied. That is, we have proven (\ref{TEquilibrium}) and the torsional Einstein relation.

From this we immediately obtain the 
expected signs of the parity even viscosities and thermal conductivity
\begin{align}
	\zeta \geq 0, \qquad
	\eta \geq 0, \qquad
	\Sigma_T \leq 0 .
\end{align}
Meanwhile the remaining terms place new restrictions on the energy and entropy 
coefficients
\begin{align}\label{restrictions}
	\Sigma_\mu = \tilde \Sigma_M = 0, \qquad&
	\begin{pmatrix}
		 \partial_\mu \zeta_M - \partial_ M \tilde \zeta_\mu 
\\
		\partial_M \tilde \zeta_T - \partial_T \zeta_M \\
		\partial_T \tilde \zeta_\mu - \partial_\mu \tilde 
\zeta_T
	\end{pmatrix}
	=
	\begin{pmatrix}
		\frac{1}{T} \chi_M  \\
		- \frac{1}{T^2} \Sigma_M + \frac{\mu}{T^2} \chi_M \\
		\frac{1}{T^2} \tilde \Sigma_\mu - \frac{1}{T} \tilde 
\chi_T - \frac{\mu}{T^2} \tilde \chi_\mu
	\end{pmatrix},
	\nonumber \\
	\begin{pmatrix}
		\partial_T \tilde \zeta_G \\
		\partial_\mu \tilde \zeta_G \\
		\partial_M \tilde \zeta_G
	\end{pmatrix}
	&=
	\begin{pmatrix}
		- \tilde \zeta_T + \frac{1}{T} \tilde \Sigma_T + 
\frac{1}{T^2} \tilde \Sigma_G - \frac{\mu}{T^2} \tilde \chi_G \\
		- \tilde \zeta_\mu + \frac{1}{T} \tilde \Sigma_\mu + 
\frac{1}{T} \tilde \chi_G \\
		- \zeta_M + \frac{1}{T} \Sigma_M
	\end{pmatrix} .
\end{align}
Note that the final set of constraints is obtained only by considering fluids on backgrounds with temporal torsion $\df T^0$, or equivalently, a nontrivial Luttinger potential.

We seek the most general solution to these constraints. Begin by 
eliminating the entropy coefficients by taking the curl of the third 
equation and plugging in the second
\begin{align}\label{PDE}
	\begin{pmatrix}
		 \partial_{\mu} \Big( \frac{1}{T} \Sigma_M \Big) - 
\partial_M \Big( \frac{1}{T}\tilde \Sigma_{ \mu} \Big) \\
		 \partial_M \Big( \frac{1}{T} \tilde \Sigma_T \Big) - 
\partial_T \Big( \frac{1}{T} \Sigma_M \Big) \\
		\partial_T \Big( \frac{1}{T} \tilde \Sigma_{ \mu} 
\Big) - \partial_{ \mu} \Big( \frac{1}{T} \tilde \Sigma_T \Big)
	\end{pmatrix} 
	=
	\begin{pmatrix}
		\frac{1}{T} \big( \chi_M + \partial_M \tilde \chi_G 
\big) \\
		- \frac{1}{T^2} \big( \Sigma_M + \partial_M \tilde 
\Sigma_G ) + \frac{\mu}{T^2} \big( \chi_M + \partial_M \tilde \chi_G 
\big) \\
		\frac{1}{T^2} \big( \tilde \Sigma_\mu + \partial_\mu 
\tilde \Sigma_G \big) - \frac{1}{T} \big( \tilde \chi_T + \partial_T 
\tilde \chi_G \big) - \frac{\mu}{T^2} \big( \tilde \chi_\mu + 
\partial_\mu \tilde \chi_G \big) 
	\end{pmatrix} .
\end{align}
Since the left hand side is the curl of a vector, the right hand side 
is divergenceless and it appears as if we might obtain another 
constraint. However one may check that this is automatically satisfied 
by virtue of the constraints (\ref{restrictions1}) and the 
thermodynamic identities (\ref{magnetoThermo}).

We may simplify the partial differential equations (\ref{PDE}) by a 
substitution that isolates the energy response's dependence on the 
equation of state and $\tilde \Sigma_G$
\begin{align}
	&\tilde \Sigma_T = - \frac{1}{T} \tilde \Sigma_G + \frac{\mu}
{T} \frac{Ts + \mu n}{B} + T^2 \tilde g_T, \qquad
	\tilde \Sigma_\mu = - \frac{Ts + \mu n}{B} + T^2 \tilde g_\mu 
, \qquad
	\Sigma_M = T^2 g_M, \nonumber \\
	&\qquad\qquad\qquad \implies \qquad
	\begin{pmatrix}
		\partial_\mu g_M - \partial_M \tilde g_\mu \\
		\partial_M \tilde g_T - \partial_T g_M \\
		\partial_T \tilde g_\mu - \partial_\mu \tilde g_T
	\end{pmatrix}
	=
	\begin{pmatrix}
		0 \\
		0 \\
		0
	\end{pmatrix} .
\end{align}
Since $( \tilde g_T , \tilde g_\mu , g_M )$ is curl free, 
it must be 
the gradient of some function
\begin{align}
	\tilde g_T = \partial_T \tilde g ,\qquad
	\tilde g_\mu = \partial_\mu \tilde g , \qquad
	g_M = \partial_M \tilde g .
\end{align}
The transverse energy response is then determined by the equation of state and the two free functions $\tilde \Sigma_G$ and $\tilde g$. In what follows, the formulas will be significantly simplified by the redefinition of coefficients
\begin{align}
	\tilde \Sigma_G &= \frac{\mu}{B} \left( Ts + \mu n 
\right) + T^2 c_{RL} - 2 M_E , \qquad
	\tilde g = \frac{M_E}{T^2}  \,.
\end{align}

In summary, the most general set of constitutive relations compatible with equations of motion and thermodynamics is given by an equation of state $p$, three parity even coefficients $\zeta, \eta, \Sigma_T$ and three parity odd coefficients $\tilde \eta, c_{RL},$ and $M_E$, all of which are arbitrary functions of the thermodynamic variables
\begin{gather}
	\mathcal S = - \zeta \theta,
	\qquad \qquad \qquad \qquad
	\mathcal T^{\mu \nu} = - \eta \sigma^{\mu \nu} - \tilde \eta \tilde \sigma^{\mu \nu} ,
	\nonumber \\
	\mathcal J^\mu = - \frac s B \varepsilon^{\mu \nu} ( \nabla_\nu - G_\nu ) T - \frac{q}{B} \varepsilon^{\mu \nu} ( \nabla_\nu - G_\nu ) \mu + \varepsilon^{\mu \nu} ( \nabla_\nu - G_\nu ) M, \nonumber \\
	\mathcal E^\mu = \Sigma_T \big( \nabla^\mu 
- G^\mu \big) T - T c_{RL} \varepsilon^{\mu \nu} (\nabla_\nu - G_\nu )T - \frac{\epsilon + p}{B} \varepsilon^{\mu \nu} ( \nabla_\nu - G_\nu ) \mu + \varepsilon^{\mu \nu} ( \nabla_\nu - 2 G_\nu ) M_E ,
\end{gather}
where 
\begin{align}
	\zeta \geq 0 ,
	&&\eta \geq 0,
	&&\Sigma_T \leq 0.
\end{align}
As we shall see, these are the bulk viscosity, shear viscosity, and thermal conductivity, which indeed should carry the signs given above.

\subsubsection{Physical Interpretation}\label{interpretation}
This completes the entropy current analysis, so let's begin to untangle what these results mean physically. To make them clearer, lets select a fluid frame (the so-called Landau frame) where $\mathcal Q = \mathcal E =0$. That is, we are defining temperature and chemical potential out of equilibrium so that physical charge and energy density match their local thermodynamic values: $n_\mu j^\mu = q ( T , \mu_Q , m)$ and $n_\mu \varepsilon^\mu = \epsilon ( T , \mu_Q , m )$. We then have $\mathcal S = \mathcal P$ and the currents are
\begin{align}\label{summary}
	\rho'^\mu &= 0 , \nonumber \\
	j'^\mu &= q' u'^\mu - \frac {s'} {B'} \varepsilon^{\mu \nu} ( \nabla_\nu - G_\nu)  T - \frac {q'}{ B'} \varepsilon^{\mu \nu} (\nabla_\nu - G_\nu ) \mu' + \varepsilon^{\mu \nu} ( \nabla_\nu - G_\nu ) M' ,\nonumber \\
	\varepsilon'^\mu &= \epsilon' u'^\mu + \Sigma'_T ( \nabla^\mu - G^\mu ) T - T c'_{RL} \varepsilon^{\mu \nu} ( \nabla_\nu - G_\nu ) T \nonumber \\
	&\qquad \qquad - \frac{\epsilon' + p'}{B'} \varepsilon^{\mu \nu} ( \nabla_\nu - G_\nu ) \mu'+ \varepsilon^{\mu \nu} ( \nabla_\nu - 2 G_\nu ) M'_\epsilon,
	\nonumber \\
	T'^{\mu \nu} &= ( p' - M' B' - \zeta' \theta') h^{\mu \nu} - \eta' \sigma'^{\mu \nu} - \tilde \eta' \tilde \sigma'^{\mu \nu} .
\end{align}
We thus see that $\zeta, \eta$, and $\tilde \eta$ are the bulk, shear, and Hall viscosities respectively, $c_{RL}$ is a thermal Hall coefficient, and $M_E$ an energy magnetization. 

The primes are to remind us that these currents are defined at $g'=2$ and $s'=0$, where the lowest Landau level limit is regular. To obtain the true currents we need to transform back to the physical values of the spin and $g$-factor using (\ref{gFactorTransf}) and (\ref{sTransf}). To do so we first note that in passing back to the physical electromagnetic field $A_\mu$ we find the drift velocity becomes
\begin{align}
	u^\mu = u'^\mu - \frac{g-2}{4m B} \varepsilon^{\mu \nu} ( \nabla_\nu - G_\nu )B .
\end{align}
The magnetic field is also altered to $B = B' + \mathfrak s R$, but this does not alter our formulas to the order in derivatives that we are working. The magnetic field dependent shift to $A_t$ also alters the chemical potential
\begin{align}
	\mu = \mu'- \frac{g-2}{4m} B .
\end{align}

Finally, the physical currents are given by\footnote{In this computation, one must also remember that these currents are the comoving ones measured relative to $u^\mu$, not $u'^\mu$. For instance $\varepsilon^\mu = \tau^\mu{}_I \mathring u^I$ while $\varepsilon'^\mu = \tau'^\mu{}_I \mathring u'^I$.}
\begin{align}\label{comovingTranslation}
	\rho^\mu &= \rho'^\mu + \frac 1 2 \mathfrak  s \varepsilon^{\mu \nu} ( \nabla_\nu - G_\nu ) q , \nonumber \\
	j^\mu &= j'^\mu + \frac{g-2}{4m} \varepsilon^{\mu \nu} ( \nabla_\nu - G_\nu ) q, \nonumber \\
	\varepsilon^\mu &= \varepsilon'^\mu - \frac{g-2}{4m} B j'^\mu + \frac{g-2}{4mB} T'^\mu{}_\nu \varepsilon^{\nu \lambda} ( \nabla_\lambda - G_\lambda) B, \nonumber \\
	T^{\mu \nu} &= T'^{\mu \nu} - \left(  \frac{g-2}{4m} Bq + \frac 1 2 \mathfrak  sq \omega \right) h^{\mu \nu} - \frac 1 2 \mathfrak  s q \tilde \sigma^{\mu \nu}
\end{align}
where $\omega = \frac 1 2 \varepsilon^{\mu \nu} \omega_{\mu \nu} = \overset u \varepsilon_{\mu \nu} \nabla^\mu u^\nu$.
Note that in passing to the physical spin $\mathfrak s$, we pick up a momentum current that is pure magnetization and tied to the variation in the spin density $\mathfrak sq$.
Using these formulas, we find
\begin{align}
	\rho^\mu &=  \frac 1 2 \mathfrak s \varepsilon^{\mu \nu} ( \nabla_\nu - G_\nu ) q , \nonumber \\
	j^\mu &= q u^\mu - \frac {s} {B} \varepsilon^{\mu \nu} ( \nabla_\nu - G_\nu)  T - \frac {q}{ B} \varepsilon^{\mu \nu} (\nabla_\nu - G_\nu ) \mu + \varepsilon^{\mu \nu} ( \nabla_\nu - G_\nu ) M ,\nonumber \\
	\varepsilon^\mu &= \epsilon u^\mu + \Sigma_T ( \nabla^\mu - G^\mu ) T - T c_{RL} \varepsilon^{\mu \nu} ( \nabla_\nu - G_\nu ) T \nonumber \\
	&\qquad \qquad - \frac{\epsilon + p}{B} \varepsilon^{\mu \nu} ( \nabla_\nu - G_\nu ) \mu+ \varepsilon^{\mu \nu} ( \nabla_\nu - 2 G_\nu ) M_\epsilon,
	\nonumber \\
	T^{\mu \nu} &= \left( p - M B - \zeta \theta- \frac 1 2 \mathfrak s q \omega \right) h^{\mu \nu} - \eta \sigma^{\mu \nu} - \tilde \eta \tilde \sigma^{\mu \nu} .
\end{align}
where
\begin{align}\label{shift}
	&\epsilon = \epsilon' - \frac{g-2}{4m}Bq' ,
	&&M = M' + \frac{g-2}{4m} q',
	&&M_\epsilon = M'_\epsilon - \frac{g-2}{4m} M' B \nonumber \\
	&\tilde \eta = \tilde \eta' + \frac 1 2 \mathfrak s q',
	&&c_{RL} = c'_{RL}- \frac{g-2}{4m} \frac{s'}{T}.
\end{align}
Here primed objects are evaluated at the primed chemical potential, say $p'=p'(T,\mu',B)$. Those unprimed coefficients not listed in (\ref{shift}) are trivially related to their primed counterparts in the manner $p(T, \mu ,B) = p'(T, \mu', B)$.

The physics of this should be fairly clear: the shift to the energy density corresponds to the energy cost of a magnetic moment in a magnetic field, the shift to $M$ simply keeps track of the increased magnetic moment assigned to each particle, while the shift to the Hall viscosity accounts for our underestimation of the spin density of the liquid. In all, the form of the constitutive relations (\ref{summary}) does not change except for the mass magnetization current that is introduced in the presence of a varying spin density $\mathfrak sq$ and the vorticity induced pressure $\frac 1 2 \mathfrak s q \omega$, also fixed to the spin density of the liquid.
Of course, one will have to perform the shift (\ref{shift}) in a hypothetical calculation of the coefficients in the lowest Landau level limit $m\rightarrow 0$.

Finally, recall that we have been working with the currents measured by observers moving with the drift velocity $u^\mu = e^\Phi \begin{pmatrix} 1 & \frac{\varepsilon^{ij} E_j}{B} \end{pmatrix}^T$. The lab currents (the components of $\tilde \tau^\mu{}_I$) must be obtained from (\ref{nonCovCur}). However, in the massless limit most of the modifications drop out and we find to first order in derivatives\footnote{Note that in passing to $\varepsilon^\mu_\text{lab}$ one introduces terms proportional to the {\it unimproved} mass current $\tilde \rho^\mu$ since one should not improve the energy current. Hence these terms are zero.}
\begin{align}\label{currents}
	\rho^t &= 0,
	\qquad \qquad
	\rho^i = \frac 1 2 \mathfrak  s e^\Phi \varepsilon^{ij} \partial_j (e^{-\Phi}q) ,
	\nonumber \\
	j^t &= e^\Phi q,
	\qquad \qquad
	\varepsilon_\text{lab}^t = e^\Phi  \epsilon, \nonumber \\
	j^i &= e^\Phi \varepsilon^{ij} \left( \frac{q}{B} 
\Big(E_j - \d_j (e^{-\Phi} \mu )\Big) - \frac{s}{B}\d_j(e^{-\Phi}T) + 
\d_j (e^{-\Phi} M  )\right),\nonumber \\
	\varepsilon^i_\text{lab} &= e^\Phi \varepsilon^{ij} \bigg(  
\frac{\epsilon + p_\text{int}}{B} \left( E_j - \d_j (e^{-\Phi}  \mu  ) \right) - 
M \d_j (e^{-\Phi}  \mu ) 
	-  T c_{RL} \partial_j ( e^{-\Phi} T ) 
	+ e^\Phi \partial_j ( e^{-2 \Phi} M_\epsilon )   \bigg) \nonumber \\
	&\qquad \qquad + \Sigma_T e^\Phi \partial^i \left( e^{-\Phi} T \right) ,
	\nonumber \\
	T_\text{lab}^{i j} &= \left( p_\text{int} - \zeta \theta + \frac 1 2 \mathfrak  s q \omega + \frac{\mathfrak s}{B} E^k \partial_k q \right) h^{i j} - \frac{\mathfrak s}{B} E^{(i} \partial^{j)} q - \eta \sigma^{i j} - \tilde \eta \tilde \sigma^{i j} .
\end{align}

\subsubsection{St\v reda Formulas}
One notable feature of these formulas is the currents that persist 
in thermal equilibrium. We now turn to these, deriving a set of St\v 
reda-like formulas for two-dimensional fluids. Throughout we shall consider flat space $h^{ij} =\delta^{ij}$ with a perturbing Luttinger potential $n_t = e^{-\Phi}$, $n_i =0$. First note that in thermal equilibrium we have $  T \partial_i ( \frac{\mu}{T} ) = 
 e^\Phi E_i$ and $\partial_i T = T G_i$, where $G_i = 
\partial_i \Phi$ is the external force exerted by the potential 
$-\Phi$. The equilibrium currents are then
\begin{align}
	j^i = \varepsilon^{ij} e^\Phi \d_j ( e^{-\Phi} M ) 
\qquad
	\varepsilon^i_\text{lab} = \varepsilon^{ij} \left( - M e^\Phi\d_j ( 
e^{-\Phi} \mu ) + e^{2\Phi}\d_j (e^{-2\Phi} M_E ) \right).
\end{align}
Expressing these in terms of the externally applied fields $E_i$, $B$, 
and $G_i$ we have
\begin{align}\label{persistent}
	j^i&= \varepsilon^{ij} \big( e^\Phi \d_\mu M E_j + 
\d_B M \d_j B + (T \partial_T M + \mu \d_\mu M - M )G_j \big) 
\nonumber \\
	\varepsilon^i_\text{lab} &= \varepsilon^{ij} \big( e^\Phi ( \d_\mu M_\epsilon 
- M ) E_j + \d_B M_\epsilon \d_j B +(T \d_T M_\epsilon + \mu \d_\mu 
M_\epsilon - 2 M_\epsilon ) G_j \big) .
\end{align}

Defining equilibrium responses by
\begin{align}
	j^i &=  \varepsilon^{ij} \big( \sigma^{\text{eq}}_H 
E_j  +  \sigma^{B\text{eq}}_H \d_j B + \sigma^{G\text{eq}}_H G_j \big),
\end{align}
and using some Maxwell relations, we have
\begin{align}
	\sigma^\text{eq}_H &= \left( \frac{\d q}{\d B} \right)_{T,\mu} ,
\qquad
	\sigma^{B\text{eq}}_H = \left( \frac{\d M}{\d B} 
\right)_{T,\mu} , \nonumber \\
	\sigma^{G\text{eq}}_H &= T \left( \frac{\d s}{\d B} 
\right)_{T,\mu} + \mu \left( \frac{\d q}{\d B}\right)_{T , \mu} - M ,
\end{align}
where we have set $\Phi = 0$ in these formulas. The first is the 
well-known St\v reda formula \cite{streda1982theory}. The following two are St\v 
reda-like formulas for currents induced by inhomogeneities in $B$ and 
an external Luttinger potential. 

Similarly working with the energy current
\begin{align}
	\varepsilon^i_\text{lab} &=  \varepsilon^{ij} \big( 
\kappa^{\text{eq}}_H E_j  +  \kappa^{B\text{eq}}_H \d_j B + 
\kappa^{G\text{eq}}_H G_j \big) ,
\end{align}
we find that all equilibrium currents are determined by the 
magnetization $M$ and energy magnetization $M_E$.
\begin{align}
	\kappa^\text{eq}_H &= \left( \frac{\d M_E}{\d \mu} 
\right)_{T,B} - M ,\qquad
	\kappa^{B\text{eq}}_H = \left( \frac{\d M_E}{\d B} 
\right)_{T,\mu} , \nonumber \\
	\kappa^{G\text{eq}}_H &= T\left( \frac{\d M_E}{\d T} 
\right)_{\mu , B}  + \mu \left( \frac{\d M_E}{\d \mu} \right)_{T , B} 
- 2 M_E .
\end{align}
A similar collection of St\v reda formulas was presented in 
~\cite{Gromov:2014vla}, however they do not agree with ours due to a difference in the Luttinger potential factors.

Finally, there are also equilibrium persistent mass currents due to the spin of the electron. By the same reasoning as above, we find
\begin{align}
	\rho^i &=  \varepsilon^{ij} \big( 
\lambda^{\text{eq}}_H E_j  +  \lambda^{B\text{eq}}_H \d_j B + 
\lambda^{G\text{eq}}_H G_j \big) ,
\end{align}
where
\begin{align}
	&\lambda^\text{eq}_H = \frac 1 2 \mathfrak s \left( \frac{\partial q}{\partial \mu} \right)_{T,B},
	\qquad \qquad \lambda^{B\text{eq}}_H = \frac 1 2 \mathfrak s \left( \frac{\partial q}{\partial B} \right)_{T,\mu} , \nonumber \\
	&\lambda^{G\text{eq}}_H = \frac 1 2 \mathfrak s \left( T \left( \frac{\partial q}{\partial T} \right)_{\mu,B} +  \mu \left( \frac{\partial q}{\partial \mu} \right)_{T,B} - q \right) .
\end{align}

\subsection{Fluids in $2+1$ Dimensions}\label{sec:Fluid}

In this section we carry out the same analysis but for standard $(2+1)$-dimensional multi-constituent fluids in a perturbing external electromagnetic field. In the relativistic case, $(2+1)$-dimensional fluids are known to exhibit a rich collection of transport coefficients in the parity odd sector \cite{Jensen:2011xb} and we find the same here.

In the parity even sector, the entropy current analysis returns four sign semidefinite functions of $T, \mu_Q$, and $\mu_M$: a bulk viscosity, shear viscosity, conductivity, and thermal conductivity
\begin{align}
	\zeta \geq 0,
	&&\eta \geq 0,
	&&\sigma_E \geq 0 ,
	&&\kappa_T \leq 0,
\end{align}
as well as an unconstrained transverse thermo-electric coefficient
\begin{align}
	\sigma_T .
\end{align}
The parity odd sector on the other hand admits six arbitrary coefficients, a Hall viscosity, Hall conductivity, thermal Hall conductivity, thermo-electric Hall coefficient, magnetization and energy magnetization
\begin{align}
	\tilde \eta,
	&&\tilde \sigma_E,
	&&\tilde \kappa_T,
	&&\tilde \sigma_T ,
	&& \tilde m,
	&& \tilde m_\epsilon.
\end{align}
The magnetization determines the magnetic field induced pressure via the coefficient
\begin{align}
	&\tilde f_B =T^2 \partial_\epsilon p  \partial_T \left(  \frac{\tilde m}{T} \right) + \partial_q p \partial_Q  \left(  \frac{\tilde m}{T} \right) + \partial_\rho p \partial_M  \left(  \frac{\tilde m}{T} \right) ,
\end{align}
where $p ( \epsilon , q , \rho)$ is the pressure as a function of energy, charge and mass density. Kubo formulas for these coefficients are presented in section \ref{sec:Kubo}.

In flat backgrounds with no Luttinger potential the constitutive relations are
\begin{align}
	&\rho^0 = \rho, \qquad \qquad \rho^i = \rho u^i
	\qquad \qquad j^0 = q , 
	\qquad \qquad  \varepsilon_\text{lab}^0 = \frac{1}{2} \rho u^2 + \epsilon ,\nonumber \\
	 &j^i = q u^i + \sigma_E (  E^i + B \epsilon^{ij} u_j - T \partial^i \nu_Q ) + \tilde \sigma_E \epsilon^{ij} (  E_j + B \epsilon_{jk} u^k - T \partial_j \nu_Q ) \nonumber \\
	 &\qquad + \sigma_T  \partial^i T + \tilde \sigma_T  \epsilon^{ij} \partial_j T + \epsilon^{ij} \partial_j \tilde m , \nonumber \\
	 & \varepsilon_\text{lab}^i = \left( \frac{1}{2} \rho u^2 + \epsilon + p - \zeta \theta - \tilde f_B B \right) u^i - \eta \sigma^{ij} u_j - \tilde \eta \tilde \sigma^{ij} u_j  \nonumber \\
	 &\qquad + T \sigma_T (  E^i + B \epsilon^{ij} u_j - T \partial^i \nu_Q ) - T \tilde \sigma_T \epsilon^{ij} ( E_j + B \epsilon_{jk} u^k - T \partial_j \nu_Q ) \nonumber \\
	 &\qquad + \kappa_T  \partial^i  T + \tilde \kappa_T  \epsilon^{ij} \partial_j T -\tilde m \epsilon^{ij} (E_j + B \epsilon_{jk} u^k ) +  \epsilon^{ij} \partial_j  \tilde m_\epsilon  , \nonumber \\
	 & T_\text{lab}^{ij} = \rho u^i u^j + ( p - \zeta \theta - \tilde f_B B ) g^{ij} - \eta \sigma^{ij} - \tilde \eta \tilde \sigma^{ij} .
\end{align}
where $\rho^\mu$ is the mass current, $j^\mu$ the  charge current, $ \varepsilon_\text{lab}^\mu$ the energy current  and $ T^{ij}_\text{lab}$ the spatial stress in the lab frame, $u^i$ is the fluid velocity and the fluid shear $\sigma^{ij}$, expansion $\theta$, and vorticity $\omega$ are defined by
\begin{align}
	&\sigma_{ij} =\partial_i u_j + \partial_i u_i - \delta_{ij} \theta ,
	&&\theta = \partial_i u^i  ,
	&&\omega = \epsilon^{ij} \partial_i u_j.
\end{align}
In summarizing these results we have made the choice of fluid frame (\ref{frame}).


\subsubsection{Constitutive Relations}

The analysis proceeds much along the lines of the previous section, but a few words are due on the setup before we proceed. We are considering slightly out of equilibrium fluids to first order in a derivative expansion about flat spacetime and zero electromagnetic field. That is, the extended vielbein is $\df e^I = (dt ~ d x^a ~ 0)^T$ plus first order perturbations and
\begin{align}
	F_{\mu \nu}, 
	&&\nabla_\mu,
	&&T^I{}_{\mu \nu}
\end{align}
are all counted as first order in derivatives.

We are considering fluids with independently conserved mass and charge and so have five dynamical degrees of freedom, the temperature, a chemical potential for mass and charge each, and the fluid velocity
\begin{align}
	T , 
	&&\mu_Q , 
	&&\mu_M , 
	&&u^\mu,
\end{align}
where the velocity has been normalized so that $n_\mu u^\mu = 1$.
The equilibrium properties of the system are then completely characterized by an equation of state such as $p(T, \mu_Q , \mu_M )$, which expresses the pressure as a function of the state variables. The thermodynamic entropy, charge, mass, and energy densities are then defined by
\begin{align}\label{multComponentThermo}
	dp = s dT + q d \mu_Q + \rho d \mu_M,
	&&\epsilon + p = T s + q \mu_Q + \rho \mu_M .
\end{align}

The formulas will prove somewhat simpler if we instead take our independent variables to be
\begin{align}
	&T,
	&&\nu_Q = \mu_Q /T ,
	&&\nu_M = \mu_M /T .
\end{align}
In terms of $\nu_Q$ and $\nu_M$, the thermodynamic identities read
\begin{align}\label{thermo2}
	d p = \frac{\epsilon + p}{T} dT + T q d \nu_Q + T \rho d \nu_M,
	&&\frac{\epsilon + p}{T} =  s + \nu_Q q + \nu_M \rho .
\end{align}

Now consider the available first order data
\begin{align}
	\nabla_\mu T,
	&&\nabla_\mu \nu_Q,
	&& \nabla_\mu \nu_M,
	&&\nabla_\mu u^\nu,
	&&f_{\mu \nu},
	&&F_{\mu \nu},
	&&G_{\mu \nu} ,
\end{align}
where recall we have defined $\df T^I = \mathring u^I \df G + n^I \df f$. 
Now let's separate this data into irreducible representations of $SO(2)$.
\begin{center}
\begin{tabular}{ l | c c c}
		& Data 	  \\
	Scalar & $\theta \qquad b \qquad B \qquad \omega  $  \\
	&$( \dot T) \qquad ( \dot\nu_Q)  \qquad ( \dot\nu_M)$   \\
	Vector & $ \nabla^\mu T \qquad \nabla^\mu \nu_Q \qquad \nabla^\mu \nu_M \qquad (\alpha^\mu)$\\
		& $ e^\mu \qquad  E^\mu  \qquad G^\mu$ \\
	Symmetric Traceless Tensor & $\sigma^{\mu \nu}$ \\
\end{tabular}
\end{center}
where 
\begin{align}
	&E^\mu = {F^\mu}_\nu u^\nu ,
	&&B = \frac{1}{2} \varepsilon^{\mu \nu} F_{\mu \nu} ,\nonumber \\
	&e^\mu = {(d \overset u {\df a})^\mu}_\nu u^\nu ,
	&&b = \frac{1}{2} \varepsilon^{\mu \nu } (d \overset u {\df a})_{\mu \nu } ,\nonumber \\
	&G^\mu = - {G^\mu}_\nu u^\nu ,
	&&\theta = \nabla_\mu u^\mu ,\nonumber \\
	&\sigma^{\mu \nu} = \nabla^\mu u^\nu + \nabla^\nu u^\mu - h^{\mu \nu} \theta,
	&&\omega = \overset u \varepsilon_{\mu \nu} \nabla^\mu u^\nu ,\nonumber \\
	&\alpha^\mu = u^\nu \nabla_\nu u^\mu .
\end{align}
Recall that by $\varepsilon^{\mu \nu}$ we mean the ``spatial volume element'' $\varepsilon^{\mu \nu \lambda} n_\lambda$, while $\overset u \varepsilon_{\mu \nu} = \varepsilon_{\mu \nu \lambda} u^\lambda$, both of which are boost invariant, though the lowered index version is defined with respect to the comoving frame.

In the above we have also introduced the form $\overset u {\df a} = \mathring u_I \df e^I$. This is a boost invariant version of the Newtonian gravitational 3-potential. In components it reads
\begin{align}
	\overset u {\df a} = \df a + u_a \df e^a - \frac 1 2 u^2 \df n
\end{align}
and is simply $\df a$ measured in the comoving frame. The mass torsion $\df f$ is in fact completely determined by $d \overset u {\df a}$ and the data in the tensor $\nabla_\mu u^\nu$. To see this, use $\df f = \mathring u_I D \df e^I$ and equations (\ref{extendedShear}) and (\ref{DECOMP}) with $\omega_{\mu \nu} = \omega \overset u \varepsilon_{\mu \nu}$. One then finds
\begin{align}
	\df f &=  d \overset u {\df a} - D \mathring u_I \wedge \df e^I\nonumber \\
		&= d \overset u {\df a} - \df n \wedge \df \alpha - \omega \overset u {\df \varepsilon} .
\end{align}
We have chosen to use $d \overset u {\df a}$ as part of our independent data rather than $\df f$. Since $d \overset u {\df a}$ satisfies the Bianchi identity $d^2 \overset u {\df a} =0$, we will find below that this is the easier object to work with.

Not all this data is independent on shell. We may thus use the Navier-Stokes equation to eliminate one vector degree of freedom and one scalar each for mass conservation, charge conservation and the work-energy equation. The eliminated data is indicated by parentheses in the above table.

Finally, the most general first-order constitutive relations for the frame invariants consistent with spacetime symmetries are
\begin{align}
	&\mathcal S = - \zeta \theta - \tilde f_b b - \tilde f_B B - \tilde f_\omega \omega,\nonumber \\
	&\mathcal J^\mu = \sigma_e e^\mu + \sigma_E E^\mu + \sigma_G G^\mu  
	+ \sigma_T \nabla^\mu T + \sigma_Q \nabla^\mu \nu_Q + \sigma_M \nabla^\mu \nu_M ,\nonumber \\ 
	& \qquad \qquad +  \tilde \sigma_e  \tilde e^\mu +  \tilde \sigma_E \tilde  E^\mu +  \tilde \sigma_G \tilde  G^\mu  
	+  \tilde \sigma_T \tilde  \nabla^\mu T + \tilde  \sigma_Q \tilde  \nabla^\mu \nu_Q + \tilde  \sigma_M  \tilde \nabla^\mu \nu_M ,\nonumber \\
	&\mathcal E^\mu = \kappa_e e^\mu + \kappa_E E^\mu + \kappa_G G^\mu  
	+ \kappa_T \nabla^\mu T + \kappa_Q \nabla^\mu \nu_Q + \kappa_M \nabla^\mu \nu_M ,\nonumber \\ 
	& \qquad \qquad +  \tilde \kappa_e  \tilde e^\mu +  \tilde \kappa_E \tilde  E^\mu +  \tilde \kappa_G \tilde  G^\mu  
	+  \tilde \kappa_T \tilde  \nabla^\mu T + \tilde  \kappa_Q \tilde  \nabla^\mu \nu_Q + \tilde  \kappa_M  \tilde \nabla^\mu \nu_M ,\nonumber \\
	&\zeta^\mu = (\zeta_\theta \theta + \tilde \zeta_b b + \tilde \zeta_B B + \tilde \zeta_\omega \omega) u^\mu \nonumber \\
	&\qquad \qquad + \zeta_e e^\mu + \zeta_E E^\mu + \zeta_G G^\mu  
	+ \zeta_T \nabla^\mu T + \zeta_Q \nabla^\mu \nu_Q + \zeta_M \nabla^\mu \nu_M ,\nonumber \\ 
	& \qquad \qquad +  \tilde \zeta_e  \tilde e^\mu +  \tilde \zeta_E \tilde  E^\mu +  \tilde \zeta_G \tilde  G^\mu  
	+  \tilde \zeta_T \tilde  \nabla^\mu T + \tilde  \zeta_Q \tilde  \nabla^\mu \nu_Q + \tilde  \zeta_M  \tilde \nabla^\mu \nu_M ,\nonumber \\
	&\mathcal T^{\mu \nu} = - \eta \sigma^{\mu \nu} - \tilde \eta \tilde \sigma^{\mu \nu} .
\end{align}
In this we have defined the ``dual'' operation
\begin{align}
	\tilde v^\mu = \varepsilon^{\mu \nu} v_\nu ,
	&& \tilde w^{\mu \nu} = \varepsilon_\lambda{}^{(\mu } w^{\nu ) \lambda}
\end{align}
on vectors and symmetric two tensors. It has the properties
\begin{align}
	\tilde v_{1\mu} v_2^\mu = - v_{1 \mu} \tilde v_2^\mu ,
	&&\tilde w_{1\mu \nu} w_2^{\mu \nu} = - w_{1 \mu \nu} \tilde w_2^{\mu \nu}  .
\end{align}

\subsubsection{Entropy Current Analysis}

Now let's move on to determining those constraints that result from imposing the second law of thermodynamics
\begin{align}
	( \nabla_\mu - G_\mu ) s^\mu \geq 0.
\end{align}
Start by considering the genuine second order data in the entropy production, that is, terms that are not products of first order data
\begin{align}
	(\nabla_\mu - G_\mu ) \zeta^\mu \big|_{2 - \partial} &= \zeta_\theta \dot \theta + \tilde \zeta_\omega \dot \omega + ( \tilde \zeta_b - \tilde \zeta_e )\dot b  + ( \tilde \zeta_B - \tilde \zeta_E ) \dot B \nonumber \\
	&+ \zeta_T \nabla^2 T + \zeta_Q \nabla^2 \nu_Q + \zeta_M \nabla^2 \nu_M + \zeta_e \nabla_\mu e^\mu + \zeta_E \nabla_\mu E^\mu + \zeta_G \nabla_\mu G^\mu \geq 0 \nonumber
\end{align}
where we have used the Bianchi identities $d \df F = d \df G = d^2 \overset u {\df a} = 0$, which imply
\begin{align}
	&\varepsilon^{\mu \nu} ( \nabla_\mu - G_\mu ) E_\nu = - \dot B - B \theta ,
	&&\varepsilon^{\mu \nu} \nabla_\mu  G_\nu = 0 ,
	&&\varepsilon^{\mu \nu} ( \nabla_\mu - G_\mu ) e_\nu = - \dot b - b \theta.
\end{align}

We conclude that all the coefficients listed above must vanish and so $\zeta^\mu$ only has contributions from the remaining parity odd part
\begin{align}
	\zeta^\mu &= \tilde \zeta_B ( B u^\mu 
	+ \tilde  E^\mu ) + \tilde \zeta_b (b u^\mu + \tilde e^\mu ) +  \tilde \zeta_G \tilde  G^\mu  
	+  \tilde \zeta_T \tilde  \nabla^\mu T + \tilde  \zeta_Q \tilde  \nabla^\mu \nu_Q + \tilde  \zeta_M  \tilde \nabla^\mu \nu_M ,
\end{align}

The rest of the divergence of $\zeta^\mu$ is then
\begin{align}
	( \nabla_\mu - G_\mu ) \zeta^\mu = 
	&- \left( T \partial_\epsilon p \partial_T \tilde \zeta_B + \frac{1}{T} \partial_q p \partial_Q \tilde \zeta_B + \frac{1}{T} \partial_\rho p \partial_M \tilde \zeta_B \right) B \theta \nonumber \\
	&- \left( T \partial_\epsilon p \partial_T \tilde \zeta_b + \frac{1}{T} \partial_q p \partial_Q \tilde \zeta_b + \frac{1}{T} \partial_\rho p \partial_M \tilde \zeta_b \right) b \theta \nonumber \\
	&+ \partial_T \tilde \zeta_B \tilde E^\mu \nabla_\mu T + \partial_Q \tilde \zeta_B \tilde E^\mu \nabla_\mu \nu_Q + \partial_M \tilde \zeta_B \tilde E^\mu \nabla_\mu \nu_M \nonumber \\
	&+ \partial_T \tilde \zeta_b \tilde e^\mu \nabla_\mu T + \partial_Q \tilde \zeta_b \tilde e^\mu \nabla_\mu \nu_Q + \partial_M \tilde \zeta_b \tilde e^\mu \nabla_\mu \nu_M \nonumber \\
	&+ ( \tilde \zeta_T + \partial_T \tilde \zeta_G ) \tilde G^\mu \nabla_\mu T + ( \tilde \zeta_Q + \partial_Q \tilde \zeta_G ) \tilde G^\mu \nabla_\mu \nu_Q + ( \tilde \zeta_M + \partial_M \tilde \zeta_G ) \tilde G^\mu \nabla_\mu \nu_M \nonumber \\
	&+ ( \partial_Q \tilde \zeta_T - \partial_T \tilde \zeta_Q ) \tilde \nabla^\mu T \nabla_\mu \nu_Q + ( \partial_M \tilde \zeta_T - \partial_T \tilde \zeta_M ) \tilde \nabla^\mu T \nabla_\mu \nu_M \nonumber \\
	&+ ( \partial_M \tilde \zeta_Q - \partial_Q \tilde \zeta_M ) \tilde \nabla^\mu \nu_Q \nabla_\mu \nu_M ,
\end{align}
which is supplemented by the canonical entropy production $s^\mu = s^\mu_\text{can} + \zeta^\mu$
\begin{align}
	( \nabla_\mu - G_\mu ) s^\mu_\text{can} &= \frac{1}{T} \zeta \theta^2 + \frac{1}{2 T} \eta \sigma_{\mu \nu} \sigma^{\mu \nu} \nonumber \\
	&+ \frac{1}{T} \sigma_E (E_\mu - T \nabla_\mu \nu_Q) (E^\mu - T \nabla^\mu \nu_Q ) - \frac{1}{T^2} \kappa_T ( \nabla_\mu T - T G^\mu ) ( \nabla^\mu T - T G^\mu ) \nonumber \\
	&+ \frac{1}{T} \tilde f_b b \theta + \frac{1}{T} \tilde f_B B \theta + \frac{1}{T} \tilde f_\omega \omega \theta 
	+ \frac{1}{T} \sigma_e e_\mu E^\mu  + \frac{1}{T} \kappa_e e^\mu G_\mu + \frac{1}{T} \left( \sigma_G + \kappa_E \right) E^\mu G_\mu  \nonumber \\
	&- \frac{1}{T^2} \kappa_e e^\mu \nabla_\mu T - \sigma_e e^\mu \nabla_\mu \nu_Q + \frac{1}{T} \left( \sigma_T - \frac{1}{T} \kappa_E \right) E^\mu \nabla_\mu T + \frac{1}{T} \sigma_M E^\mu \nabla_\mu \nu_M \nonumber \\
	&+ \left( \frac{1}{T} \kappa_Q - \sigma_G \right) G^\mu \nabla_\mu \nu_Q + \frac{1}{T} \kappa_M G^\mu \nabla_\mu \nu_M \nonumber \\
	&- \left( \sigma_T + \frac{1}{T^2} \kappa_Q \right) \nabla^\mu T \nabla_\mu \nu_Q - \frac{1}{T^2} \kappa_M \nabla^\mu T \nabla_\mu \nu_M - \sigma_M \nabla^\mu \nu_Q \nabla_\mu \nu_M  \nonumber \\
	&+ \frac{1}{T} \tilde \sigma_e \tilde e^\mu E_\mu + \frac{1}{T} \tilde \kappa_e \tilde e^\mu G_\mu - \frac{1}{T} ( \tilde \sigma_G - \tilde \kappa_E ) \tilde E^\mu G_\mu \nonumber \\
	&- \frac{1}{T^2} \tilde \kappa_e \tilde e^\mu \nabla_\mu T - \tilde \sigma_e \tilde e^\mu \nabla_\mu \nu_Q - \frac{1}{T} \left( \tilde \sigma_T + \frac{1}{T} \tilde \kappa_E \right) \tilde E^\mu \nabla_\mu T \nonumber \\
	 &- \left( \tilde \sigma_E + \frac{1}{T} \tilde \sigma_Q \right) \tilde E^\mu \nabla_\mu \nu_Q - \frac{1}{T} \tilde \sigma_M \tilde E^\mu \nabla_\mu \nu_M \nonumber \\
	 &- \frac{1}{T} \left( \tilde \kappa_T + \frac{1}{T} \tilde \kappa_G \right) \tilde G^\mu \nabla_\mu T - \left( \tilde \sigma_G + \frac{1}{T} \tilde \kappa_Q \right) \tilde G^\mu \nabla_\mu \nu_Q - \frac{1}{T} \tilde \kappa_M \tilde G^\mu \nabla_\mu \nu_M \nonumber \\
	 &- \left( \tilde \sigma_T - \frac{1}{T^2} \tilde \kappa_Q \right) \tilde \nabla^\mu T \nabla_\mu \nu_Q + \frac{1}{T^2} \tilde \kappa_M \tilde \nabla^\mu T \nabla_\mu \nu_M + \tilde \sigma_M \tilde \nabla^\mu \nu_Q \nabla_\mu \nu_M .
\end{align}
In the above we have made the identifications $- \frac{1}{T} \sigma_Q = \sigma_E $ and $- \frac{1}{T} \kappa_G  = \kappa_T $ so that the entropy production due to electrical and thermal conductivity factors into a perfect square.\footnote{What one should do is demand that the quadratic form defined by these transport coefficients be degenerate and positive semidefinite (degenerate so that that equilibrium solutions exist in non-zero background fields). This immediately gives these identities.}

Demanding the second law then requires
\begin{align}
	&\zeta \geq 0 ,\qquad
	\eta \geq 0 ,\qquad
	\sigma_E \geq 0 ,\qquad
	\kappa_T \leq 0,  \qquad
	\tilde \sigma_G = \tilde \kappa_E ,\nonumber \\
	&\sigma_e = \sigma_M = \kappa_M = \kappa_e = \tilde \sigma_e = \tilde \kappa_e = 0 ,
	\qquad \qquad \kappa_E = - \sigma_G = - \frac{1}{T} \kappa_Q  = T \sigma_T ,\nonumber \\
	&
	\tilde f_\omega = \tilde f_b = 0 ,\qquad
	\tilde f_B = T^2 \partial_\epsilon p \partial_T \tilde \zeta_B + \partial_q p \partial_Q \tilde \zeta_B + \partial_\rho p \partial_M \tilde \zeta_B ,\nonumber \\
	&\begin{pmatrix}
		\tilde \zeta_T + \partial_T \tilde \zeta_G \\
		\tilde \zeta_Q + \partial_Q \tilde \zeta_G \\
		\tilde \zeta_M + \partial_M \tilde \zeta_G
	\end{pmatrix} 
	=
	\begin{pmatrix}
		\frac{1}{T} \tilde \kappa_T + \frac{1}{T^2} \tilde \kappa_G \\
		\tilde \kappa_E + \frac{1}{T} \tilde \kappa_Q \\
		\frac{1}{T} \tilde \kappa_M
	\end{pmatrix},
	\qquad \qquad
	\begin{pmatrix}
		\partial_T \tilde \zeta_B \\
		\partial_Q \tilde \zeta_B \\
		\partial_M \tilde \zeta_B
	\end{pmatrix} 
	=
	\begin{pmatrix}
		\frac{1}{T} \tilde \sigma_T + \frac{1}{T^2} \tilde \kappa_E \\
		\tilde \sigma_E + \frac{1}{T} \tilde \sigma_Q \\
		\frac{1}{T} \tilde \sigma_M
	\end{pmatrix},
	\nonumber \\
	&\begin{pmatrix}
		\partial_Q \tilde \zeta_M - \partial_M \tilde \zeta_Q \\
		\partial_M \tilde \zeta_T - \partial_T \tilde \zeta_M \\
		\partial_T \tilde \zeta_Q - \partial_Q \tilde \zeta_T
	\end{pmatrix}
	=
	\begin{pmatrix}
		\tilde \sigma_M \\
		- \frac{1}{T^2} \tilde \kappa_M \\
		\frac{1}{T^2} \tilde \kappa_Q - \tilde \sigma_T
	\end{pmatrix}.
\end{align}

To untangle the differential constraints, begin by defining
\begin{align}\label{functions}
	&\tilde f = \tilde \zeta_B,
	&&T \tilde h_T = \tilde \zeta_T + \partial_T \tilde \zeta_G,
	&&T \tilde h_Q = \tilde \zeta_Q + \partial_Q \tilde \zeta_G + T \tilde f,
	&&T \tilde h_M = \tilde \zeta_M + \partial_M \tilde \zeta_G .
\end{align}
These then read
\begin{align}\label{pde}
	&\begin{pmatrix}
		\tilde h_T  \\
		\tilde h_Q  \\
		\tilde h_M
	\end{pmatrix} 
	=
	\begin{pmatrix}
		\frac{1}{T^2} \tilde \kappa_T + \frac{1}{T^3} \tilde \kappa_G \\
		\frac{1}{T} \tilde \kappa_E + \frac{1}{T^2} \tilde \kappa_Q + \tilde f\\
		\frac{1}{T^2} \tilde \kappa_M
	\end{pmatrix},
	\qquad \qquad
	\begin{pmatrix}
		\partial_T \tilde f \\
		\partial_Q \tilde f \\
		\partial_M \tilde f
	\end{pmatrix} 
	=
	\begin{pmatrix}
		\frac{1}{T} \tilde \sigma_T + \frac{1}{T^2} \tilde \kappa_E \\
		\tilde \sigma_E + \frac{1}{T} \tilde \sigma_Q \\
		\frac{1}{T} \tilde \sigma_M
	\end{pmatrix},
	\nonumber \\
	&\begin{pmatrix}
		\partial_Q \tilde h_M - \partial_M \tilde h_Q \\
		\partial_M \tilde h_T - \partial_T \tilde h_M \\
		\partial_T \tilde h_Q - \partial_Q \tilde h_T
	\end{pmatrix}
	=
	\begin{pmatrix}
		\frac{1}{T} \tilde \sigma_M - \partial_M \tilde f\\
		\frac{1}{T} \tilde h_M - \frac{1}{T^3} \tilde \kappa_M \\
		- \frac{1}{T} \tilde h_Q + \frac{1}{T^3} \tilde \kappa_Q - \frac{1}{T}\tilde \sigma_T + \partial_T \tilde f + \frac{1}{T} \tilde f
	\end{pmatrix} .
\end{align}
This leads to several consistency relations on the four functions (\ref{functions})
\begin{align}\label{consistency}
	&\begin{pmatrix}
		\partial_Q \tilde h_M - \partial_M \tilde h_Q \\
		\partial_M \tilde h_T - \partial_T \tilde h_M \\
		\partial_T \tilde h_Q - \partial_Q \tilde h_T
	\end{pmatrix}
	=
	\begin{pmatrix}
		0\\
		0 \\
		0
	\end{pmatrix} .
\end{align}
The first comes from comparing the final component of the second equation to the first component of the third while the second follows from comparing the final component of the first equation to the second of the third. The final condition results from combining the second component of the first equation, the first component of the second, and the final component of the third.
The vector $( \tilde h_T ~ \tilde h_Q ~ \tilde h_M )^T$ is then curl free as so must be the gradient of some function $\tilde g ( T , \nu_Q , \nu_M )$
\begin{align}
	\tilde h_T = \partial_T \tilde g,
	&&\tilde h_Q = \partial_Q \tilde g ,
	&&\tilde h_M = \partial_M \tilde g .
\end{align}

\subsubsection{Summary of Results}\label{subsec:fluid_summary_multi}
This solves the full set of restrictions imposed by the second law. Before summarizing results, the following redefinition of transport coefficients will simplify the final answer
\begin{align}
	&T \tilde f \rightarrow \tilde m,
	&&T^2 \tilde g \rightarrow \tilde m_\epsilon , \nonumber \\
	&\tilde \sigma_ T \rightarrow \tilde \sigma_T + \partial_T \tilde m,
	&&\tilde \kappa_T \rightarrow \tilde \kappa_T + \partial_T \tilde m_\epsilon . \nonumber \\
\end{align}
after which frame invariants are 
\begin{align}\label{frame invariants summary}
	&\mathcal T^{\mu \nu} = - \eta \sigma^{\mu \nu} - \tilde \eta \tilde \sigma^{\mu \nu} \qquad \qquad \qquad \qquad
	\mathcal S = - \zeta \Theta - \tilde f_B B \nonumber \\
	&\mathcal J^\mu = \sigma_E \left( E^\mu - T \nabla^\mu \nu_Q \right) + \sigma_T ( \nabla^\mu T - T G^\mu ) +  \tilde \sigma_E \left( \tilde  E^\mu - T \tilde \nabla^\mu   \nu_Q \right) + \tilde  \sigma_T ( \tilde  \nabla^\mu T - T \tilde  G^\mu ) \nonumber \\
	& \qquad \qquad   - \tilde m  \tilde G^\mu + \tilde \nabla^\mu \tilde m \nonumber \\
	&\mathcal E^\mu = T \sigma_T \left( E^\mu - T \nabla^\mu \nu_Q \right) +  \kappa_T  ( \nabla^\mu T - T G^\mu )   -T \tilde \sigma_T \left( \tilde E^\mu - T \tilde \nabla^\mu \nu_Q \right) +  \tilde \kappa_T  ( \tilde \nabla^\mu T - T \tilde G^\mu )     \nonumber \\
	&\qquad \qquad  
	- \tilde m \tilde E^\mu - 2 \tilde m_\epsilon \tilde G^\mu  + \tilde \nabla^\mu \tilde m_\epsilon .
\end{align}
This gives the set of response coefficients listed at the beginning of this section.

To get a feel for these results, it's helpful to fix a fluid frame and write them for the non-covariant currents defined in (\ref{stress-energy}). We choose our frame so the physical mass, charge and energy correspond with the thermodynamic ones and the velocity is that of the mass current
\begin{align}\label{frame}
	\mathcal Q = \varrho =  \mathcal E = 0, && \mu^\mu = 0 .
\end{align} 
The frame invariants $\mathcal J^\mu$ and $\mathcal E^\mu$ are then simply the first-order deviations $\nu^\mu $ and $\xi^\mu$. In components, the currents are then
\begin{align}
	&\rho^t = \rho, \qquad \qquad  \rho^i = \rho u^i
	\qquad \qquad j^t = q , 
	\qquad \qquad  \varepsilon^t_\text{lab}  = \frac{1}{2} \rho u^2 + \epsilon ,\nonumber \\
	 &j^i = q u^i + \sigma_E (e^\Phi  E^i + B \varepsilon^{ij} u_j - T \partial^i \nu_Q ) + \tilde \sigma_E \varepsilon^{ij} ( e^\Phi E_j + B \varepsilon_{jk} u^k - T \partial_j \nu_Q ) \nonumber \\
	 &\qquad+ \sigma_T e^\Phi \partial^i ( e^{-\Phi} T ) + \tilde \sigma_T e^\Phi \varepsilon^{ij} \partial_j (e^{-\Phi} T) + e^\Phi \varepsilon^{ij} ( e^{-\Phi} \tilde m ) , \nonumber \\
	 &\varepsilon^i_\text{lab}  = \left( \frac{1}{2} \rho u^2 + \epsilon + p - \zeta \theta - \tilde f_B B \right) u^i - \eta \sigma^{ij} u_j - \tilde \eta \tilde \sigma^{ij} u_j  \nonumber \\
	 &\qquad + T \sigma_T ( e^\Phi  E^i + B \varepsilon^{ij} u_j - T \partial^i \nu_Q ) -T \tilde \sigma_T \varepsilon^{ij} ( e^\Phi E_j + B \varepsilon_{jk} u^k - T \partial_j \nu_Q ) \nonumber \\
	 &\qquad   + \kappa_T e^\Phi \partial^i ( e^{-\Phi} T ) + \tilde \kappa_T e^\Phi \varepsilon^{ij} \partial_j ( e^{- \Phi} T ) 
	  - \tilde m \varepsilon^{ij} ( e^\Phi E_j + B \varepsilon_{jk} u^k ) + e^{2 \Phi} \varepsilon^{ij} \partial_j ( e^{-2 \Phi } \tilde m_\epsilon ) , \nonumber \\
	 & T^{ij}_\text{lab} = \rho u^i u^j + ( p - \zeta \theta - \tilde f_B B  ) h^{ij} - \eta \sigma^{ij} - \tilde \eta \tilde \sigma^{ij} .\label{eq:non-cov_currents_multi}
\end{align}

In this and what follows $E^i$ is defined to be the electric field in the lab frame so that the comoving electric field used earlier is\footnote{We hope the reader will forgive the notational dissonance.}
\begin{align}
	E_\mu = F_{\mu \nu} u^\nu =
	\begin{pmatrix}
		0 & - E_j \\
		E_i & B \varepsilon_{ij} 
	\end{pmatrix}
	\begin{pmatrix}
		e^\Phi \\
		u^j
	\end{pmatrix}
	=
	\begin{pmatrix}
		- E_j u^j \\
		e^\Phi E_i + B \varepsilon_{ij} u^j
	\end{pmatrix} .
\end{align}
$\theta$ and $\sigma^{ij}$ are the curved space quantities defined in (\ref{DECOMP}) .
Explicitly, these are
\begin{align}
	\nabla^i u^j &= \partial^i u^j + \Gamma^{ji}{}_\lambda u^\lambda = \partial^i u^j + \Gamma^{ji}{}_k u^k + e^\Phi \Gamma^{ji}{}_0 , \nonumber \\
	&=  \nabla^i u^j - u^j  \nabla^i \Phi - \frac{1}{2} \dot h^{ij} + e^\Phi  \nabla^{[j}( e^{- \Phi} u^{i]} )+ \frac{1}{2} \Omega^{ ij} , \nonumber \\
	&= e^\Phi  \nabla^{(i} ( e^{- \Phi} u^{j)} ) - \frac{1}{2}  \dot h^{ij} + \frac{1}{2} \Omega^{ ij} , \nonumber \\
	\implies \qquad  \sigma^{ij} &= e^\Phi  \nabla^i ( e^{-\Phi} u^j ) + e^\Phi  \nabla^j ( e^{-\Phi} u^i ) - \dot h^{ij}  - h^{ij} \theta
\end{align}
where $ \nabla_{i}$ is the covariant derivative on a spatial slice and
\begin{align}
	&\theta = \frac{1}{\sqrt{h} e^{- \Phi}} \partial_\mu ( \sqrt{h} e^{-\Phi} u^\mu) = e^\Phi   \nabla_i ( e^{- \Phi} u^i ) + \frac{1}{2} e^\Phi h^{ij} \dot h_{ij} .
\end{align}

\subsection{Kubo Formulas}\label{sec:Kubo}

In this section we collect the relevant Kubo formulas for calculating the response coefficients found in the previous two sections. They are stated in terms of the retarded correlators
\begin{align}
	G^{i j, k l} (t , \mathbf x) &= \frac{\delta \left \langle T^{ij}_\text{lab} ( t , \mathbf x) \right \rangle}{\delta h_{kl} ( 0, \mathbf 0 )}  = \left \langle \frac{\delta T^{ij}_\text{lab} (t , \mathbf x ) }{ \delta h_{kl} (0, \mathbf 0) }\right \rangle+ \frac 1 2 i \theta ( t ) \left \langle \left[ T^{ij}_\text{lab} (t , \mathbf x) , T^{kl}_\text{lab} (0, \mathbf 0) \right] \right \rangle ,\nonumber \\
	G^{\mu, \nu}_{jj} (t , \mathbf x) &=  \frac{\delta \left \langle j^{\mu} ( t , \mathbf x) \right \rangle}{\delta A_\nu ( 0, \mathbf 0 )}  = \left \langle \frac{\delta j^\mu (t , \mathbf x)}{\delta A_\nu (0, \mathbf 0)} \right \rangle + i \theta ( t ) \left \langle \left[ j^\mu (t , \mathbf x) , j^\nu (0, \mathbf 0) \right] \right \rangle , \nonumber \\
	G^{\mu ,\nu}_{j\varepsilon}(t , \mathbf x) &= \frac{\delta \left \langle j^{\mu} ( t , \mathbf x) \right \rangle}{\delta n_\nu ( 0, \mathbf 0 )}  = \left \langle \frac{\delta j^\mu (t , \mathbf x)}{\delta n_\nu (0, \mathbf 0)} \right \rangle - i \theta ( t ) \left \langle \left[ j^\mu (t , \mathbf x) , \varepsilon^\nu_\text{lab} (0, \mathbf 0) \right] \right \rangle , \nonumber \\
	G^{\mu, \nu}_{\varepsilon \varepsilon}(t , \mathbf x) &=  \frac{\delta \left \langle \varepsilon^\mu_\text{lab} ( t , \mathbf x) \right \rangle}{\delta n_\nu ( 0, \mathbf 0 )}  =\left \langle \frac{\delta \varepsilon^\mu_\text{lab} (t , \mathbf x)}{\delta n_\nu (0, \mathbf 0)} \right \rangle - i \theta ( t ) \left \langle \left[ \varepsilon^\mu_\text{lab} (t , \mathbf x) , \varepsilon^\nu_\text{lab} (0, \mathbf 0) \right] \right \rangle ,
\end{align}
and are to be evaluated on the trivial spacetime background $\df e^I = ( dt ~ d x^a ~ 0 )$ (and in the LLL case in a constant background magnetic field). We will denote their Fourier transforms by
\begin{align}
	G(\omega, \mathbf k ) = \int dt d^2 x e^{i (\omega_+ t - \mathbf k \cdot \mathbf x)}G(t , \mathbf x) ,
\end{align}
 where $\omega_+ = \omega + i \delta$ for small, positive $\delta$.

\subsubsection{LLL Fluid}

There are a very small number of coefficients available in the lowest Landau level theory. Let's begin with the viscosities
\begin{gather}\label{viscosities}
	\eta = - \lim_{\omega \rightarrow 0} 
\frac{\Pi_{ijkl}  G^{ij,kl} (  \omega )}{2 i \omega_+} - \frac{p_\text{int}}{i \omega_+ },\qquad
	\tilde \eta = - \lim_{\omega \rightarrow 0} \frac{\tilde 
\Pi_{ijkl}  G^{ij , kl}( \omega)}{2 i  \omega_+},\nonumber \\
	\zeta =  \lim_{\omega \rightarrow 0} 
\frac{ G_{i \phantom{i , }j}^{\phantom i i , 
\phantom j j}( \omega)}{2 i \omega_+} + \frac{\kappa^{-1} + p_\text{int}}{i \omega_+} , 
\end{gather}
where we have introduced the projectors
\begin{align}
	\Pi^{ijkl} = \delta^{i(k} \delta^{l)j} - \frac 1 2 \delta^{ij} \delta^{kl} , 
	&&\tilde \Pi^{ijkl} = \frac 1 2 \left( \delta^{i ( k } \epsilon^{l) j} +\delta^{j(k} \epsilon^{l)i} \right),
\end{align}
and $\kappa^{-1} = - V \frac{\partial p_\text{int}}{\partial V}$ is the inverse compressibility at fixed chemical potential and field strength $F_{\mu \nu}$ (not fixed $B$). To derive these we have used
\begin{align}
	\delta \sigma^{ij} = i \omega_+ \delta h^{ij} - \delta^{ij} \delta \theta,
	&&\delta \theta = - \frac 1 2 i \omega_+ \delta^{ij} \delta h_{ij} ,
\end{align}
under a variation $\delta h_{ij} e^{- i \omega_+ t}$.
We recommend \cite{Bradlyn:2012ea} for a careful discussion of these Kubo formulas.

There is also an energy magnetization
\begin{align}
	&\partial_\mu M_\epsilon= \lim_{\mathbf k \rightarrow \mathbf 0} \frac{i 
\varepsilon_{ij} k^i \mathcal G^{j,t}_{\varepsilon j} (\mathbf k)}{|\mathbf k|^2} + M, 
\end{align}
and the thermal conductivities
\begin{gather}
	\Sigma_T = \frac{1}{T} \lim_{\omega \rightarrow 0} 
\frac{\delta_{ij} G^{ij}_{\varepsilon \varepsilon} (  \omega 
)}{2  i \omega_+}, \nonumber \\
	T^2 c_{RL} + \frac{\mu}{B} ( \epsilon + p) - 2 M_\epsilon = - 
\lim_{\omega \rightarrow 0} \frac{\epsilon_{ij} 
G^{ij}_{\varepsilon \varepsilon} (  \omega )}{2 i  \omega_+},
\end{gather}
which one may obtain by noting that $G_i = - i \omega_+ \delta n_i$ under a variation of the clock form.

\subsubsection{Standard Fluid}

For the standard, parity braking, $(2+1)$-dimensional fluid, the transport coefficients we have found are for the most part quite familiar and have been subjected to extensive study in the literature and calculated for a number of systems.

First of all, we again have the viscosities (\ref{viscosities}) but with $p_\text{int} \rightarrow p$.
The equations for the conductivity and thermoelectric conductivities are prototypical examples and first found in \cite{Kubo:1957uq,Kubo:1957yq}
\begin{align}
	&\sigma_E = \lim_{\omega \rightarrow 0} \frac{\delta_{ij} G^{i,j}_{jj} ( \omega_+ )}{2 i \omega_+} ,
	&& \tilde \sigma_E = \lim_{\omega \rightarrow 0} \frac{\epsilon_{ij} G^{i,j}_{jj} ( \omega_+ )}{2 i \omega_+} , \nonumber \\
	&T \sigma_T = \lim_{\omega \rightarrow 0} \frac{\delta_{ij} G^{i,j}_{j\varepsilon} ( \omega_+ )}{2 i \omega_+} ,
	&& T \tilde \sigma_T + \tilde m = \lim_{\omega \rightarrow 0} \frac{\epsilon_{ij} G^{i,j}_{j\varepsilon} ( \omega_+ )}{2 i \omega_+} .
\end{align}
Kubo formulas for the thermal conductivities were first computed in the classic work \cite{Luttinger:1964zz}, where the Luttinger potential was introduced. We find them to be
\begin{align}
	&T \kappa_T = \lim_{\omega \rightarrow 0} \frac{\delta_{ij} G^{i,j}_{\varepsilon \varepsilon} ( \omega_+ )}{2 i \omega_+} ,
	&& T \tilde \kappa_T + 2 \tilde m_\epsilon = \lim_{\omega \rightarrow 0} \frac{\epsilon_{ij} G^{i,j}_{\varepsilon \varepsilon} ( \omega_+ )}{2 i \omega_+} .
\end{align}
The attentive reader may have noted that the the thermoelectric and thermal Hall conductivities differ from the parity odd response to the chemical and Luttinger potentials by magnetizations $\tilde m$ and $\tilde m_\epsilon$ respectively, unlike the standard formulas found in \cite{Kubo:1957yq,Luttinger:1964zz}. This is because these works assumed vanishing equilibrium currents, as pointed out in the footnote below equation (4.10) of \cite{Kubo:1957yq}. In general, pure curl persistent equilibrium currents may arise, and in this case the proper relationship is that given above.

The  Kubo formulas for $\tilde \sigma_T$ and $\tilde \kappa_T$ are completed by expressions for the magnetizations
\begin{align}
	& \tilde m - T \partial_T \tilde m = - \lim_{| \mathbf k | \rightarrow 0} \frac{i \epsilon_{ij} k^i G^{j,0}_{j \varepsilon} ( \mathbf k )}{|\mathbf k |^2} ,
	&& 2 \tilde m_\epsilon - T \partial_T \tilde m_\epsilon = - \lim_{| \mathbf k | \rightarrow 0} \frac{i \epsilon_{ij} k^i G^{j,0}_{\varepsilon \varepsilon} ( \mathbf k )}{|\mathbf k |^2} .
\end{align}

\begin{appendix}

\section{A Formal Construction}\label{app:Coset}
In this appendix we motivate the definition of a Bargmann geometry, given in chapter \ref{chap:Formalism}, via a formal symmetry breaking procedure. Since this procedure has been used to motivate previous attempts to define non-relativistic geometries that are not equivalent to our own, we feel an appendix dedicated to this topic may be of some interest. Mathematical preliminaries are established in section \ref{sec:Bargmann} and the analysis is performed in section \ref{sec:Coset2}, which concludes with a comparison to previous attempts.

\subsection{The Bargmann Group}\label{sec:Bargmann}

The Bargmann group is the non-relativistic analogue of the Poincar\'e group, including time and space translations in addition to the Galilean group $Gal(d)$. It can be concretely realized by it's action on $\mathbb R^{d+1}$
\begin{align}
	&\begin{pmatrix}
		t \\
		x^i
	\end{pmatrix}
	\rightarrow
	\begin{pmatrix}
		1 & 0 \\
		- k^i & R^i{}_j
	\end{pmatrix}
	\begin{pmatrix}
		t \\
		x^j
	\end{pmatrix}+
	\begin{pmatrix}
		a^0\\
		a^i
	\end{pmatrix}.
\end{align}
However, this symmetry is realized projectively for massive fields as recollected in section \ref{sec:ActionPrinciples}. This is similar to the case of spinning fields in the pseudo-Riemannian case, but there is one key difference. For spinor representations, one can pass to the universal cover of of the gauge group, which is then linearly realized. One cannot do this in our case since realizing the projective representation (\ref{projTransf}) as a true representation of some group alters the local structure: boosts and translations commute to a phase factor.
That is, we must pass to the central extension of the Bargmann group \cite{Bargmann:1954gh}.

The centrally extended Bargmann group, which we shall simply refer to simply as the Bargmann group $Barg(d)$ from now on, may be realized by it's action on $\mathbb R^{d+1} \times S^1$ and now includes translations along the internal circle in addition to spacetime translations
\begin{align}\label{BargmannAction}
	&\begin{pmatrix}
		t \\
		x^i \\
		\varphi
	\end{pmatrix}
	\rightarrow
	\begin{pmatrix}
		1 & 0 & 0 \\
		- k^i & R^i{}_j & 0 \\
		 -\frac{1}{2} k^2 &   k_k R^k{}_j & 1
	\end{pmatrix}
	\begin{pmatrix}
		t \\
		x^j \\
		\varphi
	\end{pmatrix}+
	\begin{pmatrix}
		a^0\\
		a^i \\
		a^M
	\end{pmatrix}.
\end{align}
where $\varphi \sim \varphi + 2 \pi / m$. It should be clear from (\ref{BargmannAction}) that spacetime translations plus $U(1)_M$ phase rotations form a normal subgroup of $Barg(d)$, which has the semidirect product decomposition
\begin{align}\label{productDecomp}
	Barg(d) = \left( \mathbb R^{d+1} \times  U(1)_M\right)\rtimes_E Gal(d) .
\end{align}
where $E$ indicates that $Gal(d)$ acts on $\mathbb R^{d+1} \times  U(1)_M$ in the extended representation.
Many of the novel features of the extended vielbein formalism follow from the need to pass to the centrally extended Bargmann group. In section \ref{sec:Coset2} we shall find the Newtonian gravitational potential will arise as a gauge field $\df a$ for translations along the $U(1)_M$ factor and it's collection into the extended representation along with the spacetime vielbein $\df e^A$ is dictated by (\ref{productDecomp}).

In it's standard basis, the Lie algebra of the Bargmann group is\footnote{Our conventions are that a finite rotation is executed by $e^{- \frac i 2 J^{ab} \theta_{ab}}$, a boost by $e^{- i k_a K^a}$, time translations by $e^{-i H a^0}$, space translations by $e^{- i a^a P_a}$, and $U(1)$ rotations by $e^{- i M a^M}$.}
\begin{align}\label{bargmannAlgebra}
	&[ J^{ab} , J^{cd} ] = i \left( \delta^{bc} J^{ad}  - \delta^{ac} J^{bd} - \delta^{bd} J^{ac} + \delta^{ad} J^{bc}  \right) \nonumber \\
	&[J^{ab} , P^c ] = i \left( \delta^{bc} P^a - \delta^{ac} P^b \right)
	&&[J^{ab} , K^c ] = i \left( \delta^{bc} K^a - \delta^{ac} K^b  \right) \nonumber \\
	&[P^a , K^b] = - i\delta^{ab} M
	&& [ H , K^a ] = i P^a 
\end{align}
where $J^{ab}= J^{[ab]}$ are rotation generators, $K^a$ are boosts, $P^a$ spatial translations, $H$ temporal translations, and $M$ is the mass operator that performs $U(1)_M$ rotations.

We shall find it convenient to restate these commutation relations in a manner that makes their transformation under the Galilean group manifest. We can do this by collecting generators together in the following way
\begin{align}\label{standardBasis}
	J^{AB} =
	\begin{pmatrix}
		0 & K^b \\
		- K^a & J^{ab}
	\end{pmatrix},
	&&P_I = 
	\begin{pmatrix}
		H & P_a & M
	\end{pmatrix} .
\end{align}
Then (\ref{bargmannAlgebra}) is equivalent to
\begin{gather}\label{covBargmann}
	[ J^{AB} , J^{CD} ] = i ( h^{BC} J^{AD}- h^{AC} J^{BD} - h^{BD} J^{AC} + h^{AD} J^{BC}   ) , \nonumber \\
	[J^{AB} , P_I ] = i (  \Pi^B{}_I P^A - \Pi^A{}_I P^B  ),
\end{gather}
where $P^A = \Pi^{AI} P_I$.

This is a sensible definition since these objects do in fact transform as indicated under the adjoint action of $Gal(d)$. An arbitrary Galilean group element is of the form
\begin{align}
	h = e^{- \frac i 2 J^{AB} \Theta_{AB}},
	&&
	\text{where}
	&&
	\Theta_{AB}
	=
	\begin{pmatrix}
		0 & k_b \\
		- k_a & \theta_{ab}
	\end{pmatrix} ,
\end{align}
and $\theta_{ab}$ is totally antisymmetric.
We may then calculate how $J^{AB}$ and $P_I$ transform using the Hausdorff formula and find
\begin{align}\label{generatorTransformation}
	&h J^{AB} h^{-1} =
	\Lambda^A{}_C \Lambda^B{}_D J^{CD}, 
	&&h P_I h^{-1} = (\Lambda^{-1})^J{}_I P_J ,
\end{align}
where $\Lambda^A{}_B$ and $\Lambda^I{}_J$ are simply the vector and extended representations of the Galilean group used in the previous sections
\begin{gather}\label{exponentials}
	\Lambda^A{}_B = (e^\Theta)^A{}_B=
	\begin{pmatrix}
		1 & 0 \\
		- \mathcal K^a & R ( \theta )^a{}_b 
	\end{pmatrix}, \nonumber \\
	\Lambda^I{}_J = (e^\Theta)^I{}_J =
	\begin{pmatrix}
		1 & 0 & 0 \\
		- \mathcal K^a & R ( \theta )^a{}_b & 0 \\
		- \frac 1 2 \mathcal K^2 & \mathcal K_c R(\theta)^c{}_b & 1 
	\end{pmatrix} , \nonumber \\
	\text{where}
	\qquad \Theta^A{}_B = h^{AC} \Theta_{CB}, \qquad \Theta^I{}_{J} = \Pi^{AI} \Pi^B{}_J \Theta_{AB}, \nonumber \\
	\text{and} \qquad \qquad R(\theta)^a{}_b = (e^\theta )^a{}_b,
	\qquad \qquad \mathcal K^a= \sum_{n=0}^\infty \frac{(\theta^n k)^a}{(n+1)!} .
\end{gather}

\subsection{Bargmann Symmetry Breaking}\label{sec:Coset2}
In this section we use a formal symmetry breaking procedure to generate Galilean invariant geometry beginning with the Bargmann group. It is a purely formal construction, one does not in actuality have a physical Bargmann bundle in the real world, but it is a quick and systematic way to generate consistent transformation laws and has been used by several authors in the past to motivate constructions of Galilean invariant geometry, so we feel it is worth taking the time to present in some detail.
Our approach differs from from these in a few key respects which we will outline at the end of the section.

In this approach, one considers a group $G$ and a connection one-form  $\df \omega_\mathfrak{g}$ valued in the Lie algebra $\mathfrak g$ of $G$. Under local transformations $g \in G$, the $\mathfrak g$-connection transforms as
\begin{align}
	\df \omega_\mathfrak{g} \rightarrow g ( \df\omega_\mathfrak{g} + d) g^{-1}.
\end{align}
Here $G$ is interpreted as the local symmetry group of some physical system. We imagine breaking it to a subgroup $H$ and seek to write down locally $H$ invariant actions.

Now suppose we can decompose the connection into a sum of broken and unbroken generators $\mathfrak g = \mathfrak b \oplus \mathfrak h$
\begin{align}
	\df \omega_\mathfrak{g} = \df \omega_{\mathfrak b} +\df  \omega_{\mathfrak h}
\end{align}
such that\footnote{This is in fact automatic for compact semisimple $G$, but is not guaranteed in the case of interest $G = Barg(d)$.}
\begin{align}\label{breakingCondition}
	[b,h]= b ~ \text{for all} ~ b \in \mathfrak b ~\text{and} ~ h \in \mathfrak h.
\end{align}
If this is the case, then these are good objects with which to construct invariant actions since the broken and unbroken parts of the connection transform under $h \in H$ covariantly and as an $\mathfrak h$-connection respectively
\begin{align}\label{transformationRules}
	\df \omega_\mathfrak{b} \rightarrow h\df   \omega_\mathfrak{b} h^{-1},
	&&\omega_\mathfrak{h} \rightarrow h (\df \omega_\mathfrak{h} + d) h^{-1} .
\end{align}
One may now easily construct gauge invariant actions by using $H$-covariant derivatives and the field strength
\begin{align}
	\df R_{\mathfrak{g}} \equiv d \df \omega_{\mathfrak{g}} + \frac 1 2 [ \df  \omega_{\mathfrak{g}}  , \df \omega_{\mathfrak{g}} ],
	&&\df R_\mathfrak{g} \rightarrow h \df R_\mathfrak{g} h^{-1} .
\end{align}

As an example, let's perform this procedure on the Poincar\'e group, breaking it to the Lorentz group. In this case $\mathfrak g$ is the Poincar\'e algebra
\begin{align}\label{poincareAlgebra}
	&[ J^{ab} , J^{cd} ] = i \left( \eta^{bc} J^{ad} - \eta^{ac} J^{bd} - \eta^{bd} J^{ac}  + \eta^{ad} J^{bc}  \right) \nonumber \\
	&[J^{ab} , P_c ] = i \left(  \delta^{b}{}_{c} P^a  -\delta^{a}{}_c P^b \right)
\end{align}
which decomposes as
$\mathfrak{poin}(d) = \mathfrak t \oplus \mathfrak{so}(d,1)$, where $\mathfrak{t} = \text{Span}(P_a)$ is the closed subalgebra of translations. Due to the semidirect product structure $Poin(d) = T \rtimes_v SO(d,1)$, where $v$ denotes the vector representation, the condition (\ref{breakingCondition}) is automatically satisfied. The $\mathfrak{poin}(d)$ valued form then naturally breaks up into a translation part and a Lorentz part
\begin{align}
	\df \omega_{\mathfrak{poin}} = \df e^a P_a - \frac i 2 \df \omega_{ab} J^{ab}.
\end{align}

By virtue of the transformations $h P_a h^{-1} = (\Lambda^{-1})^b{}_a P_b$, $h J^{ab} h^{-1} = \Lambda^a{}_c \Lambda^b{}_c J^{cd}$, we find that the coefficients $\df e^a$ and $\df \omega^a{}_{b}$ then transform as indicated
\begin{align}
	\df e^a \rightarrow \Lambda^a{}_b \df e^b,
	&&\df \omega^a{}_b = \Lambda^a{}_c \df \omega^c{}_d ( \Lambda^{-1})^d{}_c + \Lambda^a{}_c d ( \Lambda^{-1} )^c{}_d .
\end{align}
The Poincar\'e curvature is then
\begin{gather}
	\df R_\mathfrak{poin} = \df T^a P_a - \frac i 2 \df R_{ab} J^{ab}, \nonumber \\
	\text{where}
	\qquad \qquad \df T^a = D \df e^a,
	\qquad \qquad \df R^a{}_b = d \df \omega^a{}_b + \df \omega^a{}_c \wedge \df \omega^c{}_b ,
\end{gather}
exactly reproducing the geometrical structure of torsionful relativistic spacetimes. We would like to emphasize however that this is a purely formal procedure used to generate a set of consistent transformation laws and define the geometry, not a derivation of the kinematics of general relativity since there is no physical Poincar\'e bundle.\footnote{In the language of the coset construction, there are no Nambu-Goldstone bosons that $\df e^a$ and $\df \omega^a{}_b$ are secretly functions of and there are no $g \in G$ transformations we are trying to realize nonlinearly.}

This in mind, let's take $G$ to be the centrally extended Bargmann group and break it to a subgroup $H$. Since we seek geometries with Galilean symmetry, $H$ must at least include the Galilean group. We must also be careful to satisfy the condition (\ref{breakingCondition}), which is automatic for compact semisimple $G$ but not for us. However, since the centrally extended Bargmann group has semidirect product structure $Barg(d) = (\mathbb R^{d+1} \times  U(1) )\rtimes_E Gal(d)$, we can guarantee that this is satisfied by taking the broken generators to include both spacetime translations $( H ~ P_a)$ and the mass generator $M$ to be in the broken subalgebra, which as we have already noted, transform among each other in the extended representation $P_I$.
If we did not include $M$, the commutation relation $[P_a , K_b] = - i \delta_{ab} M$ would violate the consistency of our procedure.

We then have $\mathfrak{g} = \mathfrak{barg}(d)$, $\mathfrak h = \mathfrak{gal} (d)$, while $\mathfrak b$ is the commutative algebra of translations and $U(1)_M$ rotations. In the standard basis (\ref{bargmannAlgebra}), (\ref{standardBasis}) for the Bargmann algebra this reads
\begin{align}\label{mauerCartanForm}
	\df \omega_{\mathfrak{barg}} = \df e^I P_I - \frac i 2 \df \omega_{AB}  J^{AB} 
\end{align}
where $\df \omega_{AB} = \df \omega_{[AB]}$.
The transformation rules (\ref{transformationRules}) along with (\ref{generatorTransformation}) immediately imply that $\df e^I$ and $\df \omega_{AB}$ transform as their index placement suggests. That is, under $h = e^{- \frac i 2 J^{AB} \Theta_{AB}}$ we have
\begin{align}\label{omegaTransf}
	\df e^I \rightarrow \Lambda^I{}_J \df e^J,
	&&\df \omega_{AB} \rightarrow \df \omega_{CD} ( \Lambda^{-1} )^C{}_A  ( \Lambda^{-1} )^D{}_B  - d \Theta_{AB}.
\end{align}
To obtain the Galilean spin connection considered previously, we merely need raise the first index on $\df \omega_{AB}$ and the result transforms as expected
\begin{align}
	\df \omega^A{}_B \rightarrow \Lambda^A{}_C \df \omega^C{}_D (\Lambda^{-1})^D{}_B + \Lambda^A{}_C d ( \Lambda^{-1})^C{}_B .
\end{align}
That data in $\df \omega^A{}_B$ and $\df \omega_{AB}$ is however equivalent.
Similarly, the field strength contains the torsion and curvature tensors
\begin{align}
	&\df R_\mathfrak{barg} = \df T^I P_I - \frac i 2 \df R_{AB} J^{AB}, \nonumber \\
	\text{where} \qquad \qquad
	&\df T^I = D \df e^I,
 	\qquad \qquad
	\df R_{AB} = d \df \omega_{AB}  +\df  \omega_{AC} \wedge \df \omega^C{}_B .
\end{align}
The coset construction thus precisely reproduces the geometry given in section \ref{sec:ExtendedVielbein}.

Finally, let's take a moment for comparison with previous applications of this construction. To our knowledge it was first applied in \cite{Brauner:2014jaa}, but with a different symmetry breaking pattern in which $M\in \mathfrak h$ and boosts are broken. This breaking pattern is consistent with (\ref{breakingCondition}) and so yields sensible results, but as noted in appendix B of \cite{Jensen:2014aia}, the geometry is rather applicable to the case of broken Galilean symmetry. The resulting data includes an invariant vector field $v^\mu$ and thus a preferred family of observers from the outset and a number of the results of Galilean invariance, such as the equality of the inertial and gravitational masses, are violated.
The analysis given in \cite{Jensen:2014aia} then proceeds using the same breaking pattern as given above and arrives at the same results we have, but the author discards the mass torsion $\df T^M$, which we have seen above breaks Galilean invariance on torsionful spacetimes and so results in a non-invariant connection. This approach has also been used to gauge the Sch\" odinger algebra as in \cite{Bergshoeff:2014uea}, though in this application the authors then impose various constraints on the curvature and torsion that we have left unfixed.
Very recently, the authors of \cite{Karananas:2016hrm} performed the above analysis including dilations, and their results indeed reduce to (\ref{mauerCartanForm}) when dilations are discarded.

\section{Symmetries of the Riemann Tensor}\label{app:Riemann}

In the main text, we required the symmetries of the Newton-Cartan Riemann tensor to derive equation (\ref{cauchyEquation}). These identities involve a few subtleties not present in the pseudo-Riemannian case, so we collect their derivations here. Since we are interested in the Ward identities on unrestricted Bargmann geometries, we will present these symmetries on spacetimes with general extended torsion $\df T^I$. They are
\begin{gather}
	  R_{(AB)\mu \nu} =   R_{AB(\mu \nu)}= 0,\label{id1} \\
	R^A{}_{[BCD]} = \frac 1 3 ( D \df T^A )_{BCD}, \label{id2}  \\
	R_{IJKL} = R_{KLIJ} + \frac 1 2 \left( (D \df T_I )_{JKL} + ( D \df T_J )_{KIL} + ( D \df T_K )_{ILJ} + (D \df T_L )_{IJK} \right)\label{id3} ,  \\
	D_{[\mu|}   R_{AB| \nu \lambda ]} = T^\rho{}_{[\mu \nu|}   R_{AB| \lambda ] \rho} . \label{id4}
\end{gather}
The derivation of (\ref{cauchyEquation}) requires only the first three of these identities, but we include the Bianchi identity for completeness. Contracting equation (\ref{id2}) with $\delta^B{}_A$, we also find
\begin{align}\label{RicciSymmetry}
	2 R_{[ \mu \nu]} = 3 \nabla_{[\mu} T^\lambda{}_{\lambda \nu]} + T^\lambda{}_{\lambda \rho} T^\rho{}_{\mu \nu}.
\end{align}

The first identity follows trivially from the definition of $\df {  R}_{AB}$ while the derivations of (\ref{id2}) and (\ref{id4}) from
\begin{align}
	D \df T^A = \df R^A{}_B \wedge \df e^B,
	&&\text{and}
	&&D \df {  R}_{AB} = 0
\end{align}
are identical to the pseudo-Riemannian case. The only identity that requires some care is (\ref{id3}), which is most easily stated when valued in the extended representation. By $R^I{}_{JKL}$ we mean the curvature two-form, valued in the extended representation of $\mathfrak{gal}(d)$, with spacetime indices pulled back to the extended representation using the Galilean invariant projector 
\begin{align}
	R^I{}_{JKL} = ( \df R^I{}_J )_{\mu \nu} \Pi^\mu{}_I \Pi^\nu{}_J .
\end{align}
Since all indices in equation (\ref{id3}) are $n^I$ orthogonal, it is simply the pullback of an equation valued in the covector representation of $Gal(d)$
\begin{gather}\label{indexSym2}
	  R_{ABCD} =   R_{CDAB} + \frac 1 2 \left( (D \df T_A )_{BCD} + ( D \df T_B)_{CAD} + ( D \df T_C )_{ADB} + (D \df T_D )_{ABC} \right) ,
\end{gather}
where we have used $\df R_{IJ} = \df {  R}_{AB} \Pi^A{}_I \Pi^B{}_J$ and defined $D\df T_A$ by $D \df T_I = \Pi^A{}_I D \df T_A$.
However, the proof of this identity is most naturally carried out in it's extended form. 

One should take care here, since $D \df T_A$ is not the exterior derivative of the (in general Galilean non-covariant) two-form $\df T_A = ( \df T^M ~ \df T_a)$. In components
\begin{align}
	D \df T_A = 
	\begin{pmatrix}
		d \df T^M - \df \varpi^b \df \wedge T_b ,&
		d \df T_a - \df T_b \wedge \df \omega^b{}_a + \df T^0 \wedge \df \varpi_a
	\end{pmatrix}
\end{align}
whereas the exterior derivative of $\df T_A$ is
\begin{align}
	\begin{pmatrix}
		d \df T^M - \df \varpi^b \wedge \df T_b ,&
		d \df T_a -\df  T_b \wedge \df \omega^b{}_a
	\end{pmatrix}.
\end{align}
They are the same if and only if the temporal torsion vanishes, in which case we also have $\df T^I = \Pi^{AI} \df T_A$ and $\df T_A$ is indeed covariant. The notation $D\df T_A$ is still a useful one though since it's index raises to $D\df T^A$, which is truly the exterior derivative of the two-form $\df T^A$
\begin{align}
	D\df T^A = h^{AB} D \df T_B ,
\end{align}
as one may explicitly check in components.

To prove (\ref{id3}) we begin with
\begin{align}
	D \df T^I = \df R^I{}_J \wedge \df e^J ,
\end{align}
which written in tensor notation reads
\begin{align}
	(\df R_{IJ})_{[\mu \nu} e^J_{\lambda ]} = \frac 1 3 (D \df T_I)_{\mu \nu \lambda}.
\end{align}
Now let's pull this back to an equation involving only extended indices using $\Pi^\mu{}_I$. One may check by an explicit computation in components that
\begin{align}
	e^I_\mu \Pi^\mu{}_J = \delta^I{}_J + n^I a_J
\end{align}
where $a_I = ( a_A ~ -1 )$.\footnote{One can check by hand that $a_I$ indeed transforms covariantly as indicated.} Since $\df R_{IJ} n^J = 0$, the second term drops out and we find
\begin{align}
	R_{I[JKL]} = \frac 1 3 ( D \df T_I )_{JKL} .
\end{align}
This is simply the extended index version of (\ref{id2}), which one can obtain from here by noting that both sides are $n^I$ orthogonal in all their indices. (\ref{id3}) then follows exactly as in the pseudo-Riemannian case by repeated applications of this equation along with
\begin{align}
	R_{(IJ)KL} = R_{IJ(KL)} = 0 .
\end{align}
\end{appendix}

\bibliographystyle{JHEP}
\bibliography{GalileanRefs}

\end{document}